\documentclass[%
  reprint,
  superscriptaddress,
  floatfix,
  amsmath,
  nofootinbib,
  amsmath,amssymb,
  aps,
  prd,
]{revtex4-2}

\usepackage{dcolumn}
\usepackage{bm}
\usepackage[dvipsnames]{xcolor}
\usepackage{placeins}
\usepackage{array}
\usepackage[normalem]{ulem}
\usepackage{graphicx}
\graphicspath{./Images/}
\usepackage{wrapfig}
\usepackage{float}
\usepackage{caption, subcaption}
\usepackage[export]{adjustbox}
\usepackage[toc, page]{appendix}

\usepackage{upgreek}
\usepackage[bottom]{footmisc}
\usepackage{hyperref}
\usepackage{dblfloatfix}

\usepackage{makecell,tabularx}
\setcellgapes{3pt}

\definecolor{kAzure}{HTML}{2671D4}
\definecolor{kSeaBlue}{HTML}{050DA3}
\definecolor{kDarkBlue}{HTML}{0C0463}
\definecolor{kLava}{HTML}{CC1104}
\definecolor{kIndigo}{HTML}{9446B8}

\hypersetup{colorlinks=true,linktoc=all,pdftitle={HNL in Beamlines},
  linkcolor=kSeaBlue,
  citecolor=kAzure,
  urlcolor=kSeaBlue}

\newcommand{\Lagr}{\ensuremath{\mathcal{L}}}

\newcommand{\ord}[1]{\ensuremath{\mathcal{O}\left({#1}\right)}}

\begin{document}

\title{Modelling heavy neutral leptons in accelerator beamlines}

\newcommand{\oxford}{University of Oxford, Department of Physics, Oxford, OX1 3PJ United Kingdom}
\newcommand{\warwick}{University of Warwick, Department of Physics, Coventry, CV4 7AL United Kingdom}

\newcommand{\JAIPCP}{AIP Conf. Proc.}
\newcommand{\JANNREV}{Annu. Rev. Nuc. Part. Sci.}
\newcommand{\JAPJ}{Astrophys. J.}
\newcommand{\JCPC}{Comput. Phys. Commun.}
\newcommand{\JEPJC}{EPJ \textbf{C}}
\newcommand{\JEPJST}{EPJ Special Topics}
\newcommand{\JFIP}{Front. Phys.}
\newcommand{\JIJMPE}{Int. J. Mod. Phys. \textbf{E}}
\newcommand{\JIN}{Instruments}
\newcommand{\JJETP}{JETP}
\newcommand{\JJHEP}{J. High Energy Phys.}
\newcommand{\JJINST}{JINST}
\newcommand{\JJPG}{J. Phys. \textbf{G}.}
\newcommand{\JMNRAS}{Mon. Not. R. Astron. Soc.}
\newcommand{\JNATURE}{Nature}
\newcommand{\JNIMA}{Nucl. Instrum. Methods \textbf{A}}
\newcommand{\JNP}{Nat. Phys.}
\newcommand{\JNPB}{Nuc. Phys. \textbf{B}}
\newcommand{\JNJP}{New. J. Phys.}
\newcommand{\JNUCB}{Nuc. Phys. \textbf{B}}
\newcommand{\JPLB}{Phys. Lett. \textbf{B}}
\newcommand{\JPOS}{PoS}
\newcommand{\JPPNP}{Prog. Part. Nuc. Phys.}
\newcommand{\JPRB}{Phys. Rev. \textbf{B}}
\newcommand{\JPRC}{Phys. Rev. \textbf{C}}
\newcommand{\JPRD}{Phys. Rev. \textbf{D}}
\newcommand{\JPREP}{Phys. Rep.}
\newcommand{\JPRL}{Phys. Rev. Lett.}
\newcommand{\JPTEP}{Prog. Theor. Exp. Phys.}
\newcommand{\JZHETP}{Sov. Phys. JETP}


\author{Komninos-John Plows}    \affiliation{\oxford} \email{komninos-john.plows@mansfield.ox.ac.uk}
\author{Xianguo Lu}       \affiliation{\warwick}   \email{xianguo.lu@warwick.ac.uk}


\begin{abstract}
  Heavy Neutral Leptons (HNLs) with masses \ord{0.1-1\,\,\text{GeV}/c^{2}} are promising candidates for the simultaneous explanation of the smallness of the observed neutrino masses as well as the matter-antimatter asymmetry in the observable Universe. These particles can be produced in the decay of hadrons typically produced in a neutrino beamline used for oscillation experiments, and have sufficient lifetime to propagate to a near detector, where they decay to observable particles. For the approximation of a single new mass eigenstate mixing with the Standard Model via the lepton mixing matrix, a simulation framework based on the \verb|GENIE| event generator has been developed. This module is designed to facilitate searches for HNL through a unified, minimal interface employing a detailed treatment of the kinematics and dynamics of massive unstable neutrinos, with a transparently organised suite of physics effects tracking the HNL from its production to its decay. These mechanisms are expounded on in the current work, underlining the rich landscape for novel, non-trivial physics that has already been identified in previous literature. This framework is an ongoing effort to provide a consistent and comprehensive description of heavy neutrinos from particle decays. We highlight use cases and future applications of interest to the accelerator neutrino community.
\end{abstract}

\maketitle

\section{Introduction} \label{SECT_Introduction}
Neutrino oscillations \cite{PonteFirstOscPaper, PonteOscPaper} were proposed early in the history of neutrino physics (see for example \cite{oscHistory} and references therein). 
However, it was the discovery of the deficit in solar neutrino events in the Davis experiment \cite{DavisSolarNeutrino} that first gave indications for physics beyond the Standard Model. 
The discovery of neutrino flavour conversion first in solar \cite{SNOOscDiscovery} and atmospheric \cite{SKOscDiscovery} neutrinos, and then in reactor antineutrinos \cite{KLNDOscDiscovery}, confirmed the hypothesis that neutrinos oscillate between flavour eigenstates \cite{FittingStatus2020}, which runs contrary to the Standard Model expectation that all neutrinos, being left-chiral, are massless.
The oscillation parameters (mass splittings, mixing angles, and CP violating phase $\delta_{CP}$) have since become the subject of precise measurement campaigns from a plethora of detectors from different baselines past \cite{SNOOscDiscovery, SKOscDiscovery, GALLEX, GNO, Kamiokande, Borexino, K2K, MINOS, OPERA, LSND, KARMEN, miniBooNE, CHOOZ, PaloVerde} and present \cite{KLNDOscDiscovery, MINOSPlus, T2KOscillation, NOvAOscillation, uBooNEExcess, DayaBay, RENO, DoubleCHOOZ}.
Further experiments are being designed to increase the precision to which the oscillation parameters are known \cite{SBNReview, DUNEPhysicsOverview, JUNOOverview, HyperKOverview}, and a comprehensive effort to produce global fits to extant data is underway \cite{FittingStatus2020}.
\par At the same time, the masses of the active neutrinos, albeit nonzero, are very small. Direct measurements using beta decay at KATRIN (\cite{Katrin0.8eVNature}) currently yield a bound on the ``electron neutrino mass" of $\left(\sum_{i} \left|U_{ei}\right|^{2}m_{\nu_{i}}^{2}\right)^{1/2} \lesssim 0.8\,\,\text{eV}$, where $U$ is the $3\times3$ PMNS matrix and $i$ runs over the three known mass eigenstates. 
At the same time, indirect bounds on the sum of the masses $\sum_{i} m_{\nu_{i}}$ within a 7-parameter cosmological model ($\Lambda$CDM + $\sum_{i} m_{\nu_{i}}$) are even more stringent. 
Constraints on the cosmic microwave background from Planck 2018 data are currently at $\sum_{i} m_{\nu_{i}} < 0.26\,\,\text{eV}$, dropping to $<0.13\,\,\text{eV}$ when combined with measurements of the scale of baryon acoustic oscillations \cite{PDG2022}.
Indirect bounds on neutrino mass via the ``effective mass" $|m_{\beta\beta}| = \left|\sum_{i} U_{ei}^{2}m_{\nu_{i}}\right|$ may also be obtained from neutrinoless double-beta decay experiments \cite{FormaggioDirectMass}, under the assumption that neutrinoless double-beta decay is mediated by active Majorana neutrinos only. 
\par To explain the non-zero albeit small observed neutrino masses, a seesaw mechanism of mass generation is usually invoked \cite{SeesawAndOscillations, Abada_ReviewOfSeesaws, ColliderTestsOfSeesaws}.
Seesaw mechanisms posit new neutrino degrees of freedom, which carry no charge under electroweak symmetry $SU(2)_{L}\times U(1)_{Y}$: that is, the new neutrino states are sterile.
Notably, both Dirac and Majorana fields allow for viable seesaws. 
The mass eigenstates associated with these new fields, whose scale is dependent on the details of the seesaw mechanism, are termed heavy neutral leptons, or HNL.
These mix with the light neutrino mass states into the observed flavour eigenstates $\nu_{e, \mu, \tau}$, and therefore can enter into interactions involving visible Standard Model particles.
The smallness of the active neutrino masses implies, for a wide variety of seesaw models \cite{AsakaPhysLetB, LeptogenesisTypeI, BrdarTypeI}, that either the HNL Yukawa couplings are very small, or that their masses are much larger than currently accessible experimental scales.
However, so-called ``low-scale seesaw" models allow for HNL with masses in the range \ord{100\,\,\text{MeV}/c^{2} - 1\,\,\text{TeV}/c^{2}} \cite{Boyarsky2019,Shaposhnikov_2007,BERNABEU1987303,NeutrinosAndColliders,WhitePaperSterileNus,Ibarra2011}.
\par Apart from the ability to explain the existence and smallness of observed neutrino masses, HNL can also mediate the observed matter-antimatter asymmetry in the Universe \cite{BaryogenesisMixing, BaryogenesisNeutrinoOsc}.
Moreover, there are scenarios (\cite{GorbunovNuMSM, AsakaPhysLetB}) in which an HNL of keV-scale mass can act as a warm-dark-matter component, explaining at least part of the dark matter content in the Universe.
In fact, a detected unidentified emission line at $E_{\gamma} \simeq 3.55\,\,\text{keV}$ in stacked galactic spectra obtained with the \textit{XMM-Newton} instrument \cite{3.5keVLine} had been proposed to result from the decay of a $7.1\,\,\text{keV}/c^{2}$ mass HNL, but subsequent searches for the line in different astronomical features have not found significant evidence for an excess at $3.55\,\,\text{keV}$ \cite{LineFail1, LineFail2}.
This realisation has motivated searches for HNL at accelerator facilities by means of searches for HNL decays \cite{uBooNEMuPiHNL, Porzio2019, HPSandHNL, Goodwin2022, T2KMainHNLSearch, SKAtmHNL, Acciarri2021, LHCbHNL}, searches for anomalous peaks in the spectra of leptons from meson decays \cite{E949, PIENU_Ue42, PIENU_Um42, PS191_first, PS191_second, NA62} (see \cite{SnowmassHNLOverview} for an overview), as well as significant interest for searches in future experiments \cite{HyperKDesignDoc, DUNEBSMOverview, DuneNDBSM, Ballett2017, SHiPSensitivity, SHiPKaons, LeptonFDHNL, SnowmassForwardLHCHNL, Cerci_2022, FASERHNL}.
In addition to these searches, a series of precision measurements of the SM predictions for the decays of known particles, such as $\mu, \tau, \pi, K$, can be used to derive bounds on the HNL parameter space (see \cite{BrymanModelIndependentBounds} for an overview).
\par In the case of decay searches, it is crucial to accurately model the flux of HNL at the detector, as well as maintain precise control on the backgrounds to detection channels.
Standard Model neutrino measurements make use of a suite of ``standard" neutrino event generators.
These support wide-ranging analyses from sensitivity studies all the way through to cross section measurements to oscillation analyses (see \cite{UlrichOverview} for an overview).
This is not the case with new-physics signatures, only a limited selection of which is implemented in the tools currently available to the community \cite{EventGeneratorsOverviewTHISISWHYWEDOHNLGENIE}.
With the exception of the \verb|ACHILLES| \cite{Achilles} and \verb|DarkNews| \cite{DarkNews} generators, as well as of a partial implementation in the \verb|GENIE| \cite{GENIEMainPaper, GENIEManual} generator, there have been few efforts to incorporate HNL physics into event generators.
The \verb|GENIE| neutrino event generator is an event generator that is used by existing and future experiments.
It leverages a ROOT-based \cite{ROOT, ROOTMaster} C++ framework, including a series of mature flux and geometry drivers \cite{GENIEv3Updates}, to take advantage of object-oriented techniques in the creation of neutrino events.
\verb|GENIE| currently offers an implementation of short-lived HNL produced in Standard Model neutrino upscattering from interactions with the nucleus \cite{DarkNeutrino}, motivated by \cite{DarkNeutrinoPaper}, that however does not model long-lived HNL produced in the beamline.
\par For searches individually implementing HNL physics (see e.g. \cite{T2KMainHNLSearch, uBooNEMuPiHNL}), additional efforts to interface with the beamline simulation and implement the detector geometry are needed.
Additionally, HNL simulations must correctly describe the production kinematics of the HNL taking into account their non-negligible mass, as well as the propagation and decay of these particles.
Since HNL are unstable, this implies properly implementing the probability of decay inside (or immediately upstream) of the detector, as well as describing the angular dependence of the decay products induced by the polarisation of the HNL.
\par In this work, we not only take care of the beamline interface and detector integration in a fully controllable manner, but also investigate the kinematics and propagation/decay of HNL in detail.
Our long-lived HNL implementation starts from a general description of ``parent" particles (such as pseudoscalar mesons) travelling in a beamline.
It is a general framework that constructs HNL from the decay of these parents, calculates the decay rates to all kinematically accessible final states, and returns the decay products of HNL decays to those final states that the user wants to simulate, inside a general detector volume the user provides as input.
The choice of which channels are desirable, as well as the beamline simulation and the detector description, is fully configurable.
This has the advantage of factorising out the complexity of the beamline simulation and detector geometry, focusing instead on the core physics of massive neutrinos while simultaneously providing output that is usable for a wide array of beamlines and detector setups.
This allows for the use of this module for optimisation studies in a broad range of parameters of interest.
This framework takes care of all kinematical calculations, and it is our goal to support present and future HNL models via a replaceable module defined by the HNL production and decay rates.
We implement this framework in \verb|GENIE|; a quick guide to running the \verb|GENIE| implementation can be seen in Appendix \ref{appdx: code}.
\par The framework we introduce not only handles the various physics effects of massive neutrinos in a general way, but also produces standrd \verb|GENIE| output, enabling experiments to directly use its results in their Monte Carlo production chain.
\par This paper is organised as follows. In Sections \ref{SECT_model} and \ref{SECT_overview}, we will detail the specific model treated in this implementation, and present a brief overview of the HNL simulation chains in previous searches. 
The model can be decomposed into the physical production mechanism from hadron decays following proton-target interactions, and the Lagrangian from which the production and decay rates into the kinematically accessible channels are calculated.
In Section \ref{SECT_simulation}, we will describe in detail the implementation of HNL production, propagation, and decay in the \verb|BeamHNL| module of \verb|GENIE| v3.
Next, in Section \ref{SECT_discussion}, we will expand on avenues for further implementation of novel physics effects, and routes for extension to more general HNL models. 
Finally, in Section \ref{SECT_conclusion}, we will shortly discuss possible use cases for the generation app that could be of interest to the community, for HNL searches in present and future experiments.

\section{The HNL model} \label{SECT_model}
\begin{figure}[ht]
  \centering
  \includegraphics[width=0.47\textwidth]{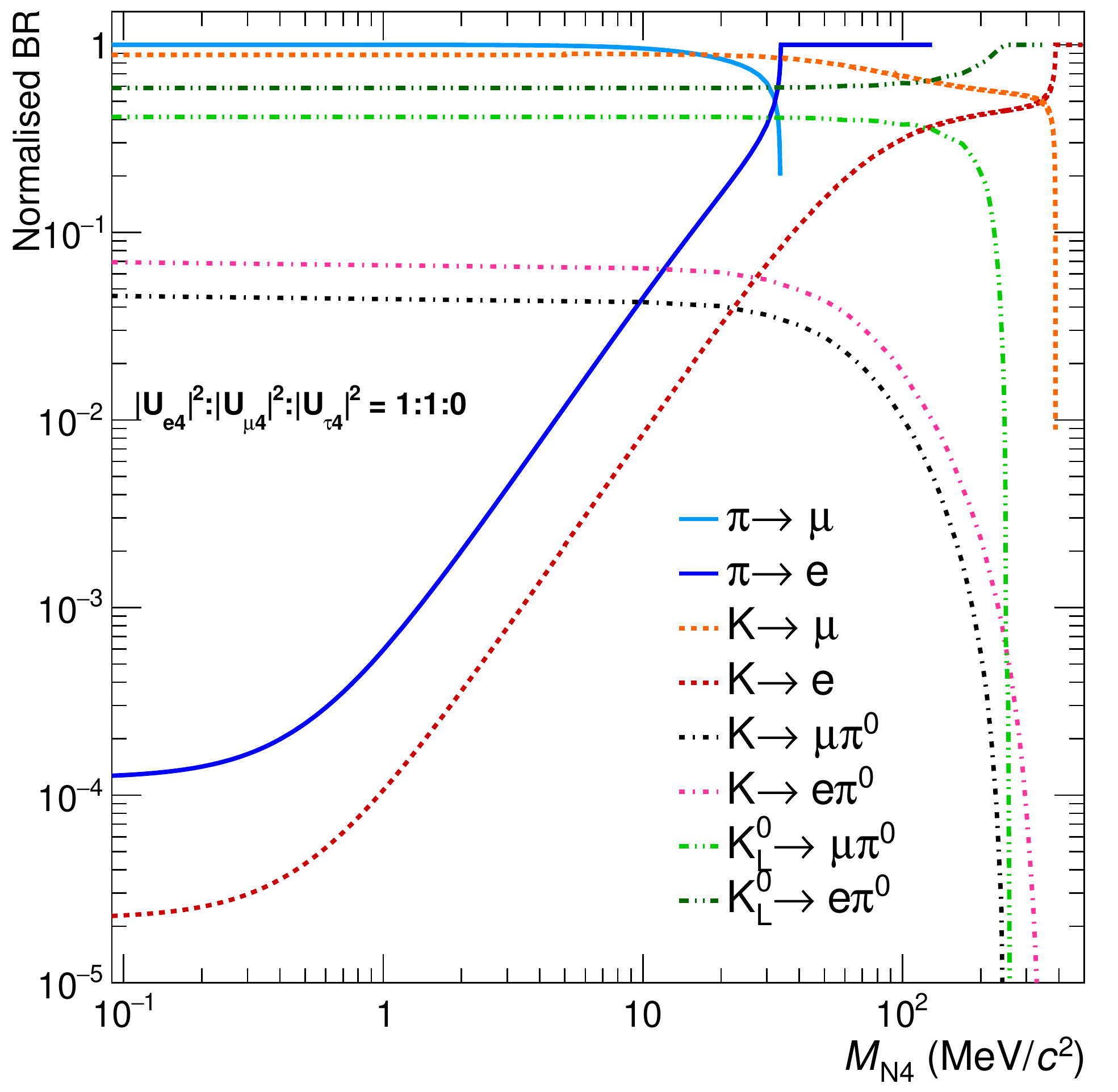}
  \caption{Normalised branching ratios for HNL production, $|U_{\tau 4}|^{2} = 0$.}
  \label{fig:prodBR}
\end{figure}
In the framework of a seesaw-mechanism neutrino mass generation, an extended mass matrix $\mathbb{M}$ sources the light neutrino masses \cite{BrdarTypeI, Abada_ReviewOfSeesaws, GorbunovNuMSM}.
Since the particle content of a seesaw mechanism generally includes new degrees of freedom in the neutrino sector (either Dirac states $\nu_{\textrm{R}}$ or Majorana states $\nu_{\textrm{L}}^{c}$, with $\textrm{L},\textrm{R}$ indicating the chirality of the neutrino fields), the full lepton mixing matrix $U$ has more than 3 rows:
\begin{equation}
  U \in \mathbb{C}^{(3+n)\times(3+n)},\quad\text{for}\,\,n\,\,\text{new neutrino fields.}
\end{equation}
These new fields are singlets under the Standard Model gauge group, which is to say, they are \emph{sterile} neutrinos.
\par The most general Lagrangian admits both Dirac and Majorana neutrino mass terms \cite{Drewes2013},
\begin{equation}\label{eq: mass_terms}
  \Lagr \supset -\frac{1}{2}\begin{pmatrix} \overline{\nu_{\textrm{L}}} &\overline{\nu_{\textrm{R}}^{c}} \end{pmatrix}\underbrace{\begin{pmatrix} 0 &m_{\textrm{D}} \\ m_{\textrm{D}}^{T} &m_{\textrm{M}} \end{pmatrix}}_{\mathbb{M}}\begin{pmatrix} \nu_{\textrm{L}}^{c} \\ \nu_{R} \end{pmatrix} + \,\,\text{h.c.}
\end{equation}
Diagonalising the full matrix in Eq. (\ref{eq: mass_terms}) yields $3 + n$ eigenvalues corresponding to the 3 known light neutrinos, and $n$ new neutrino fields.
Heavy Neutral Leptons are mass eigenstates, with masses $m_{\textrm{N}} \gg 1\,\,\text{eV}/c^{2}$; they are therefore \emph{nearly sterile}, and can interact with the Standard Model particle content through their mixing into the known neutrino flavour eigenstates $\alpha = e,\mu,\tau$,
\begin{equation}\label{eq: general_nu_mixing}
  \nu_{\alpha} = \sum_{i=1,2,3}U_{\alpha i}\nu_{i} + \sum_{j=4}^{3+n}U_{\alpha j}N_{j}.
\end{equation}
A direct consequence of introducing new degrees of freedom is that the $3\times 3$ PMNS mixing matrix $U^{\text{PMNS}} \subset U$ deviates from unitarity \cite{AntuschNonUnitaryMixing, PMNSNonUnitarityTypeI, T2KNonUnitarity}. 
From solar and atmospheric neutrino oscillation analyses it is known that there are two non-zero mass splittings \cite{NuFitPub} $\Delta m^{2}_{21} \simeq 7.4\times 10^{-5}\,\,\text{eV}^{2},\,\,\left|\Delta m^{2}_{31}\right|\simeq 2.5 \times 10^{-3}\,\,\text{eV}^{2}$.
Though the sign of $\Delta m^{2}_{31}$ remains unknown, leading to two possible neutrino hierarchies, the fact remains that at least two mass eigenstates $\nu_{i}$ must have $m_{\nu_{i}} \neq 0$ (it is currently unknown whether the lightest mass eigenstate is massive or not).
On the other hand, the most minimal HNL scenario, in which the lightest active neutrino is massless, corresponds to the case $n=2$.
In the most minimal extension to the Standard Model that admits neutrino masses (the $\nu$MSM, \cite{AsakaPhysLetB, GorbunovNuMSM, Shaposhnikov_2007}), only Majorana mass terms are added; however, there exist other seesaw mechanism implementations where Dirac masses are considered (see e.g. \cite{InverseDiracAndMajorana, LinearDiracSeesaw}).
\par Following the approach of \cite{T2KMainHNLSearch, uBooNEMuPiHNL} we will work in the simplified framework of considering one extra neutrino state ($n=1$), with the effective heavy neutrino $N_{4}$ controlled by the mixings $\left|U_{\alpha 4}\right|^{2}$.
This approach is valid in the case of two HNLs with similar masses, assuming no coherent oscillations between them.
(For a review of HNL oscillations, see for example \cite{TastetPol, HNLOsc} and references therein).
The HNL in this minimal picture has no transition magnetic moment.
There is, however, significant interest in the properties of neutrinos with transition magnetic moments, in the so-called ``dipole portal" (the reader may refer to \cite{NDipoleUpscattering, DipoleHNL, DipoleNuTail}).
The general mixing for the HNL into flavour eigenstates Eq.~(\ref{eq: general_nu_mixing}) is now written as
\begin{equation}\label{eq: fourth_nu_mixing}
  \nu_{\alpha} = \sum_{i=1,2,3}U_{\alpha i}\nu_{i} + U_{\alpha 4}N_{4}.
\end{equation}
\par In principle, the above information is sufficient to describe the production of $N_{4}$ from various different mechanisms.
These include active-to-sterile neutrino oscillations \cite{SterileOscillations}, upscattering from neutrino-nucleus interactions in a material \cite{IceCubeDoubleBangs}, decays of charged hadrons \cite{ColomaDUNEHNL, SKAtmHNL}, or direct production at colliders \cite{Colliders1, Colliders2, Colliders3, Colliders4}.
\par We have implemented in \verb|GENIE| the case of hadron decay into HNL, and the subsequent HNL decay into SM particles, within the framework of the effective field theory described in \cite{ColomaDUNEHNL}. 
This theory describes the interactions of HNL with mesons directly at the diagram level, leading to HNL production and decay rates that are valid for HNL mass ranges up to about $1\,\,\text{GeV}/c^{2}$ and for energies of \ord{\text{GeV}}.
Generally, it is possible to study the production and decays of HNL with larger masses, where production from heavy meson decays such as $D, D_{s}$ as well as from $\tau$ and baryons is possible \cite{Bondarenko_HNLPheno}.
In this work, we have implemented HNL production from pions and kaons, as well as from muons, which are the dominant particles produced in current accelerator neutrino beamlines; we consider HNL with masses up to the kaon mass, for which there exist 10 lepton-number-conserving channels.
We summarise the method used to obtain HNL production and decay rates in Appendix \ref{appdx: QM}.
\par At the end of this Section, we provide for convenience the production and decay channels in tabular form.
In Table \ref{tab: prodChannels}, we list the HNL production channels, along with the mass thresholds for channel activation, the ``kinematic scaling" factor
\begin{equation}\label{eq: kineScaling}
  \mathcal{K} = \frac{1}{\left|U_{\alpha 4}\right|^{2}}\frac{\Gamma(P \rightarrow N_{4} + \ell_{\alpha} + ...)}{\Gamma(P \rightarrow \nu + \ell_{\alpha} + ...)},
\end{equation}
and the Standard Model branching ratio $\mathfrak{B}$ for each channel with $N_{4} \rightarrow \nu$.
By $\nu$ we mean the appropriate combination of $U_{\alpha i}\nu_{i}$ with $\alpha$ being the correct flavour for the channel.
In Table \ref{tab: decayChannels} we list the decay channels, their mass thresholds, the expressions used to calculate the decay rate for a Dirac HNL (see Appendix \ref{appdx: QM} for definitions of the expressions $\mathcal{C}, \mathcal{D}_{\ell}, \mathcal{F}, \mathcal{G}, \mathcal{P}_{\ell}, \mathcal{S}_{P\ell}, \delta_{a}^{b}$, and the multiplier used for the Majorana decay rates for each channel).
\par
\begin{widetext}
  \begin{figure*}[!t]
    \centering
    \begin{subfigure}[b]{0.32\textwidth}
      \centering
      \includegraphics[width=\textwidth]{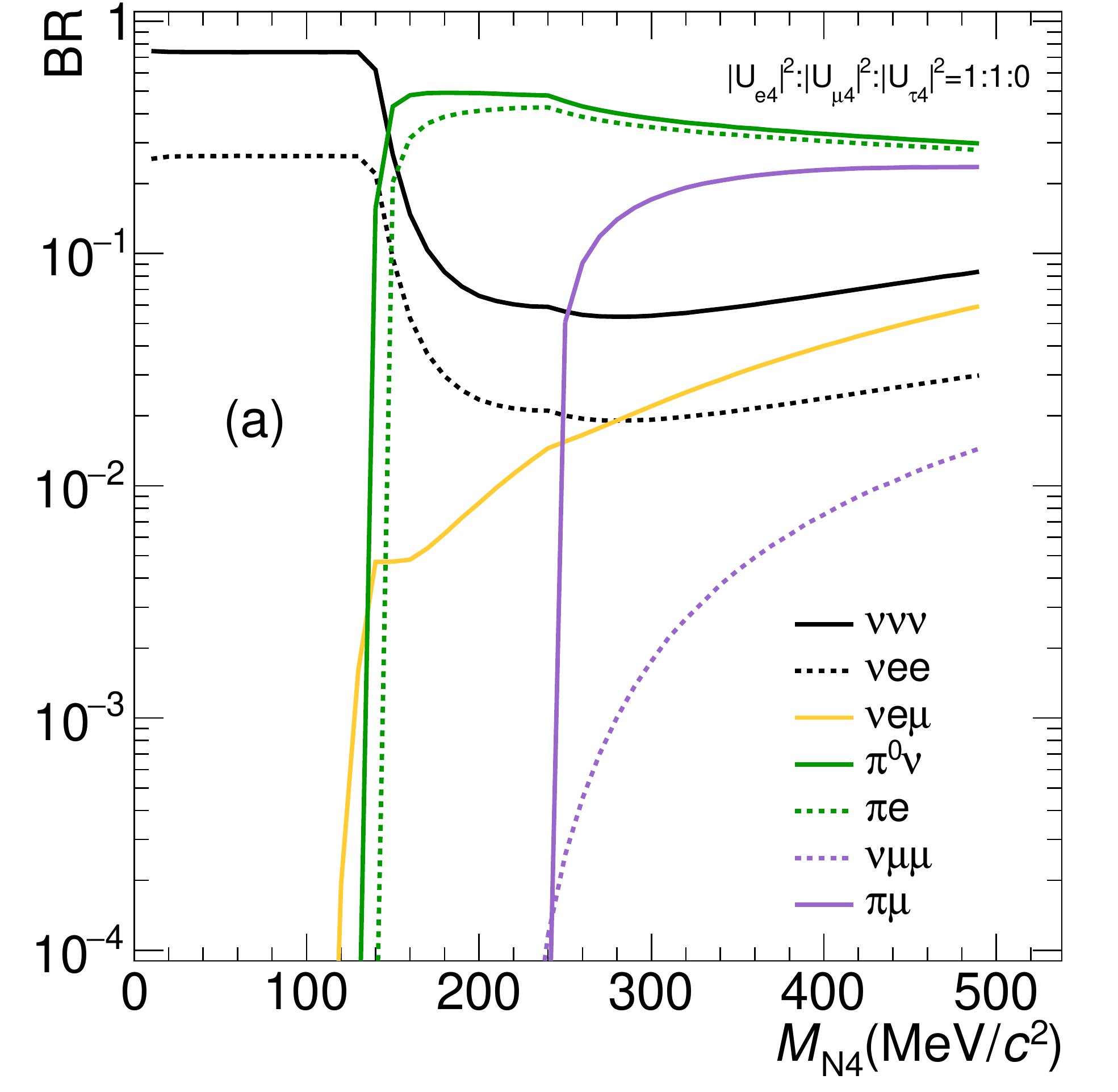}
    \end{subfigure}
    \hfill
    \begin{subfigure}[b]{0.32\textwidth}
      \centering
      \includegraphics[width=\textwidth]{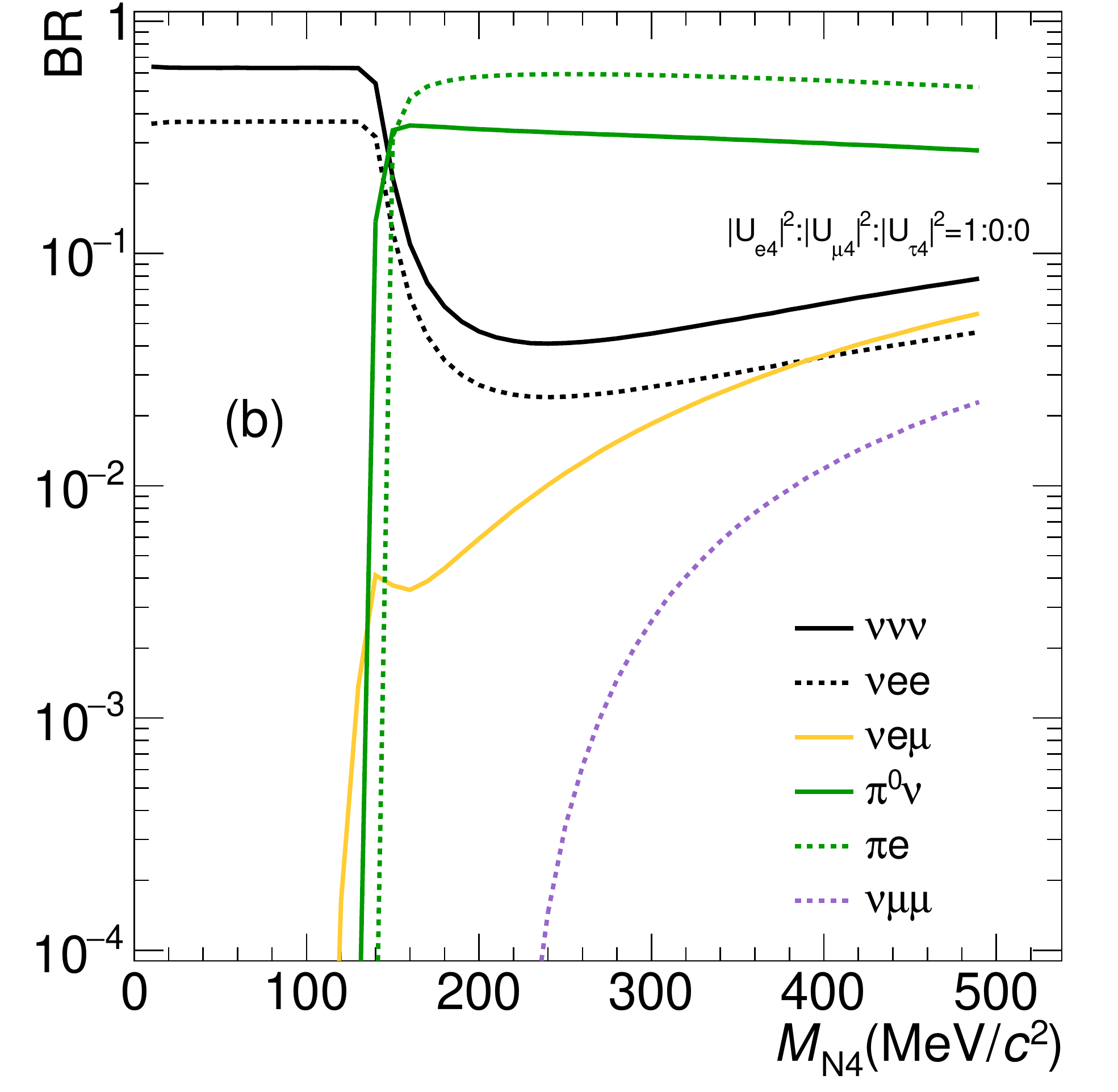}
    \end{subfigure}
    \hfill
    \begin{subfigure}[b]{0.32\textwidth}
      \centering
      \includegraphics[width=\textwidth]{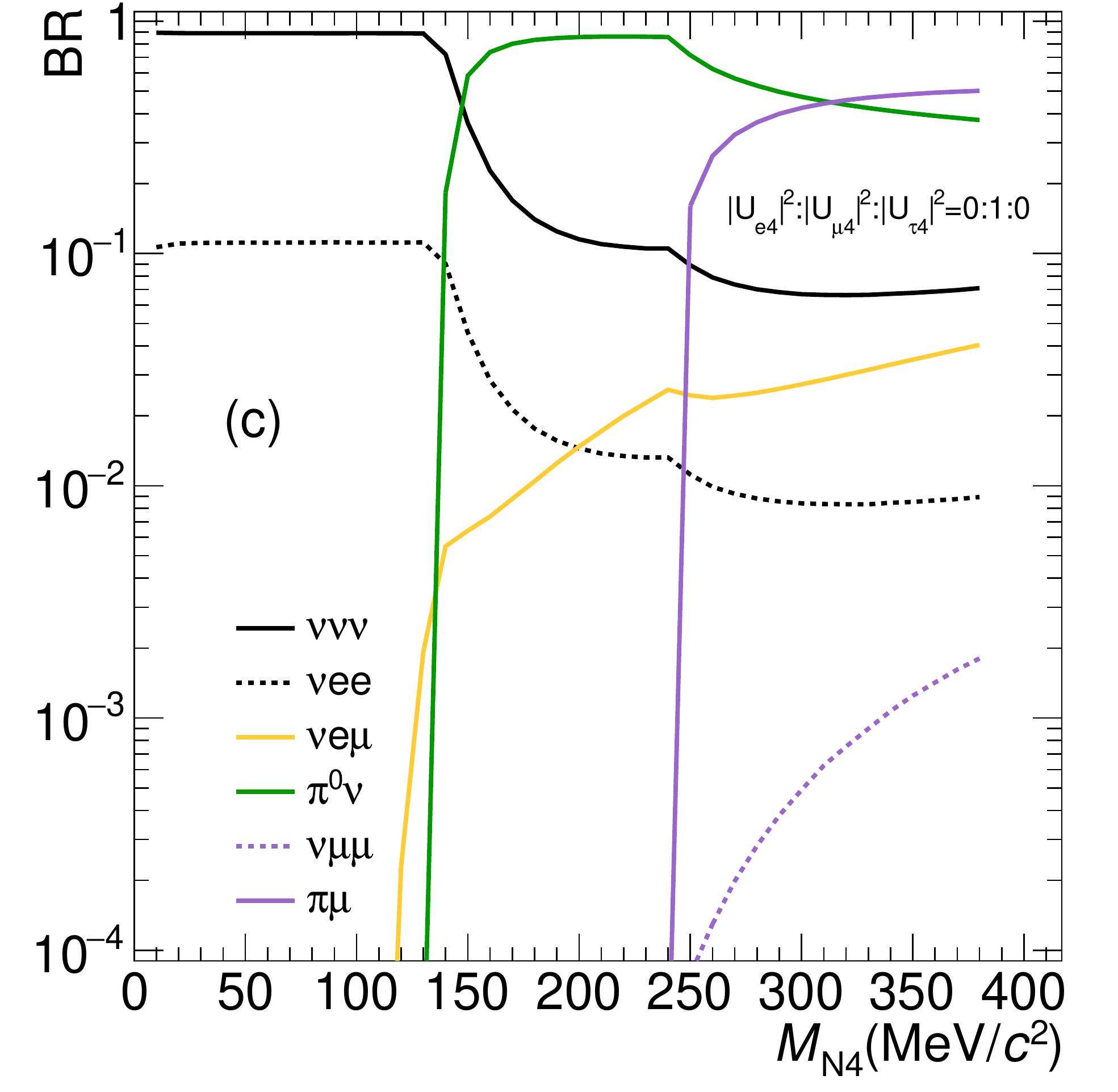}
    \end{subfigure}
    \setlength{\abovecaptionskip}{-1pt}
    \caption{Branching ratios for HNL decays, $\left|U_{\tau 4}\right|^{2} = 0$.}
    \label{fig:BR}
    \vspace*{-2em}
  \end{figure*}
\end{widetext}

 The production and decay ``normalised branching ratios'' of HNL are shown in Figs. \ref{fig:prodBR} and \ref{fig:BR}.
The normalisation of these ``branching ratios'' is over all the available production (decay) channels with an HNL in the final (initial) state.
This means that in Fig. \ref{fig:prodBR}, the sum of normalised branching ratios for each different parent is equal to 1, which is of course unphysical.
The physical case of an HNL mixing weakly with the light neutrino sector implies that, to obtain HNL production rates, all these normalised branching ratios should be corrected by the overall factor \cite{Shrock1981}
\begin{equation}\label{eq: pseudounitarity}
  \frac{\sum_{\alpha = e,\mu,\tau}\left|U_{\alpha 4}\right|^{2}}{1 - \sum_{\alpha = e,\mu,\tau}\left|U_{\alpha 4}\right|^{2}} \simeq \sum_{\alpha = e,\mu,\tau}\left|U_{\alpha 4}\right|^{2},
\end{equation}
where the denominator corresponds to the contribution from the active neutrino components.
\par Fundamentally, there is a difference between the behaviour of two-body production channels (e.g. $K \rightarrow N_{4} + \mu$, shown as ``$K \rightarrow \mu$" in the legend - the same convention is applied throughout Fig. \ref{fig:prodBR}) and three-body channels (e.g. $K \rightarrow N_{4} + \mu + \pi^{0}$). 
This is due to the breakdown of the helicity suppression mechanism that dominates the two-body production modes.
In the Standard Model, the masslessness of the neutrino suppresses decays to electron + neutrino compared to muon + neutrino by a factor $\ord{m_{\mu}^{2}/m_{e}^{2}}$ due to the neutrino being a helicity state as well as a left-chiral state.
HNL, with masses $\gtrsim \ord{100\,\,\text{MeV}/c^{2}}$, are not helicity eigenstates, and this suppression drops off with increasing mass, as shown in \cite{Shrock1981}.
Three-body decays, on the other hand, have no helicity suppression mechanism, and the HNL production rate is instead controlled just by the reduction of phase space. 
\par For the decay branching ratios, one generally has semileptonic two-body decays and (semi)leptonic three-body decays; however, the semileptonic three-body decays $N_{4} \rightarrow \ell^{\mp} \pi^{\pm} \pi^{0}, N_{4} \rightarrow \nu \pi^{0} \pi^{0}$ have negligibly small branching ratios for $M_{\textrm{N}4} < m_{\textrm{K}}$, confirming the argument made in \cite{ColomaDUNEHNL} that these channels are essentially mediated by the emission of an on-shell $\rho$ meson, $N_{4} \rightarrow \ell^{\mp} \rho^{\pm}, \rho^{\pm} \rightarrow \pi^{\pm} \pi^{0}$ (and similarly for the two-neutral-pion decay).
This assumption was checked using the Mathematica packages \verb|FeynRules| \cite{FeynRulesManual}, \verb|FeynArts| \cite{FeynArts}, and \verb|FeynCalc| \cite{FeynCalc1, FeynCalc2, FeynCalc3} and the Lagrangian published on the \verb|FeynRules| model database\footnotemark[1] \cite{FeynRulesWebsite}.
The BR for the decay $N_{4} \rightarrow \ell^{\mp} \pi^{\pm} \pi^{0}$ are $\lesssim 10^{-5}$, hence are irrelevant for the mass range $M_{\textrm{N}4} < m_{\textrm{K}}$. 
%
\section{Previous implementations of HNL simulation} \label{SECT_overview}
To fully exploit the wealth of parameter space that is accessible to accelerator neutrino experiments in searching for HNL, it is necessary to develop some commonly accepted prescription to produce well-understood and validated simulations of HNL.
Of especial interest is the ability to adequately describe any phenomena that arise from the massive nature of the HNL, particularly in the kinematics, as this could lead to powerful methods to remove background contributions, such as with a suitable timing trigger (see e.g. \cite{Porzio2019, Ballett2017}) or using appropriate final-state observables \cite{T2KMainHNLSearch, uBooNEMuPiHNL}. 
The suite of neutrino event generators commonly used by neutrino experiments to model neutrino-nuclear interactions \cite{GENIEMainPaper, GiBUU, NuWro, NEUT} have in common the desirable trait of each having a unified interface for all the various physics of Standard Model neutrinos \cite{EventGeneratorsOverviewTHISISWHYWEDOHNLGENIE}.
This allows for systematic and robust study of the complex nuclear environment that governs these interactions.
\setlength{\skip\footins}{5pt plus 10pt} 
\footnotetext{We used the relevant Feynman diagrams (see Fig. \ref{fig:pipi0ell} in Appendix \ref{appdx: QM}) to get the matrix element for the $N_{4} \rightarrow \ell^{\pm} \pi^{\pm} \pi^{0}$ decay, and obtained a differential decay rate $\textrm{d}^{2}\Gamma/\textrm{d}E_{\ell}\textrm{d}E_{\uppi\pm}$, which is integrated using Simpson's rule on the nested integral (see \cite{DoubleSimpson} Lecture 24).}
\par For BSM physics, though there has been considerable activity (particularly in the context of colliders) in unifying the production pipeline on the theory and phenomenology side \cite{SnowmassHNLOverview, Ruiz_2021, ColomaDUNEHNL, Degrande_2016}, progress for the few-GeV range of neutrinos is less rapid.
Codes to handle the production of HNL from a neutrino beam and their decays to visible final states, with varying degrees of complexity, have been implemented using either existing software frameworks such as \verb|PYTHIA| \cite{SHiPSensitivity, GorkavenkoPythiaHNL} and \verb|GEANT4| \cite{ColomaDUNEHNL}, purpose-built new simulation tools \cite{NuShock, DuneNDBSM, Porzio2019, DarkNews}, or applying an appropriate weighting procedure to existing beam simulations \cite{T2KMainHNLSearch}.
\begin{widetext}

  \begin{table*}[t!]
    \centering
    \makegapedcells
    \setlength\tabcolsep{8pt}
    \begin{tabularx}{0.7\linewidth}{c c c c}
      \Xhline{1.5pt}
      Channel &Threshold $(\text{MeV}/c^{2})$ &$\mathcal{K}$ \cite{Shrock1981, Ballett2020} &SM $\mathfrak{B}$ \\
      \Xhline{0.8pt}
      $\pi^{\pm}\rightarrow N_{4}+\mu^{\pm}$ &$33.91$ &$\mathcal{P}_{\mathcal{\ell}}\left(\delta_{m_{\uppi}}^{m_{\upmu}},\delta_{m_{\uppi}}^{M_{\textrm{N}4}}\right)$ &$0.999877$ \\
      $\pi^{\pm}\rightarrow N_{4}+e^{\pm}$ &$139.06$ &$\mathcal{P}_{\ell}\left(\delta_{m_{\uppi}}^{m_{\textrm{e}}},\delta_{m_{\uppi}}^{M_{\textrm{N}4}}\right)$ &$1.23\times10^{-4}$ \\
      $K^{\pm}\rightarrow N_{4}+\mu^{\pm}$ &$388.02$ &$\mathcal{P}_{\ell}\left(\delta_{m_{\textrm{K}}}^{m_{\upmu}},\delta_{m_{\textrm{K}}}^{M_{\textrm{N}4}}\right)$ &$0.6352$ \\
      $K^{\pm}\rightarrow N_{4}+e^{\pm}$ &$493.16$ &$\mathcal{P}_{\ell}\left(\delta_{m_{\textrm{K}}}^{m_{\textrm{e}}},\delta_{m_{\textrm{K}}}^{M_{\textrm{N}4}}\right)$ &$1.582\times 10^{-5}$ \\
      $K^{\pm}\rightarrow N_{4}+\mu^{\pm}+\pi^{0}$ &$253.04$ &$\mathcal{S}_{\textrm{K}^{+}\upmu}\left(M_{\textrm{N}4}\right)$ &$3.18\times 10^{-2}$ \\
      $K^{\pm}\rightarrow N_{4}+e^{\pm}+\pi^{0}$ &$358.18$ &$\mathcal{S}_{\textrm{K}^{+}\textrm{e}}\left(M_{\textrm{N}4}\right)$ &$4.82\times 10^{-2}$ \\
      $K^{0}_{L}\rightarrow N_{4}+\pi^{\pm}+\mu^{\mp}$ &$248.45$ &$\mathcal{S}_{\textrm{K}^{0}\upmu}\left(M_{\textrm{N}4}\right)$ &$0.2718$ \\
      $K^{0}_{L}\rightarrow N_{4}+\pi^{\pm}+e^{\mp}$ &$353.60$ &$\mathcal{S}_{\textrm{K}^{0}\textrm{e}}\left(M_{\textrm{N}4}\right)$ &$0.3878$ \\
      \Xhline{1.5pt}
    \end{tabularx}
    \caption{Production channels for HNL.}
    \label{tab: prodChannels}
    \vspace*{-2em}
  \end{table*}
\end{widetext}
There is an increasing need, then, to aid analyses aiming to look for HNL on two fronts: on the one hand, different analyses should be able to be compared straightforwardly for reproducibility and clarity; on the other, analysis resources should be conserved as much as possible, focusing on leveraging the unique capabilities of next-generation detectors for rare searches.
The current landscape of individual implementations of HNL in each experiment certainly stands to gain from such capacity, not least in the case of interfacing the work of theorists, phenomenologists, and experimentalists who are working toward the same goal. 
\par A natural expansion into the requirements outlined above is to adapt this new physics into commonly used generators, and in the case of HNL produced by neutrino upscattering has already been initiated both in \verb|GENIE| \cite{DarkNeutrino} and in the recent \verb|DarkNews| generator \cite{DarkNews}.
For the case of HNL from hadron decays, the event generation philosophy is different: instead of producing neutrino interactions based on the \emph{same} flux as for Standard Model neutrinos, HNL decays are produced from an \textit{a priori} very different flux, which results from the kinematics and dynamics of the massive neutrino; apart from the trivial scaling caused by the mixings to the SM, there is a dynamic shape distortion sourced by the generally slower, unstable particles that must reach the detector and decay inside without traversing it.
Furthermore, because HNL in this paradigm are produced directly from the hadron spectrum engendered in proton-target interactions, it is not sufficient merely to rescale the amount of recorded protons-on-target (POT) to estimate HNL event rates, or to reweight the SM neutrino flux; a full description of each HNL must be capable of tracking how many POT are being considered to avoid normalisation errors in the analysis downstream.
\section{Implementation in \small{GENIE} v3} \label{SECT_simulation}
\begin{figure}
  \centering
  \includegraphics[width=0.5\textwidth]{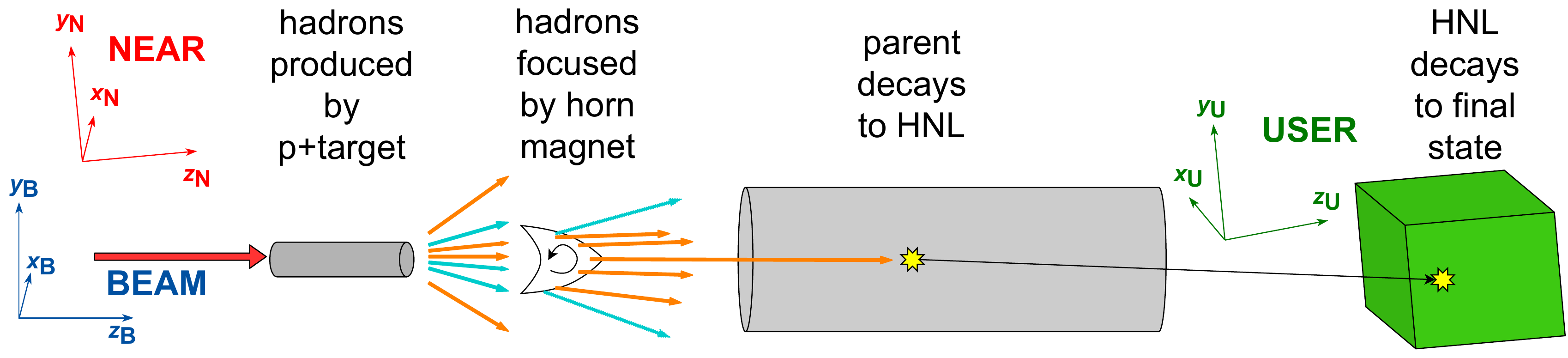}
  \caption{Cartoon of a typical neutrino beamline. See text for details.}
  \label{fig:beamline_cartoon}
  \vspace*{-2em}
\end{figure}
We extend the effort being built by contributing a detailed simulation of HNL produced by hadron decays inside a neutrino beamline within the \verb|GENIE| framework, and their subsequent decays in a detector, with a view particularly towards present and future HNL search efforts \cite{DUNEBSMOverview, FASERHNL, Ballett2017, HyperKOverview, DarkQuestHNL}.
\par We start by describing three coordinate systems, \textit{de facto} used in an accelerator neutrino experiment.
A sketch of a typical neutrino beamline outlining these coordinate systems, and showing how an HNL is typically produced, is shown in Fig. \ref{fig:beamline_cartoon}.
\par First, a ``NEAR" frame $(x_{\textrm{N}},y_{\textrm{N}},z_{\textrm{N}})$ defines the beamline coordinates.
This is typically set such that the target hall building has its floor parallel to $y=\text{const}$ and such that the origin O is within the target.
Second, a ``BEAM" frame $(x_{\textrm{B}},y_{\textrm{B}},z_{\textrm{B}})$, normally obtained by rotating $(x_{\textrm{N}},y_{\textrm{N}},z_{\textrm{N}})$ downwards in the $(y_{\textrm{N}},z_{\textrm{N}})$ plane.
\begin{widetext}
  \begin{table*}[t!]
    \centering
    \makegapedcells
    \setlength\tabcolsep{8pt}
    \begin{tabularx}{0.8\linewidth}{c c c c}
      \Xhline{1.5pt} 
      Channel &Threshold (MeV$/c^{2}$) &Decay rate (Dirac) \cite{ColomaDUNEHNL} &$\Gamma_{\text{Maj}}/\Gamma_{\text{Dirac}}$ \\
      \Xhline{0.8pt}
      $N_{4} \rightarrow \nu\nu\nu$ &$0$ &$\frac{G_{\textrm{F}}^{2}M_{\textrm{N}4}^{5}}{192\pi^{3}}\sum_{\alpha}|U_{\alpha 4}|^{2}$ &$2$ \\
      $N_{4} \rightarrow \nu e^{\pm} e^{\mp}$ &$1.02$ &$\frac{G_{\textrm{F}}^{2}M_{\textrm{N}4}^{5}}{192\pi^{3}}\cdot\left(\mathcal{C}_{\textrm{e}}\left(M_{\textrm{N}4}\right)+\mathcal{D}_{\textrm{e}}\left(M_{\textrm{N}4}\right)\right)$ &$2$ \\
      $N_{4} \rightarrow \nu e^{\pm} \mu^{\mp}$ &$106.17$ &$\frac{G_{\textrm{F}}^{2}M_{\textrm{N}4}^{5}}{192\pi^{3}}\cdot|U_{\textrm{e}4}|^{2}\cdot\mathcal{F}\left(\delta_{M_{\textrm{N}4}}^{m_{\upmu}}\right)$ &$2\frac{|U_{\textrm{e}4}|^{2} + |U_{\upmu 4}|^{2}}{|U_{\textrm{e}4}|^{2}}$ \\
      $N_{4} \rightarrow \pi^{0} \nu$ &$134.98$ &$\frac{G_{\textrm{F}}^{2}M_{\textrm{N}4}^{3}f_{\uppi}^{2}}{32\pi}\sum_{\alpha}|U_{\alpha 4}|^{2}\left(1 - \delta_{M_{\textrm{N}4}}^{m_{\uppi 0}}\right)^{2}$ &$2$ \\
      $N_{4} \rightarrow \pi^{\pm} e^{\mp}$ &$140.08$ &$\frac{G_{\textrm{F}}^{2}M_{\textrm{N}4}^{3}f_{\uppi}^{2}\left|V^{\text{CKM}}_{\text{ud}}\right|^{2}}{16\pi}\mathcal{G}\left(\delta_{M_{\textrm{N}4}}^{m_{\textrm{e}}}, \delta_{M_{\textrm{N}4}}^{m_{\uppi}}\right)$ &$2$ \\
      $N_{4} \rightarrow \nu \mu^{\pm} \mu^{\mp}$ &$211.32$ &$\frac{G_{\textrm{F}}^{2}M_{\textrm{N}4}^{5}}{192\pi^{3}}\cdot\left(\mathcal{C}_{\upmu}\left(M_{\textrm{N}4}\right)+\mathcal{D}_{\upmu}\left(M_{\textrm{N}4}\right)\right)$ &$2\frac{|U_{\textrm{e}4}|^{2}+|U_{\upmu 4}|^{2}}{|U_{\upmu 4}|^{2}}$ \\
      $N_{4} \rightarrow \pi^{\pm} \mu^{\mp}$ &$245.23$ &$\frac{G_{\textrm{F}}^{2}M_{\textrm{N}4}^{3}f_{\uppi}^{2}\left|V^{\text{CKM}}_{\text{ud}}\right|^{2}}{16\pi}\mathcal{G}\left(\delta_{M_{\textrm{N}4}}^{m_{\upmu}}, \delta_{M_{\textrm{N}4}}^{m_{\uppi}}\right)$ &$2$ \\
      \Xhline{1.5pt}
    \end{tabularx}
    \caption{Decay channels for HNL. The pion decay constant $f_{\uppi} = 130\,\,\text{MeV}$.}
    \label{tab: decayChannels}
  \end{table*}
\end{widetext}
The $z_{\textrm{B}}$ axis is typically parallel to the direction of the neutrino beam.
Third, a ``USER" frame $(x_{\textrm{U}},y_{\textrm{U}},z_{\textrm{U}})$, attached to a detector some distance away from the target.
For example, take the case of DUNE \cite{LBNF_CDR, DUNE_ND_CDR}, assuming the Liquid Argon component to be in the on-axis configuration and at a distance of $575\,\,\text{m}$ from the origin O, and that the LBNF beam is rotated $5.8^{\circ}$ downwards, the positions of the detector centre in NEAR, BEAM, and USER coordinates are, respectively:
\begin{align} \label{eq:DUNENDCoords}
  \begin{cases}
    (x_{\textrm{N}},y_{\textrm{N}},z_{\textrm{N}}) = (0, -58.11, 572.06)\,\,\text{m}, \\
    (x_{\textrm{B}}, y_{\textrm{B}}, z_{\textrm{B}}) = (0, 0, 575)\,\,\text{m}, \\
    (x_{\textrm{U}}, y_{\textrm{U}}, z_{\textrm{U}}) = (0,0,0).
  \end{cases}
\end{align}
Similarly, for the MINER$\nu$A inner detector (tracker + nuclear targets) \cite{TheMINERvADetector}, the NEAR, BEAM, and USER positions are taken to be
\begin{align}
  \begin{cases}
    (x_{\textrm{N}}, y_{\textrm{N}}, z_{\textrm{N}}) = (-0.25, -60.35, 1022.74)\,\,\text{m}, \\
    (x_{\textrm{B}}, y_{\textrm{B}}, z_{\textrm{B}}) = (-0.25, -0.66, 1024.52)\,\,\text{m}, \\
    (x_{\textrm{U}}, y_{\textrm{U}}, z_{\textrm{U}}) = (0, 0, 6.44)\,\,\text{m}.
  \end{cases}
\end{align}
\par For a detailed explanation of the convention used for positioning, see Appendix \ref{appdx: coords}.
\par It is instructive at this point to write down the ideal factorisation for such an endeavour:
\begin{enumerate}
\item A dynamic HNL flux prediction, taking into account the detailed kinematic effects of particles travelling with $\beta < 1$;
\item A sophisticated decay library, including effects of HNL polarisation;
\item A complete array of bookkeeping tools, including POT counting.
\end{enumerate}
A pre-eminent position must be given to the quite general flux and geometry drivers implemented in \verb|GENIE|, which allow this implementation to take advantage of the full complexity of a hadron beam simulation and geometry description, making this tool suited to studies at various stages of maturity of an experiment from its very early stages all the way through to its most mature stages.
Conceptually, the \verb|GENIE| \verb|BeamHNL| module may consume an arbitrarily complex simulation of hadrons in some beamline as input, along with an arbitrarily complex detector description, and output HNL decay events within the desired detector volume, in user-defined coordinates.
This underlines the nature of this tool as a self-contained kit to handle HNL kinematics and dynamics from production up until decay, with no assumptions \textit{a priori} about the nature of the experiment it is being applied to.
\subsection{HNL production in beamline}
The starting off point for deriving a neutrino flux is a precalculated spectrum of particles, such as is typically produced from proton interactions on a target in a neutrino beamline.
These particles (such as $\pi^{\pm}, K^{\pm}, \mu^{\pm}$) are then usually focused by a series of magnets to perform some charge selection.
The particles then propagate downstream, decaying in some suitably long decay volume to neutrinos (or particles that eventually decay to neutrinos). 
They are referred to as ``parents" if they decay directly to neutrinos, or more generally ``ancestors" if one of their decay products is a parent.
One or more particle absorbers are normally placed in between the decay volume and the detector, so that the only particles that survive to reach the detector are neutrinos. 
This same production philosophy applies to HNL, which by Eq. (\ref{eq: fourth_nu_mixing}) have a probability $\propto \left| U_{\alpha 4}\right|^{2}$ to be the mass state that corresponds to the $\nu_{\alpha}$ flavour state made during neutrino production.
\par Experiments simulate the spectrum for each parent species using a description of their beamlines and suitable external experimental data (\cite{NA49, NA61}) to constrain the simulation uncertainty \cite{NuMIBeamFlux}.
There are various formats the output of this simulation can be stored in; we have chosen to adapt the \verb|dk2nu| format \cite{Dk2NuProposal} developed for the Fermilab Intensity Frontier experiments to a ``flat dk2nu" format that mirrors the \verb|dk2nu| tree structure, without containing any complex classes.
This was done to minimise the build complexity for \verb|GENIE|. 
However, other flux formats can readily be converted into the format required for this simulation, and an example input flux with the necessary structure has been provided in the \verb|$GENIE/src/contrib/beamhnl| directory accompanying our module.
\par One application of this general format is the ability to use this module not only for accelerator neutrino experiments such as \cite{DUNEBSMOverview, SBNReview, DarkQuestHNL}, but also in the context of higher-energy collider neutrino experiments with detectors lying downstream of the neutrino production point \cite{SHiPSensitivity, FASERHNL}.
This is particularly interesting, because at higher energies heavy mesons (such as $D, D_{s}$), which can decay to HNL heavier than the kaon mass, are produced copiously enough to have a strong sensitivity to HNL with $M_{\textrm{N}4} > m_{\textrm{K}}$.
Equally, the same format could be adapted for use with atmospheric neutrino fluxes \cite{SKAtmHNL, AtmoLLP} to probe HNL of extraterrestrial origin. We comment further on this in Section \ref{SECT_discussion}.
\par We have used the production branching ratios from \cite{Shrock1981, Ballett2020} for the evaluation of HNL production probabilities, for the most commonly produced parent types: $\pi^{\pm}, K^{\pm}, K^{0}_{L}$ (directly from the beamline), and $\mu^{\pm}$ (from hadron decay themselves). 
Further hadron types are not yet a part of our simulation, but are compelling targets to include.
\par To conserve computing resources, we force every parent in the simulation to decay to HNL; that is equivalent to using the uncorrected branching ratios from Fig. \ref{fig:prodBR}.
This is corrected in the simulation bookkeeping by applying Eq. (\ref{eq: pseudounitarity}) on the production probability of the HNL.
\par A salient feature of the HNL production is that, depending on the mass of the HNL, the acceptance of the detector as seen by the HNL at its production point can change significantly, owing to the fact that HNL travel with $\beta_{\textrm{N}4} < 1$.
A more detailed review of massive-particle kinematics can be found in \cite{rubbia_2022}. \\
\par
\begin{figure}
  \centering
  \includegraphics[width=0.4\textwidth]{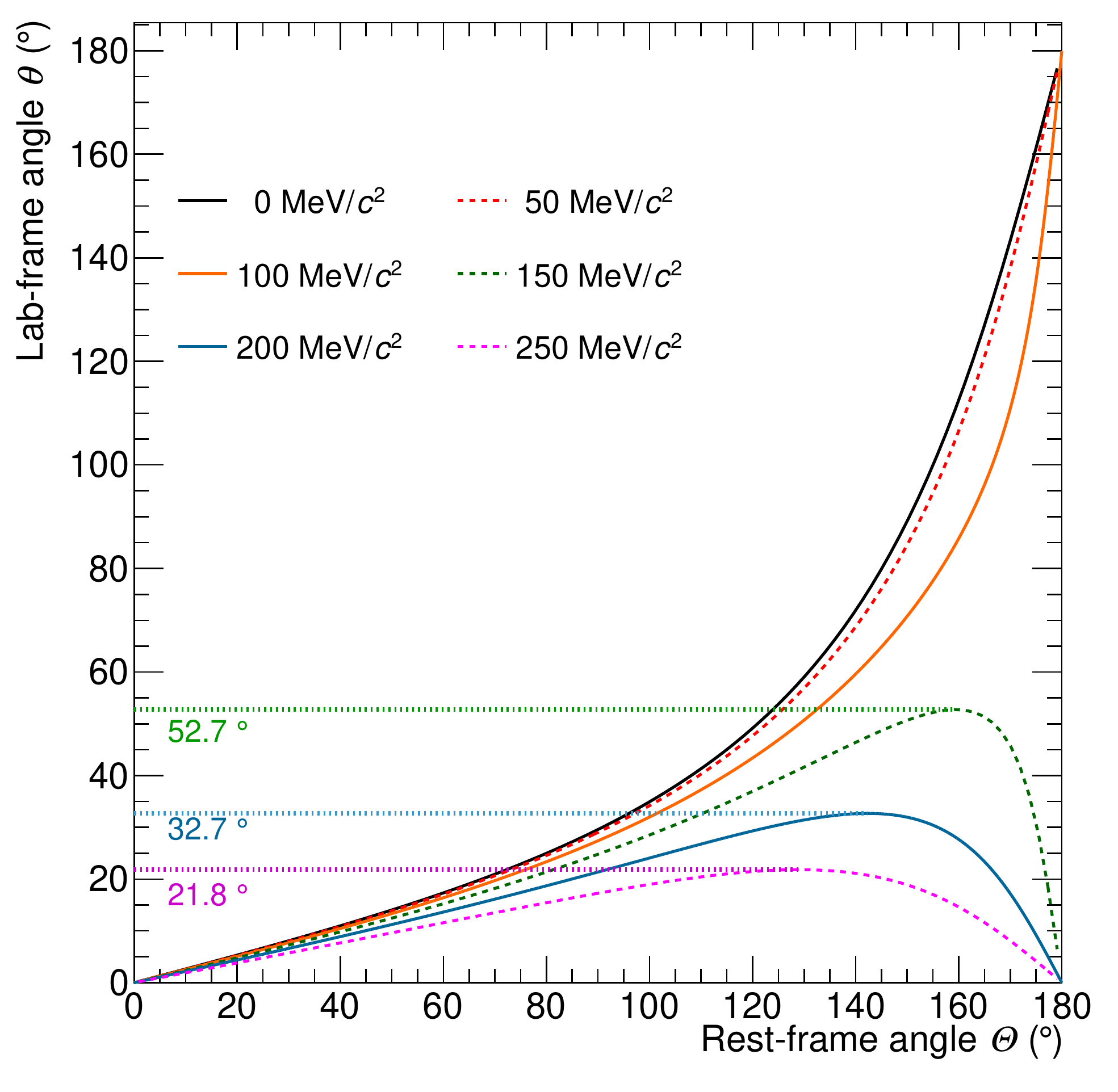}
  \caption{Collimation effect for HNL in $K^{\pm} \rightarrow N_{4}\mu^{\pm}$, $E_{\textrm{K}^{\pm}} = 1\,\,\text{GeV}$. For $M_{\textrm{N}4}$ heavy enough, there is no backward emission in the lab frame.}
  \label{fig:collimation}
\end{figure}

 Consider a parent $P$ of energy $E_{\textrm{P}}$, decaying leptonically to a neutral lepton $L$ that can either be light ($\nu$) or heavy ($N_{4}$) and a charged lepton $\ell$. 
The energy of $L$ in the rest frame of $P$ is then
\begin{equation}
  E_{\textrm{L}}^{*} = \frac{m_{\textrm{P}}^{2} - m_{\ell}^{2} + m_{\textrm{L}}^{2}}{2m_{\textrm{P}}},
\end{equation}
Suppose, without loss of generality, that an observer sees $P$ propagate along the $z$ axis, $\boldsymbol{p}_{\textrm{P}} = (0, 0, p_{\textrm{P}})$. 
Then the angle $\Theta$ at which $L$ is emitted with respect to the $z'$ axis in the rest frame can be related to the emission angle $\theta$ in the observer's frame, as
\begin{equation}\label{eq: collimation_effect}
  \tan\theta = \frac{q_{\textrm{L}}\sin\Theta}{\gamma_{\textrm{P}}\left(\beta_{\textrm{P}}E_{\textrm{L}}^{*} + q_{\textrm{L}}\cos\Theta\right)},
\end{equation}
where $q_{\textrm{L}}$ is the rest-frame momentum of $L$, and $\beta,\gamma$ are the relativistic parameters of $P$ in the lab frame.
Figure \ref{fig:collimation} shows $\theta$ as a function of $\Theta$ for a kaon parent, with $E_{\textrm{K}} = 1\,\,\text{GeV}$, for various HNL masses produced in the decay $K^{\pm} \rightarrow N_{4} + \mu^{\pm}$.
\par The cases of massless neutrinos and HNL are quite different; for massless Standard Model neutrinos, $q_{\textrm{L}} = E_{\textrm{L}}^{*}$ and the resulting function $\tan^{-1}\left[\gamma_{\textrm{P}}^{-1}\sin\Theta\left(\beta_{\textrm{P}}+\cos\Theta\right)^{-1}\right]$ is monotonically increasing.
This means that, for any arbitrary emission angle $\theta_{0}$ in the lab frame, there exists some suitable pre-image $\Theta_{0}$ in the rest frame, for any parent velocity.
In a Standard Model calculation, \emph{any} parent can produce a neutrino that is accepted by a detector, \emph{for any} parent momentum and relative position of the neutrino production vertex to the detector.
\par In contrast, for HNL with masses large enough, it is not generically true that any $\theta_{0}$ can be achieved; this places increased importance on the kinematics of the parent $P$.
The greater the angle between $P$'s momentum $\boldsymbol{p}_{\textrm{P}}$ and the relative separation $\boldsymbol{\mathcal{O}}$ between neutrino production vertex and detector, and the greater the HNL mass, the smaller the acceptance becomes, generally speaking; for large enough angles and masses, the HNL could not be accepted at all, regardless of its rest-frame emission angle.
This same collimation effect can, at the same time, account for \emph{increased} acceptance of HNL with respect to a Standard Model neutrino, if $P$ is well-collimated enough.
This happens because Eq. (\ref{eq: collimation_effect}) now reaches a maximum and tends towards $0$ for large $\Theta$: in other words, backwards-emitted HNL are swept forwards by the Lorentz boost into the lab frame, and end up accepted by the detector.
In general, this \emph{acceptance correction} respective to a Standard Model neutrino is calculated on an event-by-event basis, and applied as an additional weight to each HNL.
One can also obtain an estimate for the neutrino energy at the detector, again setting $p_{\textrm{L}}^{x} = 0$ without loss of generality
\begin{align}\label{eq: boostFactor}
  \begin{split}
    &\Rightarrow E_{\textrm{L}}^{*} = \gamma_{\textrm{P}}E_{\textrm{L}} - \gamma_{\textrm{P}}\beta_{\textrm{P}}p_{\textrm{L}}\cos\theta_{\textrm{D}}, \\
    &\Rightarrow E_{\textrm{L}} = \frac{E_{\textrm{L}}^{*}}{\gamma_{\textrm{P}}\left(1 - \beta_{\textrm{P}}\beta_{\textrm{L}}\cos\theta_{\textrm{D}}\right)} \equiv \mathcal{B}E_{\textrm{L}}^{*},
  \end{split}
\end{align}
where $\mathcal{B}$ is termed the \emph{boost factor}, and $\theta_{\textrm{D}}$ is the viewing angle between $\boldsymbol{p}_{\textrm{P}}$ and $\boldsymbol{\mathcal{O}}$.
From a Standard Model simulation standpoint, this equation simplifies to
\begin{equation}
  \mathcal{B}_{\nu} = \frac{1}{\gamma_{\text{P}}\left(1 - \beta_{\textrm{P}}\cos\theta_{\textrm{D}}\right)},
\end{equation}
which uniquely determines the lab-frame energy for the massless neutrino.
For a massive neutrino, there is a complication: Eq. (\ref{eq: boostFactor}) depends on knowledge of the lab-frame velocity through $\beta_{\textrm{L}}$.
We estimate $\beta_{\textrm{L}}$ by imposing a geometric constraint; using the worldline $(T, \boldsymbol{\mathcal{O}})$ with $T = |\boldsymbol{\mathcal{O}}|/(\beta_{\textrm{L}}c)$, we construct a candidate lab-frame momentum by boosting $(T, \boldsymbol{\mathcal{O}})$ into $P$'s rest frame and forcing the HNL momentum to point to that direction, then boosting the result back into the lab frame.
We check the distance between the point of closest approach and the detector centre; if this is too large, we decrement $\beta_{\textrm{L}}$ and repeat the procedure.
\par
\begin{figure*}
  \centering
  \includegraphics[width=\textwidth]{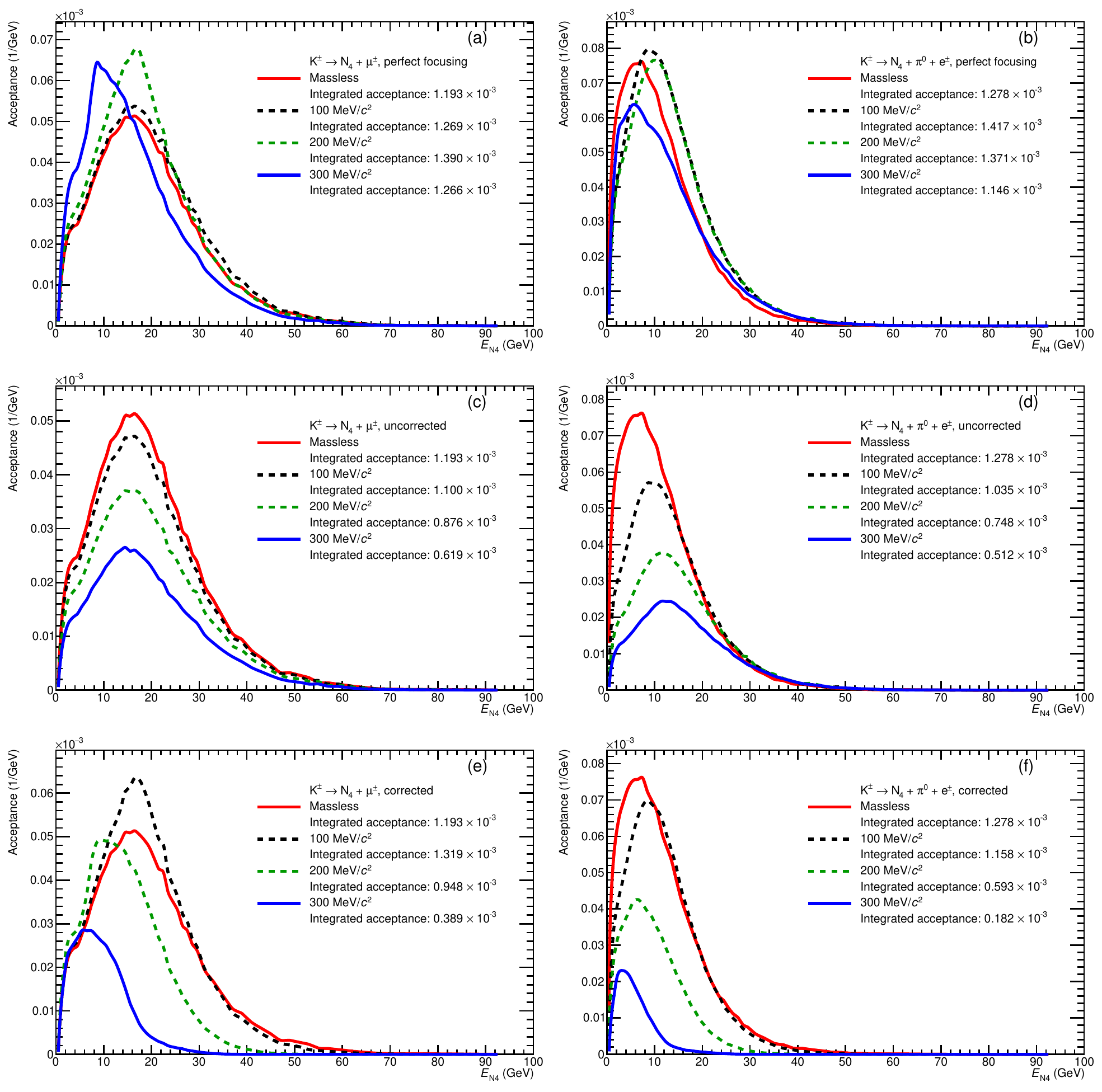}
  \caption{Acceptance for a $K^{\pm}$ decaying to HNL and either $\mu^{\pm}$ or $e^{\pm} + \pi^{0}$, for the MINER$\nu$A detector \cite{TheMINERvADetector}. Top row: Parents perfectly focused: momentum of parent always points towards detector. Middle row: parents not perfectly focused, but no acceptance correction $\mathcal{A}$ applied. Bottom row: full focusing and acceptance correction applied. \newline
    The normalisation of these curves drops considerably when focusing is not perfect; this happens because the boost factor $\mathcal{B}$ drops with opening angle between parent momentum and detector location. The effect of $\mathcal{A}$ is mainly to suppress the high-$E_{\textrm{N}4}$ tails, as hard HNL cannot deviate from their parents enough to reach the detector; also, at low enough HNL mass, $\mathcal{A}$ increases normalisation as backward-emitted HNL are accepted by the detector.}
  \label{fig:NuMI_acceptance}
\end{figure*}

 In Fig. \ref{fig:NuMI_acceptance}, we plot the differential geometrical acceptance, defined as the probability a HNL emitted isotropically in the parent's rest frame will be accepted by the detector, for the case of the MINER$\nu$A detector \cite{TheMINERvADetector} in the NuMI Medium-Energy beam \cite{NuMIBeamFlux}. 
We have plotted the acceptance for the processes $K^{\pm} \rightarrow N_{4} + \mu^{\pm}$ (left-hand side plots) and $K^{\pm} \rightarrow N_{4} + \pi^{0} + e^{\pm}$ (right-hand side plots). 
Panels (a), (b) show the differential acceptance under the assumption the parent kaons are perfectly focused, which is to say an HNL emitted with momentum collinear to the kaon momentum would definitely be accepted.
Generally, the acceptance remains about the same as for Standard Model neutrinos, though the peak shifts first to higher and then to lower energies.
This is caused by the Lorentz boost becoming more efficient for all HNL as the mass initially increases, followed by the decrease associated with the drop in boost factor $\mathcal{B}$ as the HNL's velocity drops significantly.
In panels (c), (d), we have used the kaon spectrum from the NuMI beamline simulation including realistic focusing, but not applied acceptance correction.
The shape of the acceptance remains roughly similar, but the normalisation and integrated acceptance change dramatically.
This is caused by the suppression of the boost factor $\mathcal{B}$ at both high angles and low velocities.
Finally, panels (e), (f) show the full simulation accounting for both realistic focusing and the change in suitable emission regions in the parent rest frame due to acceptance correction.
Peaks of these distributions shift decidedly to lower energy, as the higher energy HNL are more collimated with their parents and, unlike Standard Model neutrinos, are not necessarily able to reach the detector.
This effect becomes increasingly prominent as $M_{\textrm{N}4}$ goes up and the kinematic constraints become more severe.
\begin{figure}
  \centering
  \includegraphics[width=0.45\textwidth]{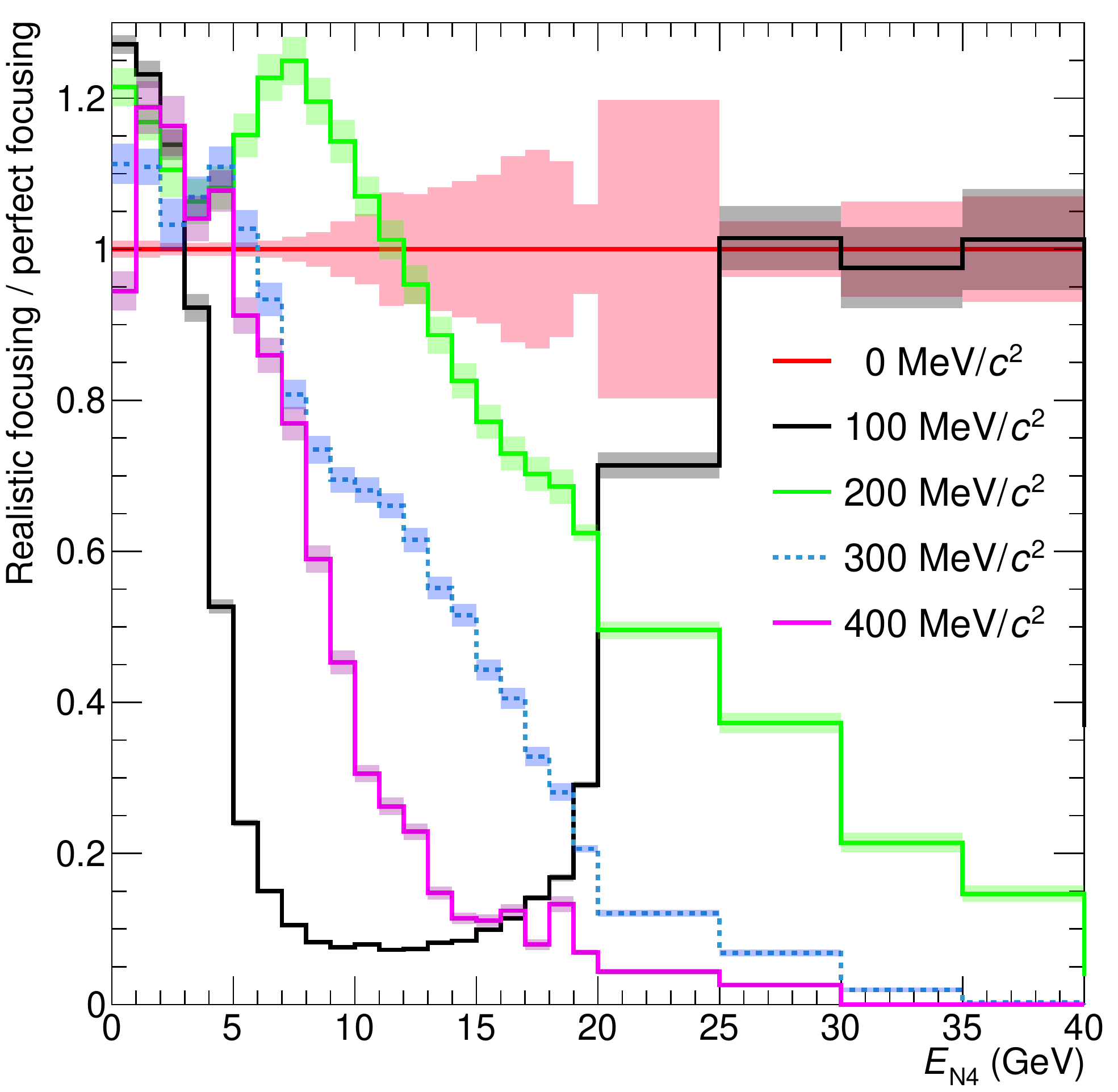}
  \caption{Effect of focusing on acceptance; flux at MINER$\nu$A (realistic focusing) / flux (perfect focusing). Error bands are purely statistical.} 
  \label{fig:ratioAcceptance}
  \vspace*{-1em}
\end{figure}
\par We further demonstrate the effect of parent focusing on the acceptance in Fig. \ref{fig:ratioAcceptance}, by simulating the flux under the assumption of perfect parent focusing or using realistic parent focusing as provided from the NuMI beamline simulation.
We show the ratios of realistic over perfect acceptance, as a function of HNL energy, for masses $0, 100, ..., 400\,\,\text{MeV}/c^{2}$.
The effect is most pronounced at $100$ and $400\,\,\text{MeV}/c^{2}$, where the thresholds of HNL production by pions and kaons are almost reached; at these masses, the collimation effect becomes most severe, and proportionally more parents can not produce HNL that are accepted by the detector, unless the parent happens to be travelling in a direction that would intersect the detector.
The effect is more pronounced at high HNL energy $E_{\textrm{N}4}$, because more energetic HNL have a stronger collimation effect due to the larger Lorentz boost.
Under realistic focusing, the HNL spectra become softer; this is due to the suppression of the boost factor with parent momentum angle.
\par
\begin{widetext}
        \begin{figure*}
          \centering
          \includegraphics[width=\textwidth]{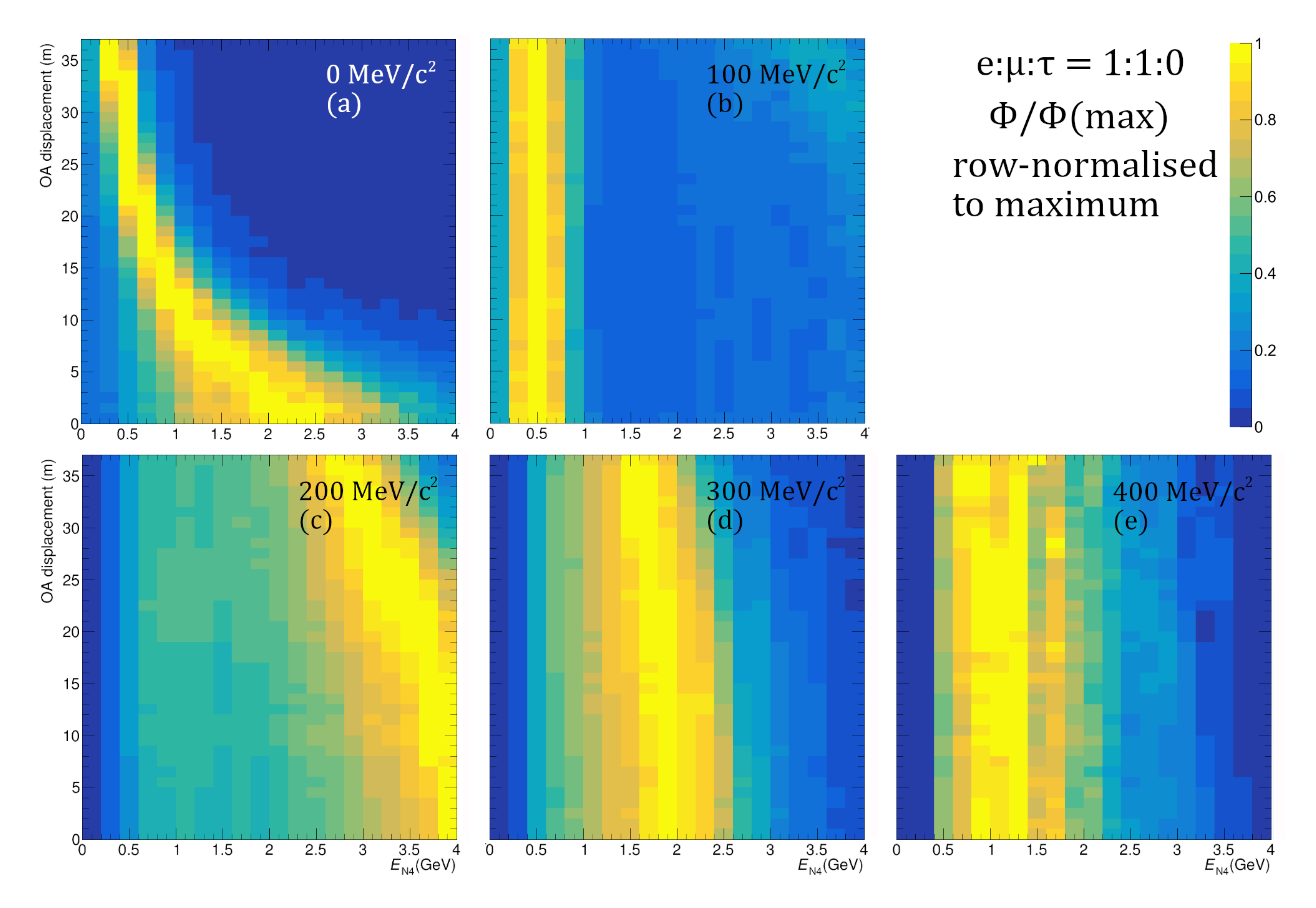}
          \caption{HNL flux shapes at DUNE PRISM, as function of HNL energy and off-axis displacement.}
          \setlength{\belowcaptionskip}{-5pt}
          \label{fig:prism}
        \end{figure*}
      \end{widetext}

 Another profound impact can be seen on off-axis neutrino spectra, relevant for PRISM-like searches.
Varying Eq. (\ref{eq: boostFactor}) over $\cos\theta$ for $\beta_{\textrm{L}} = 1$ yields the celebrated \emph{off-axis effect} for the Standard Model.
In the general case $\beta_{\textrm{L}} \neq 1$, however, the off-axis effect becomes progressively less pronounced.
In Fig. \ref{fig:pion_OA_emission}, we have chosen a pion parent, and four different values of $\beta_{\textrm{L}}$; the Standard Model case is retrieved in (a), and progressively smaller $\beta_{\textrm{L}}$ are shown in (b), (c), (d).
The various curves representing different values of $\theta_{\textrm{D}}$ collapse to a single curve.
Writing out the derivative $\partial E_{\textrm{L}}/\partial E_{\textrm{P}} = E_{\textrm{L}}^{*}\partial\mathcal{B}/\partial E_{\textrm{P}}$, one can find the peak $\mathcal{B}$ to be at
\begin{equation}
  E_{P}\big|_{\mathcal{B}\,\,\text{max}} = \frac{m_{P}}{(1 - \beta_{\textrm{L}}^{2}\cos^{2}\theta)^{1/2}}. 
\end{equation}
\begin{widetext}
  \begin{figure*}
    \centering
    \includegraphics[width=\textwidth]{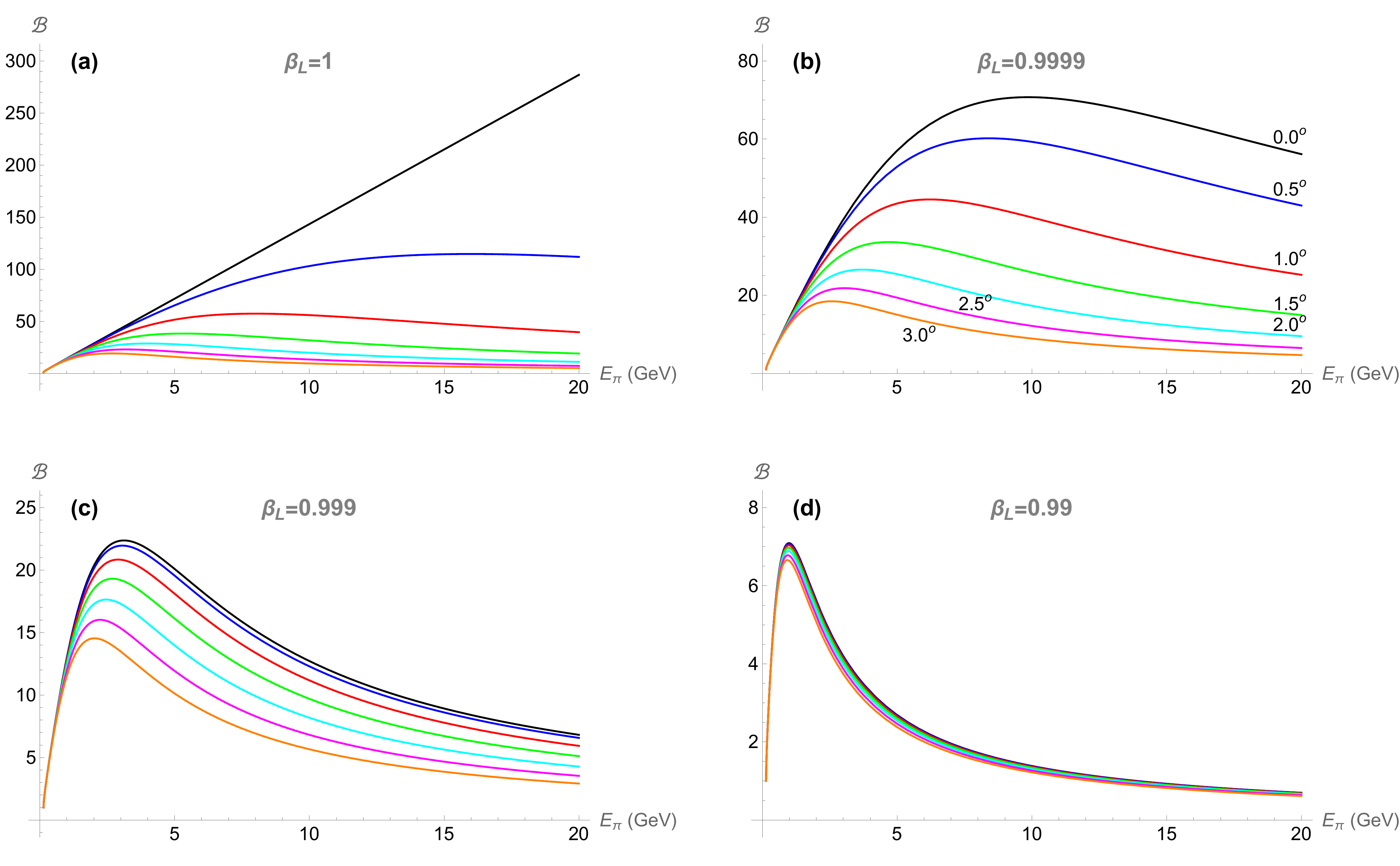}
    \caption{Boost factor $\mathcal{B}$ as function of pion energy, for different values of $\theta_{\textrm{OA}}$. (a): $\beta_{\textrm{L}} = 1$ (Standard Model); (b): $\beta_{\textrm{L}} = 0.9999$; (c): $\beta_{\textrm{L}} = 0.999$; (d): $\beta_{\textrm{L}} = 0.99$. The off axis angle $\theta_{\textrm{OA}}$ grows from top to bottom, from $0.0^{\circ}$ to $3.0^{\circ}$ in increments of $0.5^{\circ}$.}
    \label{fig:pion_OA_emission}
    \vspace*{-2em}
  \end{figure*}
  \begin{figure*}
    \centering
    \includegraphics[width=0.7\textwidth]{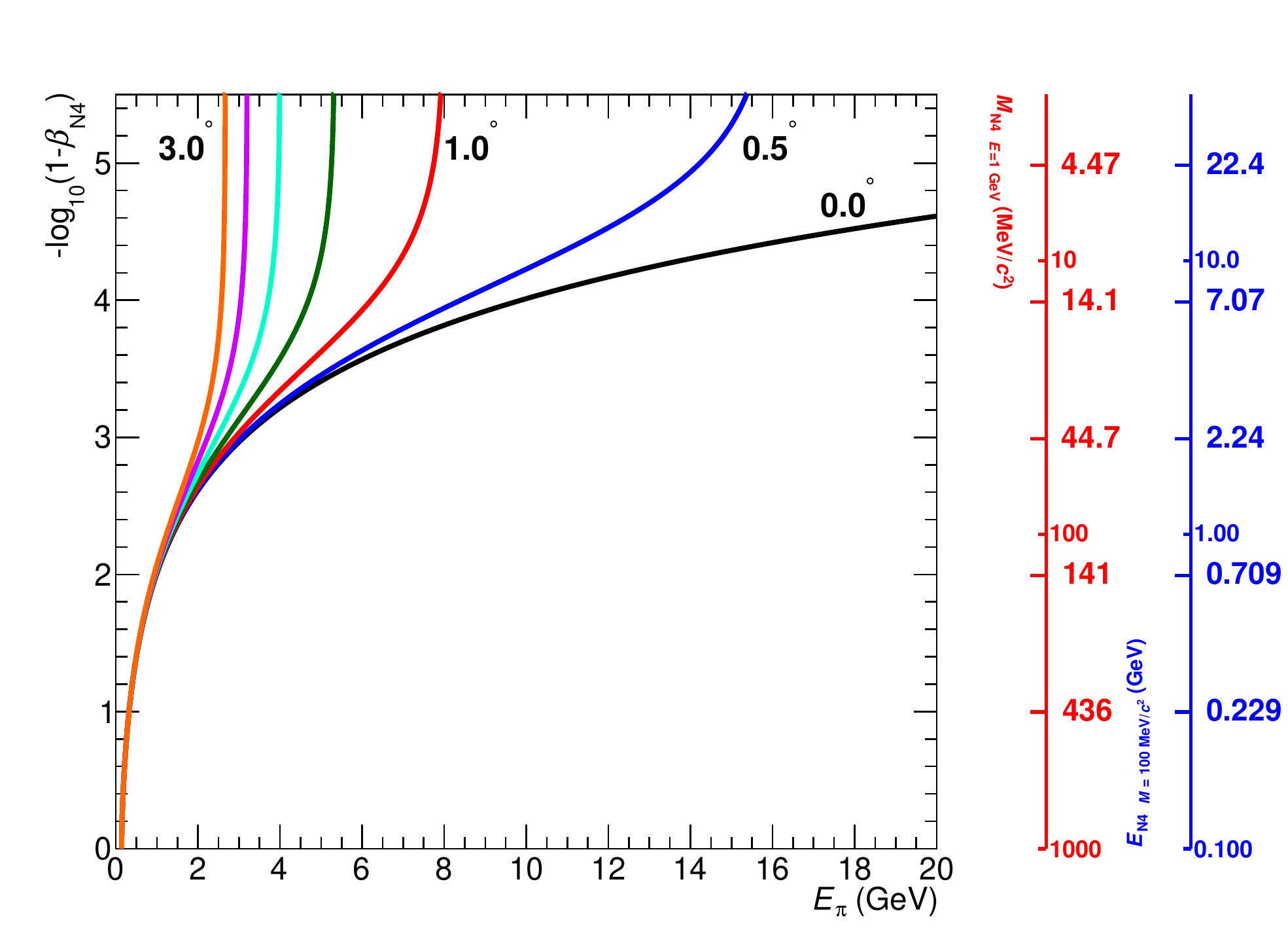}
    \caption{Pion energy at which $\mathcal{B}$ is highest, for different values of $\theta$.}
    \label{fig:pion_OA_peaks}
  \end{figure*}
\end{widetext}
In Fig. \ref{fig:pion_OA_peaks} we have plotted, for different values of $\theta_{\textrm{D}}$ and for pion parent, $E_{\textrm{P}}\big|_{\mathcal{B}\,\,\text{max}}$ (x axis) as a parametric plot of $\log_{10}\left(1-\beta_{\textrm{L}}\right)$ (y axis).
Standard Model neutrinos are the limit $y \rightarrow \infty$; additionally, we have drawn two axes for visualisation: red (left) is the mass of a $1\,\,\text{GeV}$ HNL for the given $\beta$, and blue (right) is the energy a $100\,\,\text{MeV}/c^{2}$ HNL would have at that $\beta$.
For very high $\beta$, the position of the peaks varies greatly with $E_{\uppi}$, meaning the off-axis effect is visible; this motivates, for example, the PRISM concept \cite{DUNE_ND_CDR} of utilising a narrower neutrino spectrum at high off-axis angles to constrain the neutrino flux.
However, as $\beta$ decreases, so does the variation of the neutrino spectrum (as one can also see in Fig. \ref{fig:pion_OA_emission} panel (d), with curves converging into one for $\beta = 0.99$): since $\beta$ scales inversely to $M_{N4}$, the HNL flux is expected to be \emph{less} sensitive to the off-axis effect than Standard Model neutrinos.
In principle, the spectrum of HNL is still expected to become softer \cite{DuneNDBSM}, but the effect seen would be much smaller than predicted previously.
In Fig. \ref{fig:prism}, one can see our calculation of the flux shapes\footnotemark[2] at the DUNE PRISM with $|U_{\textrm{e}4}|^{2}:|U_{\upmu 4}|^{2}:|U_{\uptau 4}|^{2} = 1:1:0$, for masses $M_{\textrm{N}4} = 0, 100, ..., 400\,\,\text{MeV}/c^{2}$.
Notice how as the off axis angle $\theta_{\textrm{OA}}$ changes, the massless neutrino flux shifts in accordance with the SM prediction, whereas the $M_{N4} = 100, ..., 400\,\,\text{MeV}/c^{2}$ HNL have a far smaller (but still present) dependence on the off axis angle.
\par
\begin{widetext}
        \begin{figure*}
          \centering
          \includegraphics[width=\textwidth]{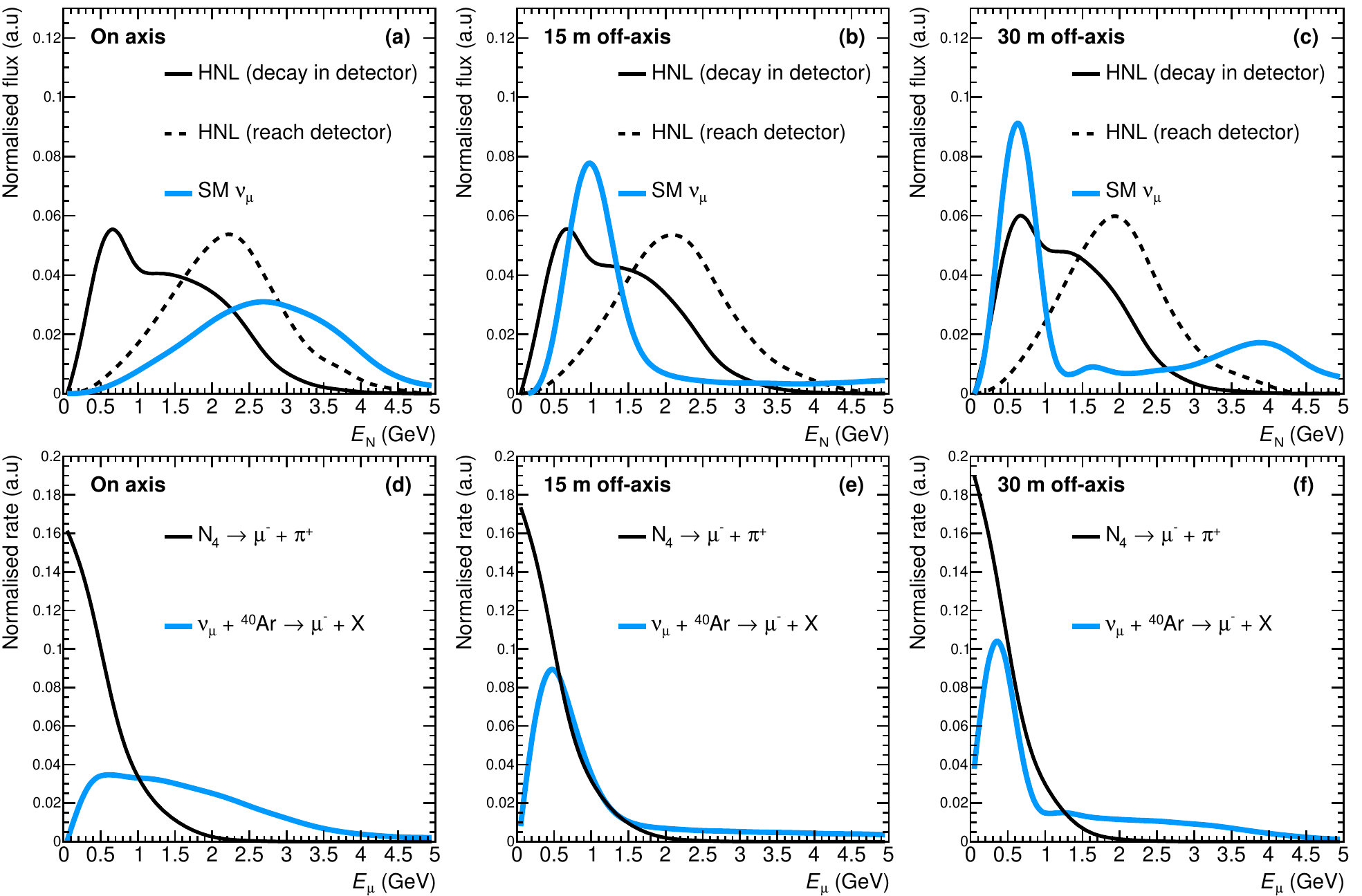}
          \caption{Example of truth-level muon spectral shapes originating from the HNL decay $N_{4} \rightarrow \pi^{+} + \mu^{-}$ ($M_{\textrm{N}4} = 300\,\,\textrm{MeV}/c^{2}$) in a 5m-side box-shaped detector at the DUNE PRISM baseline, at three off-axis displacements. Panels (a), (b), and (c) show the expected flux shape of HNL that decay inside the detector (solid black line) and that reach the detector (dashed black line), compared to the expected flux of Standard Model muon neutrinos (solid blue line). Panels (d), (e), and (f) show the muon energy from HNL decays (black) vs the expected muon spectrum from charged-current inclusive interactions of $\nu_{\mu}$ on $^{40}\textrm{Ar}$. SM interactions simulated with $\textrm{GENIE}\,\,\textrm{v}3.02.00\,\,\textrm{tune}\,\,\textrm{G}18\_02\textrm{a}\_00\_000$.}
          \label{fig:OAspectra}
        \end{figure*}
      \end{widetext}

 We have further shown in Fig. \ref{fig:OAspectra} an example of the diminished off-axis effect on the truth-level muon energy distribution from the decay $N_{4} \rightarrow \pi^{+} + \mu^{-}$ with HNL mass $M_{\textrm{N}4} = 300\,\,\textrm{MeV}/c^{2}$ and $\left|U_{\textrm{e}4}\right|^{2} = \left|U_{\upmu4}\right|^{2}$, and compared with the expectation on the equivalent distribution of charged-current muon neutrino interactions with one $\mu^{-}$ in the final state.
This would correspond to panel (d) of Fig. \ref{fig:prism}, where the flux of HNL that are crossing the detector is shown, for the bins $[0,1), [15,16)$ and $[30,31)$ in OA displacement.
      In the first three panels of Fig. \ref{fig:OAspectra}, the suppression of the off-axis effect on HNL energy distributions (black curves) at the detector, which is taken to be a 5m-side cube at the coordinates (\ref{eq:DUNENDCoords}), is apparent, especially when compared to the Standard Model muon neutrino expectation (solid blue curves).
      Note that the dashed curves, which show the HNL that are expected to cross the detector but not necessarily decay inside it, are markedly different to the solid curves which correspond to HNL actually decaying inside the detector.
      This is due to the lower velocities of softer HNL causing them to decay in the detector at a disproportionately high rate.
      Also note the difference from the solid blue curves that represent the SM expectation, in line with the suppression of the off-axis effect for slower-than-light particles as shown in Figs. \ref{fig:pion_OA_emission}, \ref{fig:pion_OA_peaks}.
      For the Standard Model muon energy distributions we have used events generated with \verb|GENIE| version 3.02.00 on argon-40, selecting only those events with precisely 1 muon in the final state.
      \footnotetext[2]{We have used the DUNE flux files available at \cite{DUNEFluxesWebsite}, for the ``Optimized 3-Horn Design with 1.5m target and Fully Engineered Horn A (Jan 2021)'' configuration.}
      \par The baseline for the detector is 575 m, which means the reference off axis transverse displacements of $5, 10, 20,$ and $30$ m are equivalent to $\theta_{\textrm{OA}} \simeq 0.5, 1.0, 2.0, 3.0^{\circ}$.
      \par Our HNL simulation produces the flux calculation based on minimal information from the beamline simulation, which is passed as an input.
      \par There are two fundamentally important inputs:
      \begin{enumerate}
      \item Parent momentum and decay vertex position in NEAR coordinates;
      \item ``Importance weight'' \cite{Dk2NuProposal, Goodwin2022} - a multiplicity factor for hadrons with very similar kinematics.
      \end{enumerate}
      Based on this information, the module assigns the appropriate decay channel, calculates the boost factor to obtain the energy of the HNL at the detector under the constraint that the HNL can reach it, and constructs the first particle in \verb|GENIE|'s particle stack that corresponds to an HNL that decays in the detector.
      We provide details of the bookkeeping in Appendix \ref{appdx: book}.
      \vspace*{1em}
      \subsection{Decay to Standard Model particles}
      \vspace*{-0.5em}
      Unlike Standard Model neutrinos, HNL are unstable and can decay directly (semi-)leptonically to SM particles.
      \par The lifetime of an HNL is generally inversely proportional to the mixing with the SM leptons $\sum_{\alpha} |U_{\alpha 4}|^{2}$ and to the HNL mass $M$. 
      Depending on the parameter space point being searched, the branching ratios for HNL decays vary as thresholds for various channels open; a few possibilities can be seen in Fig. \ref{fig:BR}. 
      Details of how we keep track of the decay channels and selects the correct one are in Appendix \ref{appdx: QM}.
      For HNL below the kaon mass, the most prevalent production channels are the two-body production channels $\pi \rightarrow N_{4} + \ell$ and, above the pion threshold, $K \rightarrow N_{4} + \ell$; for HNL above about $100\,\,\text{MeV}/c^{2}$, the main decay channels are the two-body channels $N_{4} \rightarrow \pi + \ell, N_{4} \rightarrow \pi^{0} + \nu$.
      \par Since the SM weak force couples to left-chiral particles and right-chiral antiparticles, the decays of $N_{4}$ and $\overline{N}_{4}$ have opposite angular dependencies \cite{TastetPol}; the well-known corollary is that in two-body decays only Dirac HNL have a $\cos\theta$ dependence in their decay spectra, whereas Majorana HNL do not (as has been shown explicitly in the case of neutral-mediated decays in \cite{BahaDiracVsMajPol}).
      In practical terms, charge-blind detectors that cannot distinguish between leptons and antileptons cannot search for forward-backward asymmetries in the decay distributions of HNL \cite{uBooNEMuPiHNL}.
      \par An HNL has the same intrinsic angular momentum as a Standard Model neutrino; however, its mass implies that it is \emph{partially}, rather than completely, polarised \cite{Levy}.
      Equivalently, the HNL and the other particle(s) that were produced in the decay of the pseudoscalar $P$ form a pure $J = 0$ state, but the HNL itself is not a pure state.
      Considering the leptonic production mode $P \rightarrow N_{4} + \ell'$, one can write down a polarisation vector $\mathbb{P}$ for the HNL.
      $\mathbb{P}$ has magnitude $|\mathbb{P}| < 1$ (partial polarisation) and direction collinear to the momentum $\boldsymbol{q}_{\ell'}$ where $\boldsymbol{q}$ is written in the HNL's rest frame.
      Because in $P$'s rest frame the momenta $\boldsymbol{p}_{\textrm{N}4}$ and $\boldsymbol{p}_{\ell'}$ are collinear, $\mathbb{P}$'s direction is the same as the momentum $\boldsymbol{q}_{\textrm{P}}$ of $P$ in the HNL rest frame.
      \par It is important, moreover, to keep track of where the polarisation vector $\mathbb{P}$ of the HNL is pointing, since it defines the only ``privileged direction" in the HNL rest frame, and thus the axis with which an angular dependence of the decay products $N_{4} \rightarrow X$ manifests.
      $\mathbb{P}$'s direction can be tracked at the moment of HNL production.
      \begin{widetext}
        \par The partial polarisation of $N_{4}$ also means that its differential decay rate into the generic final state $X$ is multiplied, again, by the appropriate coefficient determined by $|\mathbb{P}|$. 
        Essentially, instead of the well-known $1\mp\cos\theta$ dependence expected from the decay of fully polarised spin-$1/2$ particles, the angular dependence is modulated by a suitable polarisation modulus $H$ that is \textit{a priori} dependent on both the production and decay modes of the HNL, resulting in an angular dependence $1 \pm H\cos\theta$.
        The reader will notice that $-1 < H < 1$, i.e. the polarisation modulus can switch sign depending on the HNL mass.
        Its sign, $s_{\textrm{N}4}$, is $+1\,\,(-1)$ if the expectation value of measuring the HNL spin on the direction of its momentum is positive (negative).
        In the simple case of two-body production and two-body decay $P \rightarrow N_{4} + \ell', N_{4} \rightarrow D + \ell$, $H$ is explicitly given as \cite{Levy}
        \begin{equation}\label{eq: polModulus}
          H = - \frac{\left(m_{\ell'}^{2} - M_{\textrm{N}4}^{2}\right)\lambda^{1/2}\left(m_{\textrm{P}}^{2},M_{\textrm{N}4}^{2},m_{\ell'}^{2}\right)}{m_{\textrm{P}}^{2}\left(M_{\textrm{N}4}^{2} + m_{\ell'}^{2}\right)-\left(m_{\ell'}^{2}-M_{\textrm{N}4}^{2}\right)^{2}}\cdot\frac{\left(M_{\textrm{N}4}^{2} - m_{\ell}^{2}\right)\lambda^{1/2}\left(M_{\textrm{N}4}^{2},m_{\ell'}^{2},m_{\textrm{D}}^{2}\right)}{\left(M_{\textrm{N}4}^{2}-m_{\ell}^{2}\right)^{2}-m_{\textrm{D}}^{2}\left(M_{\textrm{N}4}^{2} + m_{\ell}^{2}\right)} = \left(s_{\textrm{N}4}\cdot|\mathbb{P}|\right)\cdot|\mathbb{D}|,
        \end{equation}
        where $\lambda(x,y,z)$ is the K\"{a}ll\'{e}n function defined in Appendix \ref{appdx: QM}, $s_{\textrm{N}4} = +1\,\,(-1)$ if $M_{\textrm{N}4}$ is greater (smaller) than $m_{\ell'}$, and $\mathbb{D}$ is the factor resulting from the decay $N_{4} \rightarrow D + \ell$.
        The factorisation of the polarisation modulus from HNL production and decay is thus made apparent.
      \end{widetext}
      \par In this work, we have implemented the simple, yet analytically calculable ``two-body-production, two-body-decay" polarisation prescription for all (Dirac) HNL decays, assigning an angular distribution $1 \pm H\cos\theta$ to the spectrum of decay products.
      However, we have implemented a switch that allows the user to turn this simple scheme off if it is desirable to do so, reverting to pure phase-space decays instead.
      Because the user has access to the truth-level four-momentum of each of the decay particles, it is in principle possible to reweight the spectra of the final state products with any desired polarisation scheme.
      The implementation of a fuller description of polarisation effects, including in the three-body decays of Majorana HNL (see for example \cite{HNL3BodyDecaysPol, Ballett2020} for discussions on Majorana HNL polarisations), remains an appealing avenue for future work.
      We comment on this in Section \ref{SECT_discussion}.
      \par
\begin{figure}[t]
        \centering
        \begin{subfigure}[b]{0.23\textwidth}
          \centering
          \includegraphics[width=\textwidth]{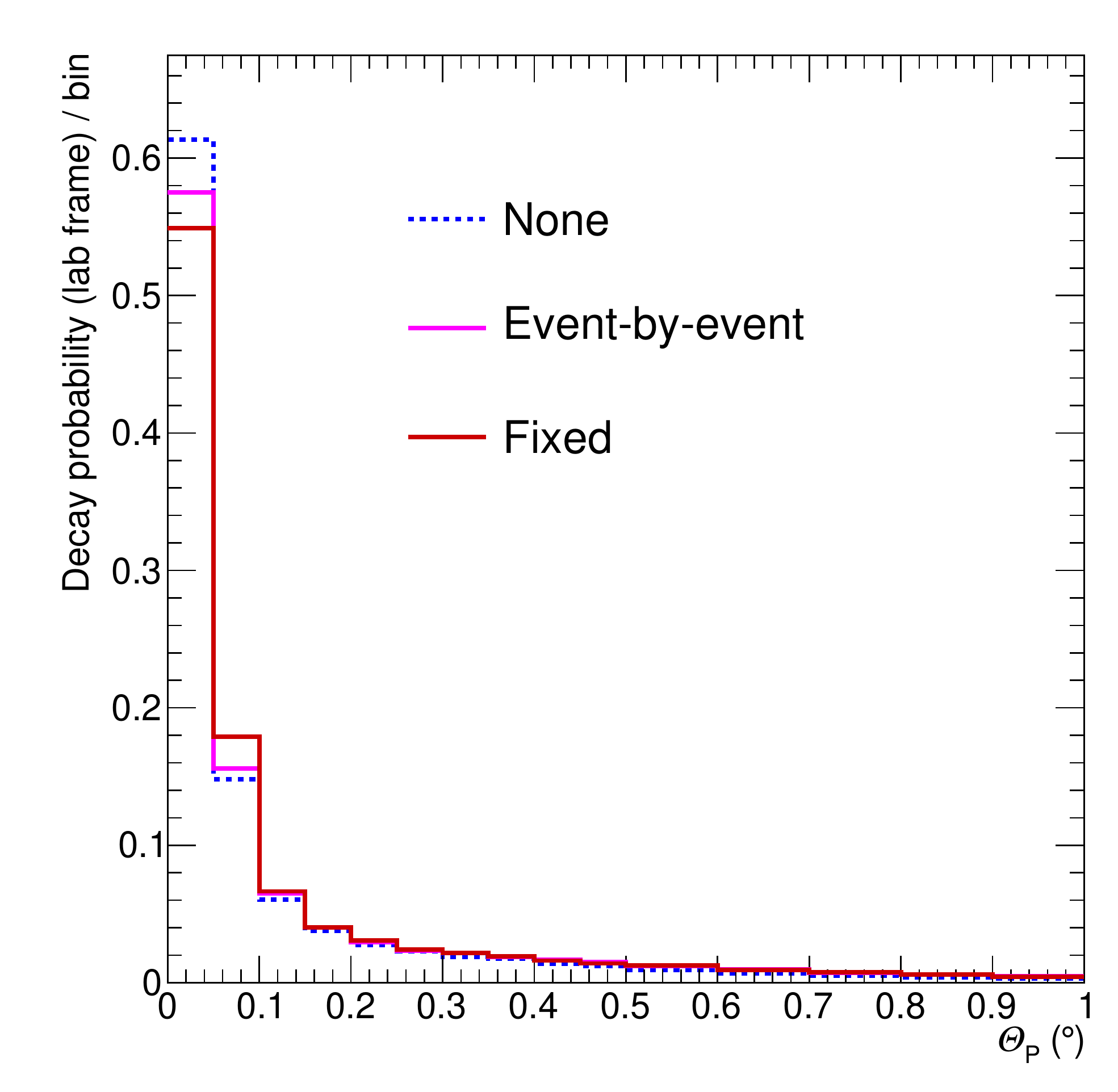}
          \caption{Lab frame}
        \end{subfigure}
        \hfill
        \begin{subfigure}[b]{0.23\textwidth}
          \centering
          \includegraphics[width=\textwidth]{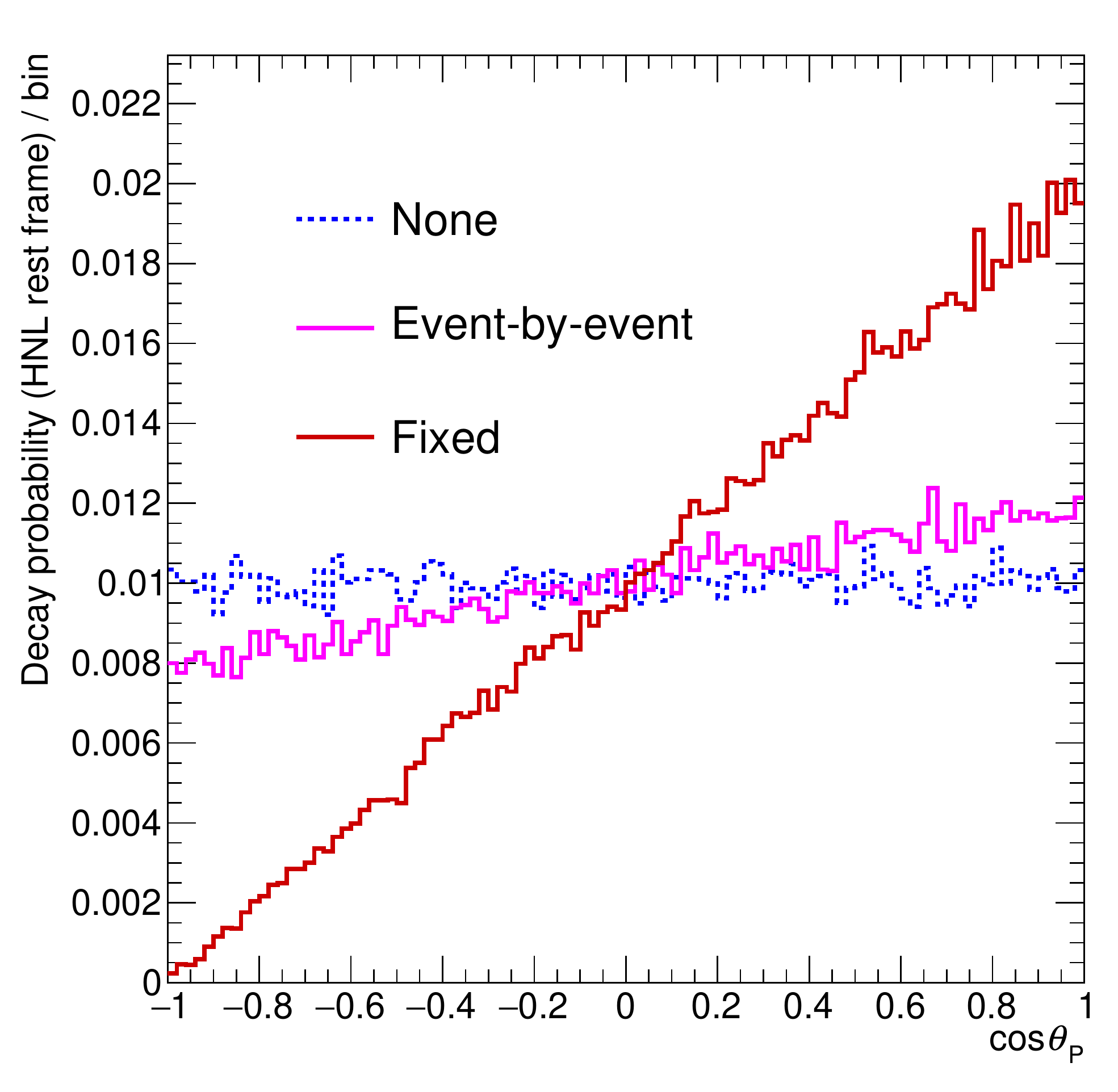}
          \caption{HNL rest frame}
        \end{subfigure}
        \caption{Angular distributions of final-state muons in $K^{+} \rightarrow N_{4}+e^{+},\,\,N_{4} \rightarrow \pi^{+}+\mu^{-}$ for Dirac HNL with $M_{\textrm{N}4} = 400\,\,\text{MeV}/c^{2}$, using NuMI flux and for a detector located at MINER$\nu$A's coordinates. See text for details.}
        \label{fig:polarisation}
      \end{figure}

\begin{figure}
        \centering
        \includegraphics[width=0.4\textwidth]{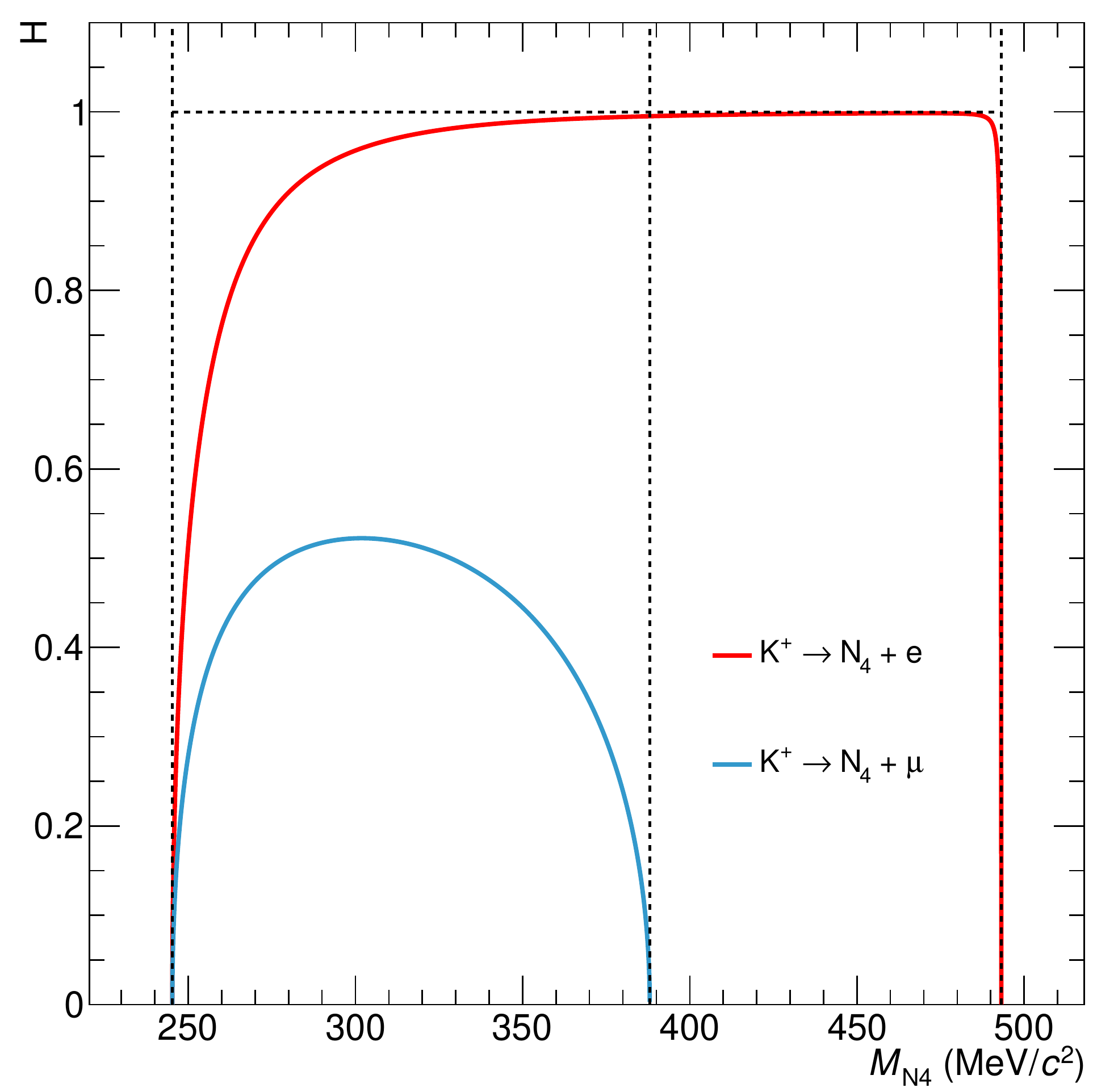}
        \caption{Polarisation modulus $H$ for the chains $K^{\pm} \rightarrow N_{4} + e^{\pm}, N_{4} \rightarrow \mu^{\mp} + \pi^{\pm}$ (red), $K^{\pm} \rightarrow N_{4} + \mu^{\pm}, N_{4} \rightarrow \mu^{\mp} + \pi^{\pm}$ (blue).}
        \setlength{\belowcaptionskip}{-5pt}
        \label{fig:polModulus}
      \end{figure}

 We have simulated in Fig. \ref{fig:polarisation} the effect of three different polarisation prescriptions for $400\,\,\text{MeV}/c^{2}$ HNL from the kaon decay $K^{+} \rightarrow N_{4} + e^{+}$, decaying as $N_{4} \rightarrow \pi^{+} + \mu^{-}$ at the MINER$\nu$A detector in the NuMI beam \cite{TheMINERvADetector}.
      The modulus $H$ is plotted as a function of $M_{\textrm{N}4}$ in Fig. \ref{fig:polModulus}, with the red curve corresponding to the production mode $K^{+} \rightarrow N_{4} + e^{+}$ and the blue curve to $K^{+} \rightarrow N_{4} + \mu^{+}$; for $M_{\textrm{N}4} = 400\,\,\text{MeV}/c^{2}$ and $K^{+} \rightarrow N_{4} + e^{+}$, it is $H \simeq 0.9961$.
      The simplest prescription in Fig. \ref{fig:polarisation} (blue curve) is the absence of any polarisation effect; the angular distribution of final-state products is isotropic in the HNL rest frame, as expected.
      In the red curve, a ``maximal scenario" of polarisation is implemented.
      The \emph{direction} of $\mathbb{P}$ is kept fixed to $\widehat{z}$, which shows up as a distribution $1 + H\cos\theta_{\mathbb{P}}$, where $\theta_{\mathbb{P}}$ is the angle between the muon and $\widehat{z}$.
      This shows that, for suitable HNL mass, there could be a significant polarisation effect if one did not account for the variation of the direction of $\mathbb{P}$.
      Finally, the magenta curve implements the realistic scenario where the direction of $\mathbb{P}$ is evaluated event-by-event according to the procedure outlined above Eq. (\ref{eq: polModulus}).
      It is immediately apparent that the polarisation is almost completely washed out; this occurs as a consequence of backwards-emitted HNL being accepted by the detector.
      For such a heavy mass, there is almost the same chance of a backwards-emitted HNL being accepted as a forwards-emitted one; this ``averages out" the polarisation effect. 
      \par The proper implementation of polarisation effects is doubtlessly going to be crucially important for lower-energy beamlines, or decays at rest, where the Lorentz transformations into the lab frame are small or identity; for the case of the $120\,\,\text{GeV}$ NuMI beam, for example, the transformation into the lab frame makes polarisation effects matter very little for the correct description of final-state kinematics, affecting the opening angle $\Theta_{\mathbb{P}}$ between the muon momentum and beam direction in the region $\Theta_{\mathbb{P}} \lesssim 0.1^{\circ}$.
      \subsection{Determination of decay vertex}
      \par The final task for the description of a single HNL decay event is its location in spacetime, given some origin; frequently, the user will input their own coordinate system, which is just the USER frame defined earlier.
      \par Given the velocity $\beta$ and lifetime $\tau$ of the HNL, one can calculate where in the detector the decay occurs, which is of unique interest to segmented detectors such as the DUNE near detector \cite{DUNE_ND_CDR}, whose different submodules may have different tracking capability, thresholds, geometries and fiducial volumes, etc.
      \par Since fast HNL also spend less time in the detector, there is an effect of the HNL velocity (and hence parent velocity) on the expected spectrum of decays; in other words, the detector dimensions and location are convolved with the flux information \cite{Chun_2019}.
      \par Heavier HNL tend to ``lag behind" Standard Model neutrinos, and thus could, for long enough baselines and small enough velocities, decay during a timing window with little to no Standard Model expected background such as the period in between beam spills, or even in between beam bunches (\cite{ShrockBeamBuckets, FNALFMMF}).
      This lends itself primarily to the development of special HNL triggers \cite{Porzio2019, Ballett2017}, and underlines the importance of the accurate determination of the position of the HNL decay, as well as the HNL energy and position relative to the USER frame.
      \par
\begin{figure*}
        \centering
        \begin{subfigure}[b]{0.38\textwidth}
          \centering
          \subfloat[$25\,\,\text{MeV}/c^{2}$, unweighted]{%
            \includegraphics[clip,width=\textwidth]{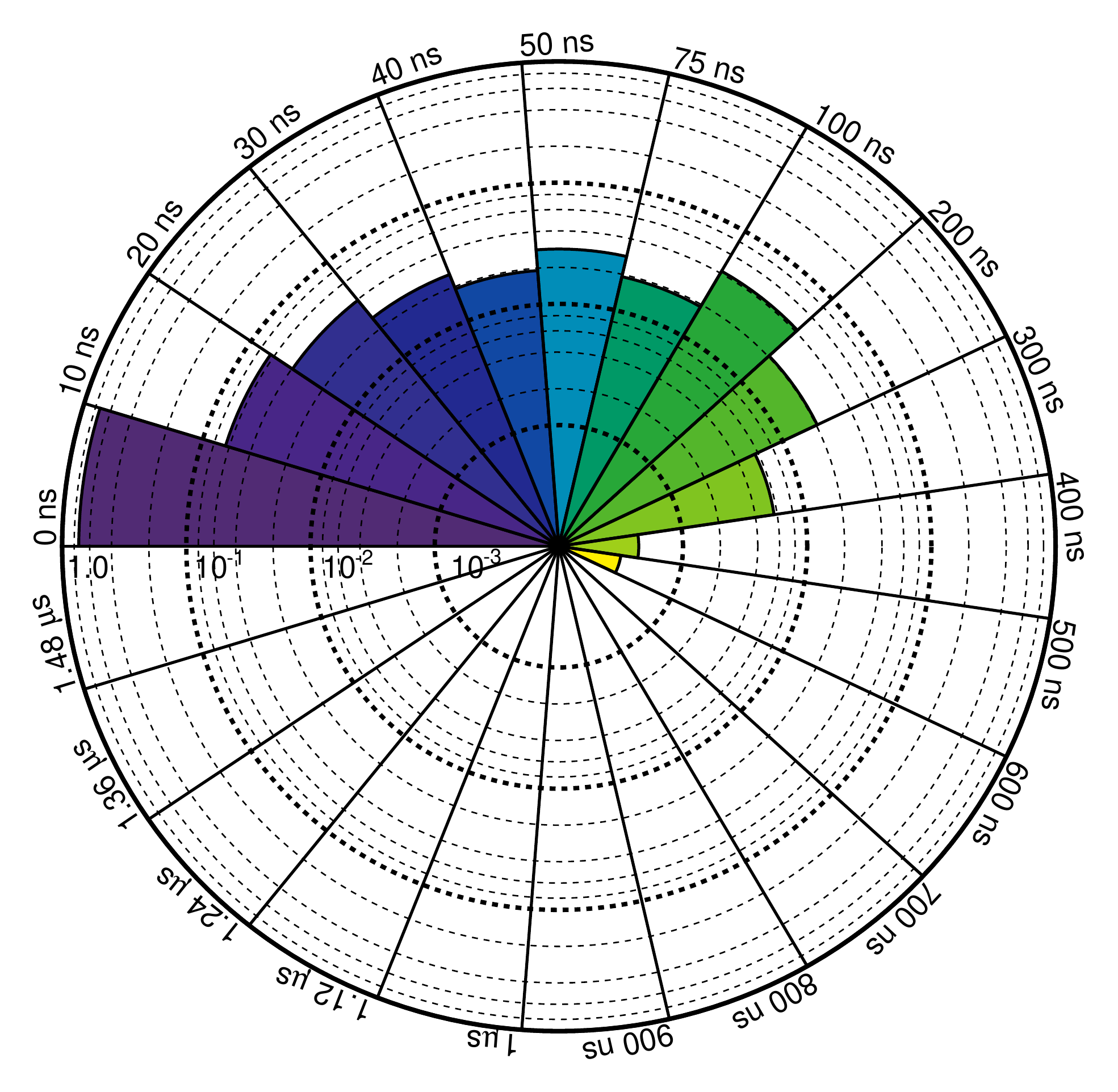}
          }

          \subfloat[$100\,\,\text{MeV}/c^{2}$, unweighted]{%
            \includegraphics[clip,width=\textwidth]{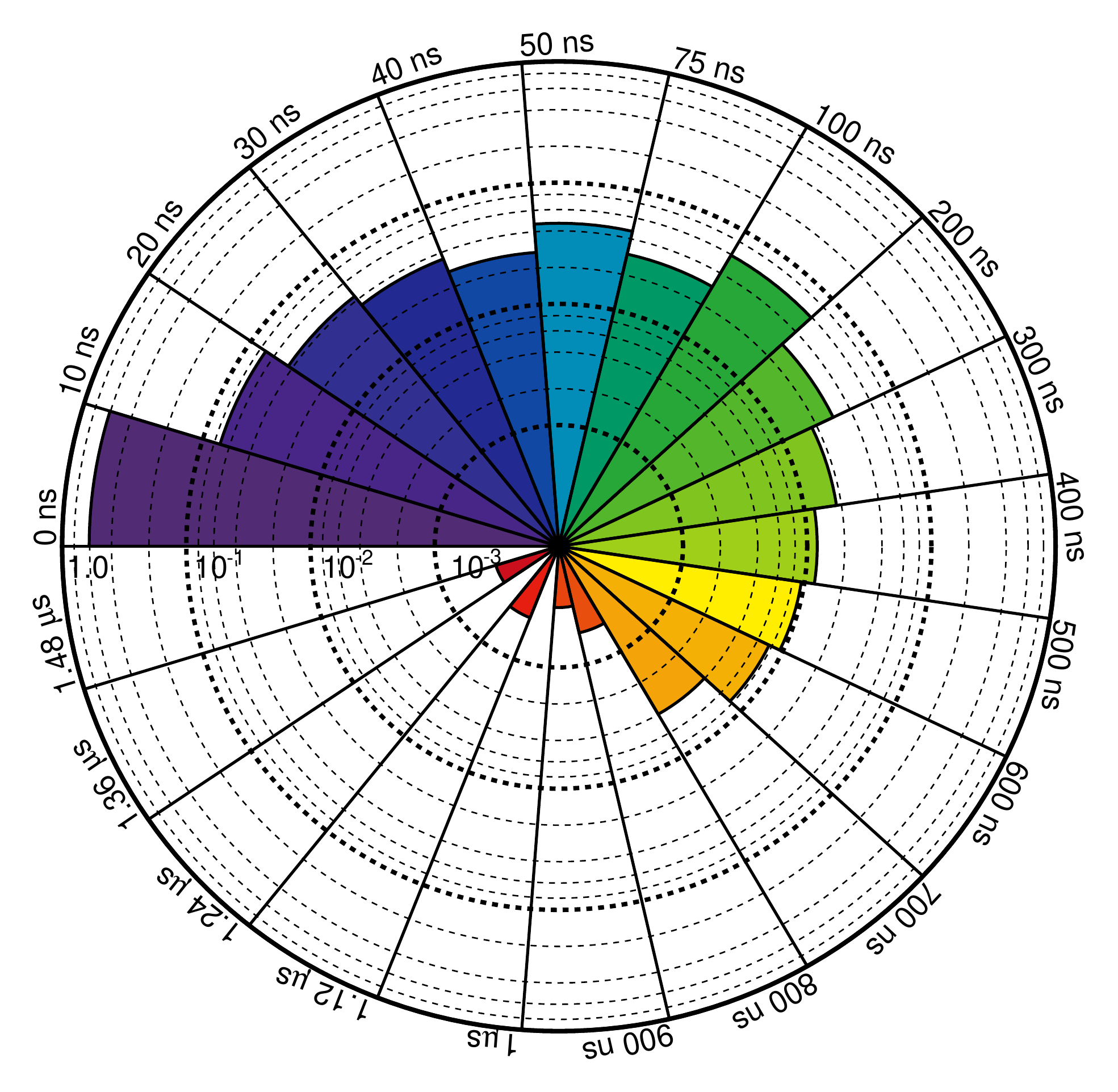}
          }

          \subfloat[$250\,\,\text{MeV}/c^{2}$, unweighted]{%
            \includegraphics[clip,width=\textwidth]{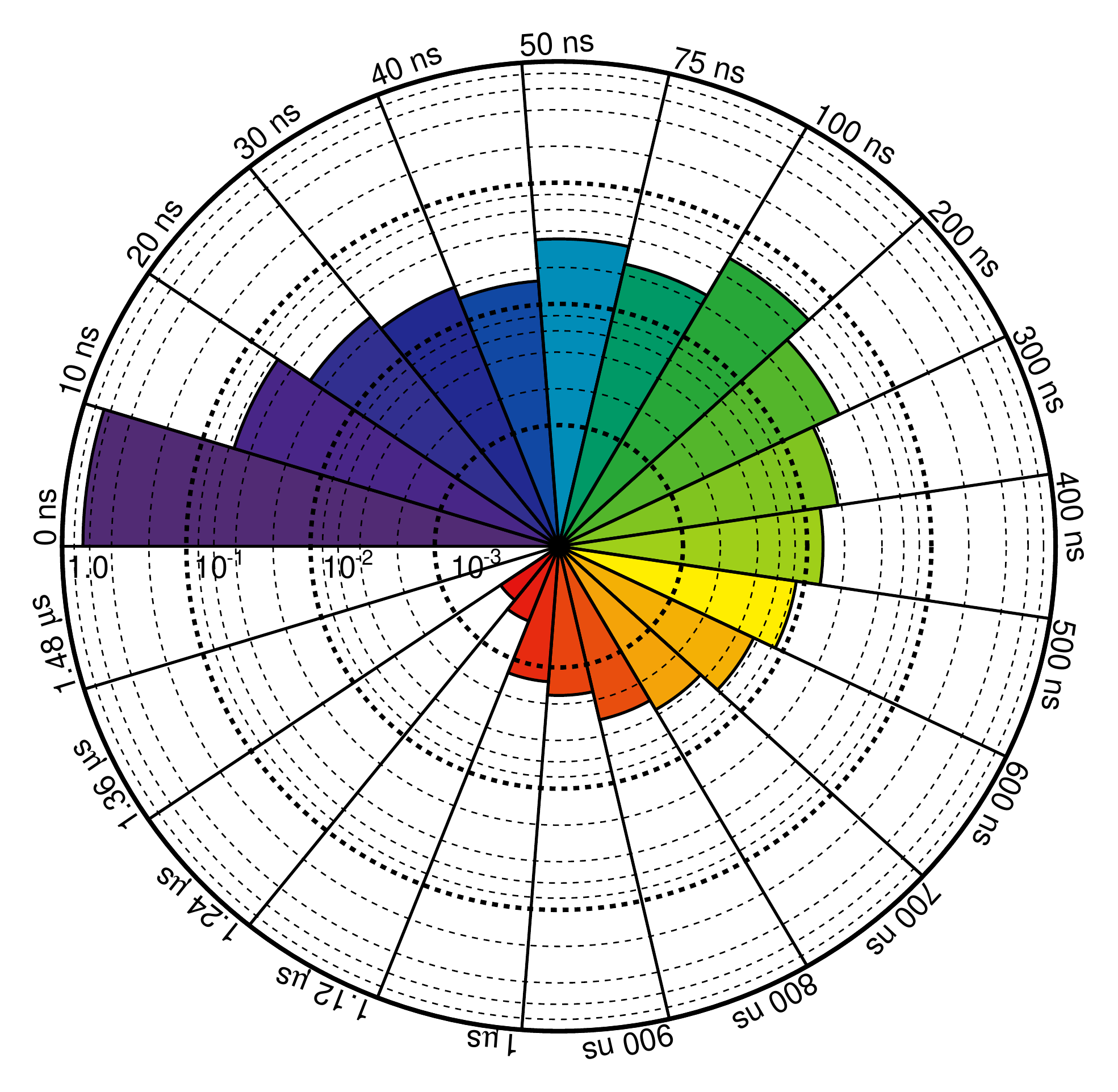}
          }
        \end{subfigure}
        \begin{subfigure}[b]{0.38\textwidth}
          \centering
          \subfloat[$25\,\,\text{MeV}/c^{2}$, weighted]{%
            \includegraphics[clip,width=\textwidth]{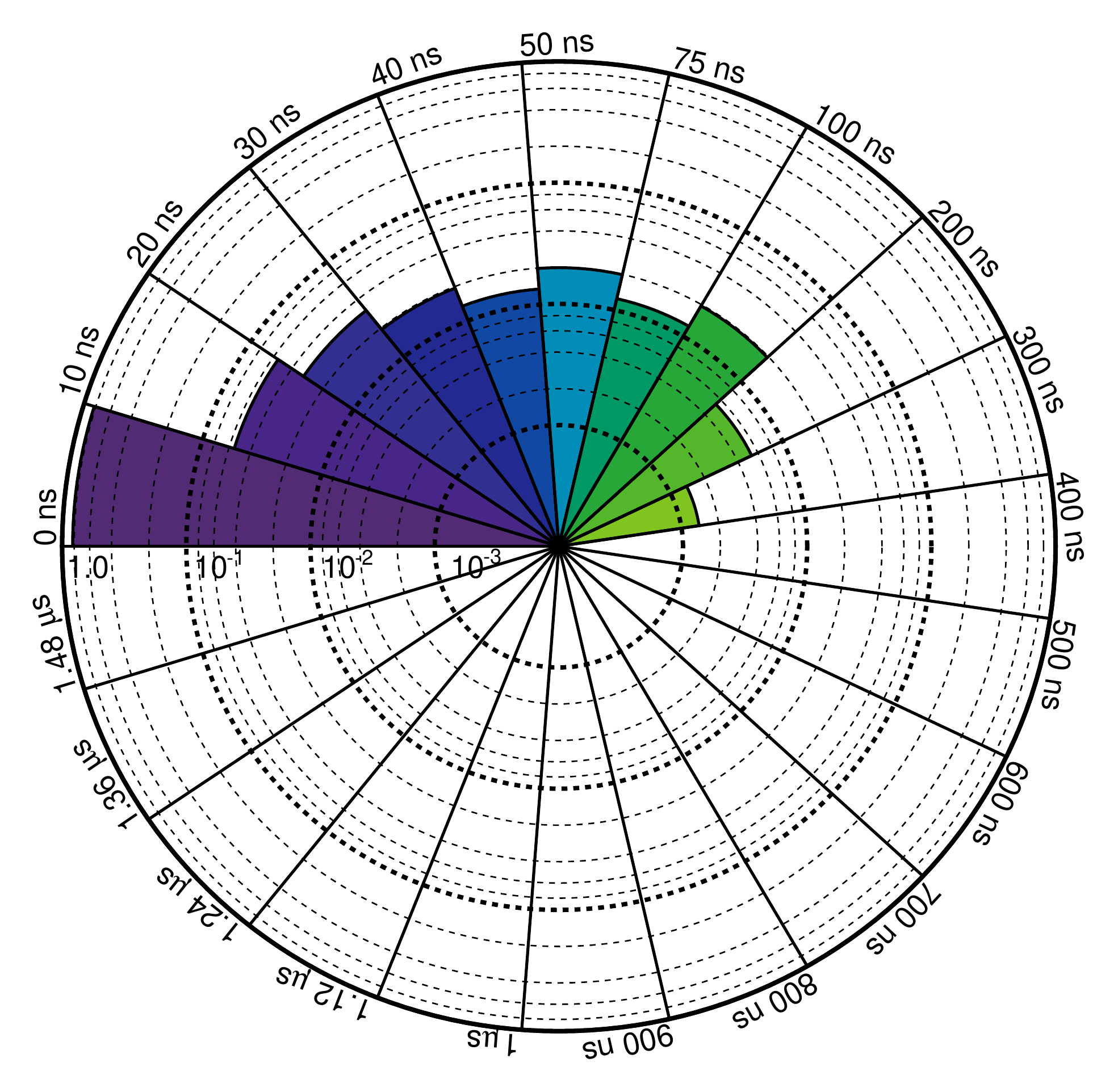}
          }

          \subfloat[$100\,\,\text{MeV}/c^{2}$, weighted]{%
            \includegraphics[clip,width=\textwidth]{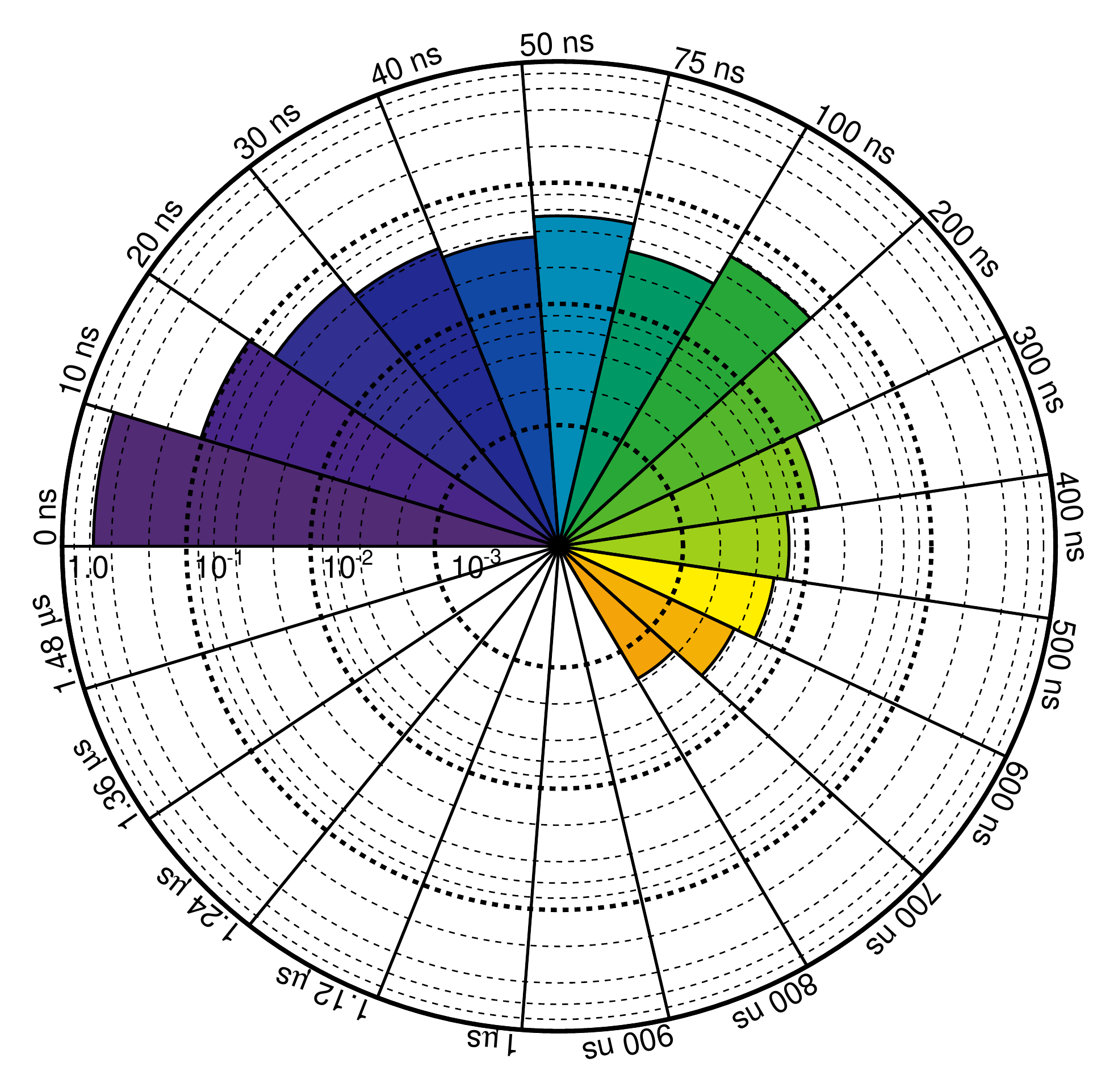}
          }

          \subfloat[$250\,\,\text{MeV}/c^{2}$, weighted]{%
            \includegraphics[clip,width=\textwidth]{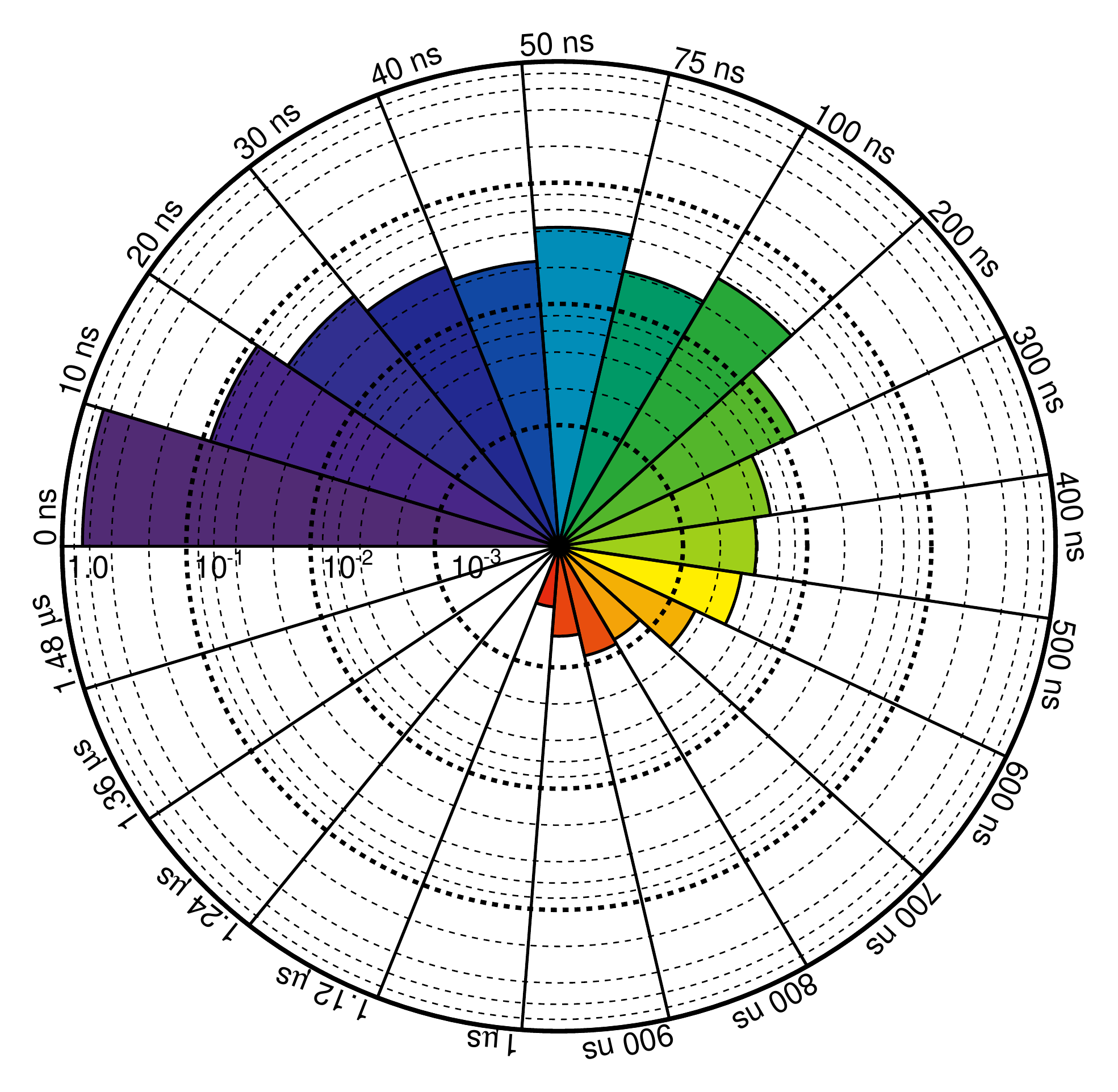}
          }
        \end{subfigure}
        \caption{Delays of HNL with respect to a SM neutrino produced by the same parent, for the MINER$\nu$A detector. Each circular slice represents one timing bin. The final timing bin is $[1.48,\,1.6]\,\upmu\text{s}$. Panels (a), (b), (c): HNL decay events as a proportion of all simulated events. Panels (d), (e), (f): HNL decay events, weighted for the probability of production, propagation, and decay, as a proportion of the sum of weights of all events. Note the logarithmic scale on the radial axis.}
        \label{fig:delays}
      \end{figure*}

 In Fig. \ref{fig:delays}, we show the distribution of the delay for HNL arrival at the MINER$\nu$A detector, compared to the arrival time of Standard Model neutrinos. 
      Each slice of a circular plot corresponds to one bin of delay, i.e. the first bin signifies delay $\Delta t \in [0, 10]\,\,\text{ns}$, and so on.
      The span of the timing bins covers the range $\Delta t \in [0, 1.6]\,\,\upmu\text{s}$, which is roughly the length of one beam spill in NuMI.
      The left panels (a), (b), (c) are filled with the proportion of HNL events in each bin over all HNL events, whereas the right panels (d), (e), (f) are filled with the HNL events, weighted for the overall probability that each HNL would be produced, propagated, and decayed inside the detector, normalised to all HNL.
      Note that the radial direction is on a logarithmic scale, and that bins are only drawn if their content is greater than $10^{-3.5}$.
      The mass $M_{\textrm{N}4}$ increases from $25\,\,\text{MeV}/c^{2}$ on the top row to $250\,\,\text{MeV}/c^{2}$ on the bottom row.
      \par As expected, heavier HNL travel slower than lighter ones, and end up arriving at the detector appreciably later. 
      Though there exist a small handful of HNL events with great delays that could end up delayed by about the length of one NuMI beam spill, they are less likely to survive long enough to reach the detector.
      This explains why the latest bins seem to ``drop off" in the weighted right column of plots.
      \par We also see that a significant proportion (about $10\%$) of HNL have a delay within the small delay bins $\sim \ord{10\,\,\text{ns}}$. 
      This implies that, for detectors that support sufficient triggering sensitivity to beam-bucket timing (ns scale), it is possible in principle to obtain a trigger for delayed HNL that arrive after the Standard Model neutrinos from the beam have traversed the detector.
      Work utilising such a trigger has already been done by the MicroBooNE collaboration \cite{uBooNEMuPiHNL}, and studied for the SBN programme \cite{Ballett2017}.
      \par The simulation, much like Standard Model \verb|GENIE| output, returns \verb|EventRecord|s that summarise the HNL decay.
      The defining features of the event record are:
      \begin{itemize}
      \item The particle stack, containing each particle's PDG code and four-momentum;
      \item The event vertex, in USER coordinates, with the time component measuring the delay $\Delta t := t\left(\text{HNL reaches}\,\text{V}\right) - t\left(\text{SM}\,\nu\,\text{reaches}\,\text{V}\right)$;
      \item The weight, containing the calculated $N_{\text{POT}}$ for this signal event;
      \item The ``event probability'', containing the HNL lifetime $\tau = 1/\Gamma_{\text{tot}}$ in units of $\text{GeV}^{-1}$.
      \end{itemize}
      We present more information about configuration and running of the module in Appendix \ref{appdx: code}.
      \section{Discussion} \label{SECT_discussion}
      \par The \verb|BeamHNL| module currently implements the modified neutrino kinematics including the collimation effect, polarisation effects, produces realistic spatial and time distributions of HNL decay vertices, and robustly calculates production / decay rates.
      This corresponds to a detailed picture of the most prominent physics effects relevant for HNL below the kaon mass in all stages from parent spectra to final state distributions.
      Despite the wealth of physics effects already presented, however, there exist avenues for extension of the functionality of the module.
      Each of the following additions to the codebase would represent a successively more complete and general description of the physics of massive neutrinos, which can provide the basis for an expansion of the validity of the code from the medium beam energy \ord{100\,\,\text{MeV} - 10\,\,\text{GeV}} range to the collider regime, with support for an increasing number of production channels and decay models that can be tested, such as the one described in \cite{ADasCollider}, or to the atmospheric regime.
      For example, it would be possible in the short term to proceed in a manner similar to \cite{Ballett2020, DipoleNuTail} to implement a spectrum of $D_{s}$ mesons and $\tau$ and their decays into HNL, unlocking a higher mass range still probe-able in the medium-energy regime.
      \par Equally, an appealing extension to the module's capability would be to generate HNL from the decays of hadrons in the atmosphere (or, generally, \emph{not} travelling along some well-defined axis).
      This would open the possibility to probe HNL of extraterrestrial origin, which raises the possibility of leveraging atmospheric neutrinos to obtain higher coverage of the parameter space at low HNL masses.
      Work towards estimating the sensitivity of Super-Kamiokande (SK) has already been undertaken in \cite{SKAtmHNL, AtmoLLP}; as an example of a natural next step, we envisage adapting the flux description provided in that work to work with our module in order to simulate HNL decays in SK and future neutrino experiments such as DUNE, Hyper-Kamiokande or JUNO.
      Atmospheric experiments may also be sensitive to HNL produced through active neutrino upscattering in the Earth; IceCube is also sensitive to such a production mechanism for HNL \cite{IceCubeHNL}.
      \par HNL may also be produced in nuclear beta decays (for example, in the Sun through $^{8}\textrm{B} \rightarrow \,^{8}\textrm{Be} + e^{+} + N_{4}$), which raises the possibility of probing the low-mass range through direct searches.
      On the solar neutrino front, the Borexino collaboration has published results of a search for the decay $N_{4} \rightarrow \nu + e^{+} + e^{-}$ \cite{BorexinoHNL}.
      This paradigm also applies to reactor HNL, where there exist limits from HNL decays in detectors (including the radiative decay $N_{4} \rightarrow \nu_{i} + \gamma$) \cite{BugeyReactorHNL, SoLidHNL}.
      As opposed to the collider or atmospheric paradigms, though, HNL from beta decays constitute a different HNL production mechanism entirely and work to incorporate this mode into our module would necessarily be more profound.
      It is thus alluring to imagine the likelihood of multiple HNL production mechanisms being incorporated for a fuller description of the physics of HNL in the mass-mixing portal.
      \par Alternatively, there has been quite some activity recently to estimate the full differential distributions of HNL decays to three bodies, including in the case for Majorana HNL \cite{HNL3BodyDecaysPol, Ballett2017, BahaDiracVsMajPol}.
      The full description of higher-multiplicity polarisation is an attractive goal for future high-precision experiments, assuming HNL are discovered; in principle, constructing a beam of low-energy HNL with small Lorentz boosts could allow for a measurement of the angular distributions of decay products \cite{HNL3BodyDecaysPol}, which could lend insight into the nature of the neutrino as a Dirac or a Majorana particle \cite{BahaDiracVsMajPol}.
      However, as can be seen in Fig. \ref{fig:polarisation}, the details of such an implementation become important, for an accelerator neutrino context, only at sub-degree detector angular resolutions.
      \par Notwithstanding the model-independent effect of HNL polarised decay, we must point out that our current treatment of HNL decays to visible particles is dependent on the effective field theory presented in \cite{ColomaDUNEHNL}. 
      The production philosophy of parent-decay $\rightarrow$ HNL propagation $\rightarrow$ HNL decay, on the other hand, is model-independent, which motivates an effort to generalise the interface such that alternative HNL decay models can be accommodated.
      In our view, this should be as general as the choice of input flux simulation or geometry description, to enable versatile usage of this module for different theories that differ only as to the decay widths they predict.
      Given that the most stringent limits on the mixing elements $\{|U_{\alpha 4}|^{2}\}$ are placed by dedicated experimental searches that look for model-dependent kinematic signatures \cite{deGouveaGlobalConstraints}, it is important to enable the implementation of models that all share the same conception of HNL production and decay, in order to maximise the utility of experimental data.
      \par
\begin{widetext}
        \begin{figure*}
          \centering
          \begin{subfigure}[b]{0.32\textwidth}
            \centering
            \includegraphics[width=\textwidth]{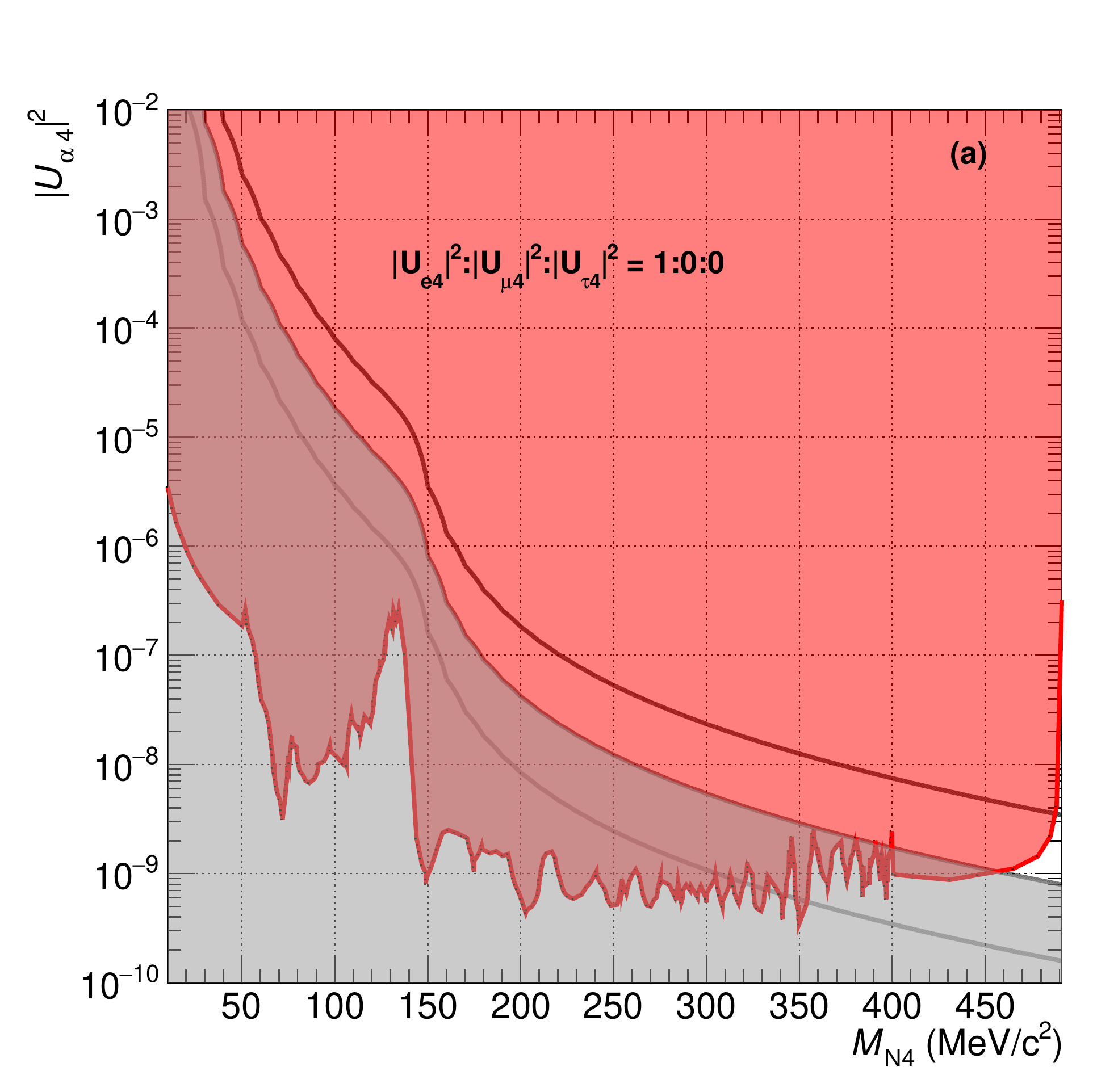}
          \end{subfigure}
          \hfill
          \begin{subfigure}[b]{0.32\textwidth}
            \centering
            \includegraphics[width=\textwidth]{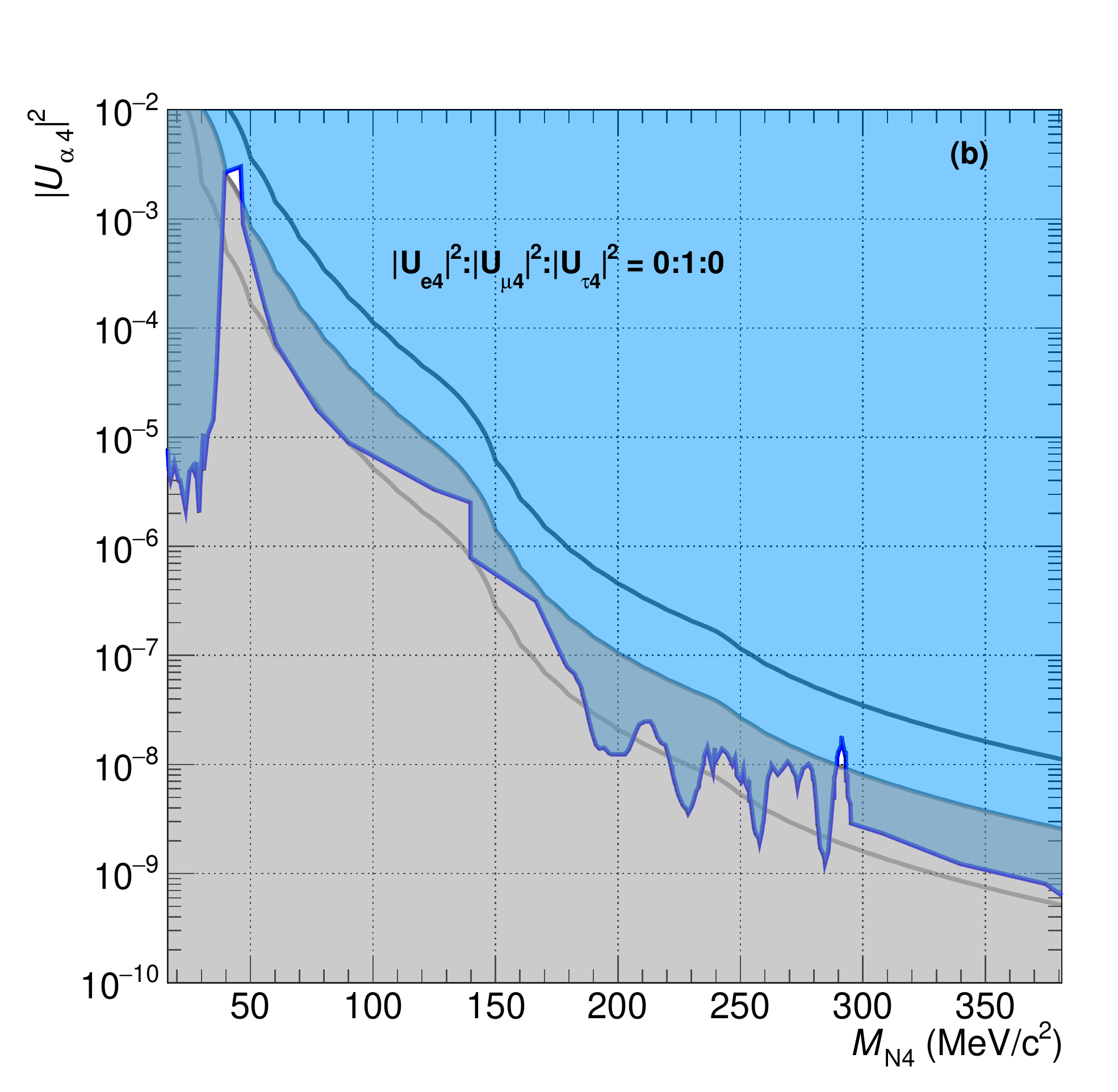}
          \end{subfigure}
          \hfill
          \begin{subfigure}[b]{0.32\textwidth}
            \centering
            \includegraphics[width=\textwidth]{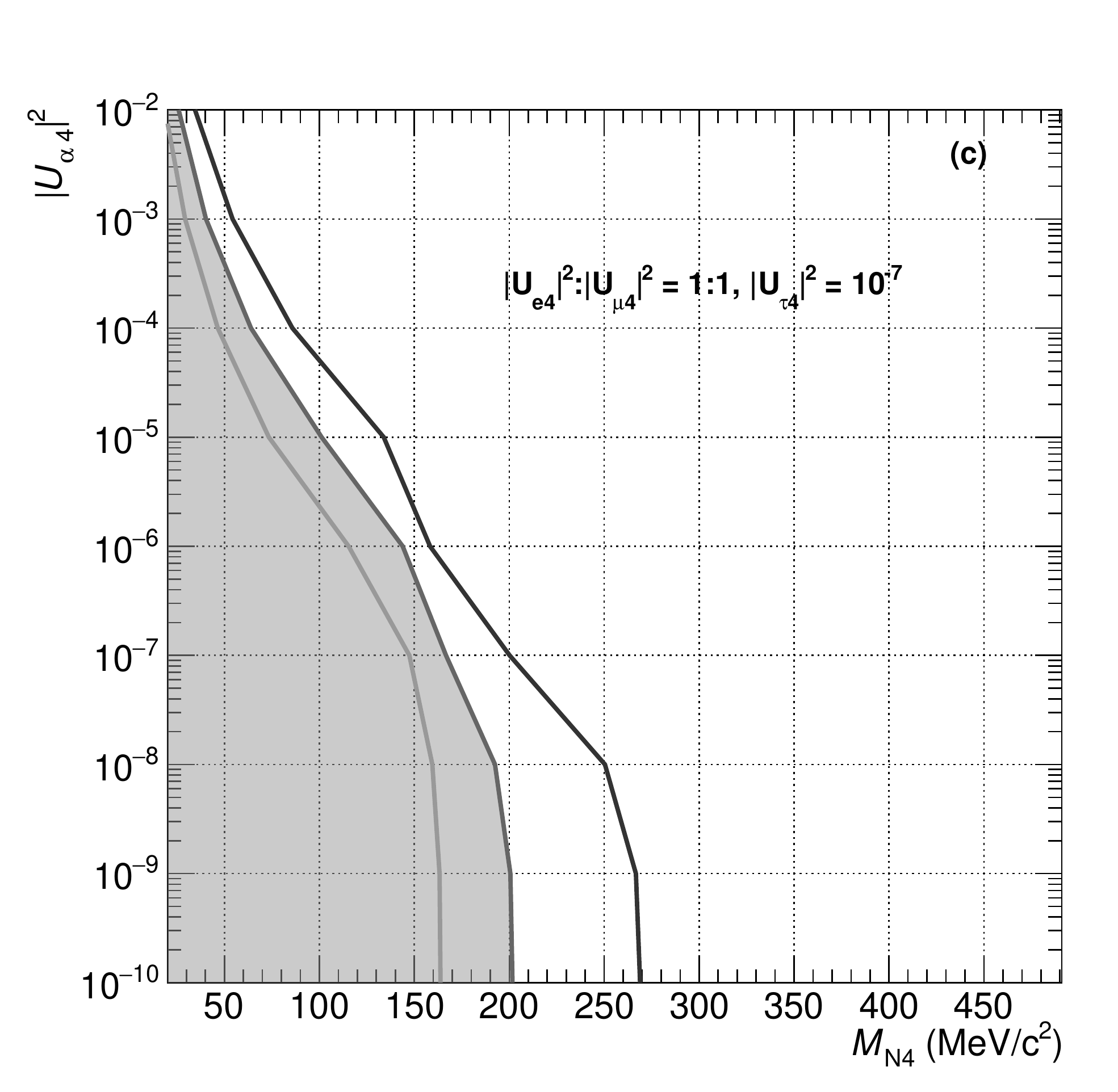}
          \end{subfigure}
          \caption{Parameter space limits for HNL searches below the kaon mass, with limits used from \cite{SnowmassHNLOverview, NA62EDominance, NA62MDominance, PIENU_Ue42, PIENU_Um42, T2KMainHNLSearch} and BBN contours shown at $\tau = 0.023, 0.1\,\,(\text{filled}), 0.5\,\,\text{s}$ \cite{BoyarskyBBN}.}
          \label{fig:limits}
          \vspace*{-2em}
        \end{figure*}
      \end{widetext}

 The HNL parameter space under the flavour mixing assumptions $\left|U_{e4}\right|^{2}:\left|U_{\mu4}\right|^{2}:\left|U_{\tau4}\right|^{2} = 1:0:0, 0:1:0$ is constrained by particle physics experiments (see for example \cite{SnowmassHNLOverview, BrymanModelIndependentBounds} and references therein).
      Complementary bounds are obtained by measuring the primordial elemental abundances that result from Big Bang Nucleosynthesis (BBN); if the lifetime $\tau$ of HNL is long enough, they decay to mesons in the primordial plasma, which in turn forces proton and neutron number densities towards equilibrium and alters the primordial elemental abundances.
      In Fig. \ref{fig:limits}, we have shown in panels (a) and (b) the currently excluded parameter space, with the coloured region corresponding to experimental limits (\cite{SnowmassHNLOverview, NA62EDominance, NA62MDominance, PIENU_Ue42, PIENU_Um42, T2KMainHNLSearch}), and the shaded grey area corresponding to a BBN constraint $\tau \leq 0.1\,\,\text{s}$ (\cite{BondarenkoBBN}).
      Also shown are the constraints for $\tau \leq 0.023\,\,\text{s}$ (\cite{BoyarskyBBN}), $\leq 0.5\,\,\text{s}$ as solid grey lines.
      \par As can be seen, for these mixing assumptions the conjunction of experimental limits with BBN almost completely covers the avaiable parameter space for a single HNL below the kaon mass.
      Note, however, that the constraint $\tau \lesssim \tau_{0}$, for $\tau_{0}$ some suitable upper bound, places constraints on the \emph{total} effective mixing $\sum_{\alpha}\left|U_{\alpha 4}\right|^{2}$, which makes the curves in Fig. \ref{fig:limits} dependent on the assumed flavour structure.
      In panel (c), we show how assuming a large enough mixing with $\tau$ modifies the BBN bounds; assuming $\left|U_{e4}\right|^{2}=\left|U_{\mu4}\right|^{2}=\left|U_{\alpha}\right|^{2}$ and $\left|U_{\tau 4}\right|^{2} = 10^{-7}$, the parameter space above $\left(M_{\textrm{N4}}, |U_{\alpha4}|^{2}\right) \sim (200\,\,\text{MeV}/c^{2}, 10^{-10})$ is completely free of BBN constraints for the same HNL lifetimes.
      We show no experimental bounds in panel (c), as these are extracted under a particular flavour mixing assumption; to the best of our knowledge, no search has explored this particular mixing scenario.
      \par A final consideration is the implication of the realistic picture where multiple HNL can oscillate into each other; it was shown in \cite{veeMultipleHNL} that this can significantly modify the decay rates into visible channels.
      There has been some discussion in \cite{veeMultipleHNL, MultipleHNL, Resurrection_belowKaon, NA62MultipleHNL} of what the implications of multiple HNL would be on the current most strict bounds on sterile-neutrino mixing into light neutrinos. 
      Specifically, if there exist multiple HNL ($n > 1$, which is required by realistic extensions to the Standard Model that explain neutrino masses), then the current most stringent limits on the HNL parameter space may be significantly relaxed.
      This occurs because, if there exists at least one other HNL $N_{5}$ such that $\Delta M_{45} \ll M_{\textrm{N}4,5}$, then the decay rates to visible final states can change drastically, weakening the current exclusion bounds that are made under the assumption of one heavy neutrino state.
      \par It is, however, possible that the experimental bounds could be relaxed in the multiple-HNL picture, once again opening the allowed window for HNL at $M_{\textrm{N}4} < m_{\textrm{K}}$ (for instance, \cite{MultipleHNL} showed that ATLAS bounds on HNL are relaxed by up to two orders of magnitude, depending on the prevalence of lepton-number violating processes and the precise ratios of the mixing elements).
      A numerical study showing how the parameter space with two HNL is still compatible with both experimental and BBN constraints was carried out in \cite{Resurrection_belowKaon}.
      Future accelerator neutrino experiments will be capable of exploring this parameter space region, which emphasises the crucial importance of consistent, well-founded modelling for HNL searches, as well as the desirability of a multiple-HNL implementation to support these efforts.
      %
      %
      \section{Summary} \label{SECT_conclusion}
      Heavy Neutral Leptons are among the most natural and minimal Beyond the Standard Model theories, which are motivated by the measurement of nonzero neutrino masses, the first evidence that the Standard Model is incomplete.
      They have very important implications for both particle physics and cosmology, which makes them prime targets for searches in accelerator neutrino experiments and beyond.
      Already at masses \ord{100\,\,\text{MeV}/c^{2}}, there is a rich phenomenology derived from the massive neutrino kinematics that is qualitatively different from massless Standard Model neutrinos.
      Any simulation treating HNL needs to carefully implement this phenomenology.
      Indeed, a definite detection of HNL would have ground-breaking implications; on top of being an incontroversible piece of evidence of Beyond the Standard Model physics and a probe for the details of neutrino mass generation, knowledge of how to produce HNL would open up the landscape for precision searches of HNL decays and, through this, the possibility for determining the neutrino's nature.
      \par Our goal is to provide a single, sophisticated simulation that incorporates this new physics in a self-consistent manner, for use with experiments past, present, and future.
      This paper has reviewed the theory of HNL production from heavy meson decays and presented the new physics effects associated with neutrino mass, comparing to the Standard Model case of massless neutrinos.
      We also commented on extant similar simulation efforts, showing how this work is complementary to the alternative production model of HNL-by-upscattering and describing the benefits associated with accepting a generalised hadron beamline simulation and detector description as inputs.
      Importantly, we have decoupled the description of the primary beam and the decay volume from the intrinsic physics of the massive neutrino in a general manner, which we presented here.
      We then discussed interesting phenomenology that is crucial to properly simulate HNL from particle decay.
      We described the philosophy and crucial ingredients of the full modelling chain, from production and propagation to decay in a detector.
      \par With our model, we have constructed a fully factorisable interface to beamline simulation and detector geometry that simulates the detailed phenomenology of HNL in a beamline.
      The module can automatically incorporate future developments in beamline modelling, as these will generally be contained in the beamline simulation provided as input.
      Likewise, as future experiments make design choices about what detector setups to use, this module allows them to study in detail the prospects for HNL detection for all the potential detectors being considered.
      \par Though this implementation leverages the \verb|GENIE| framework, the philosophy behind it is general; likewise, though the first use case could be the FNAL experiments, other accelerator neutrino contexts could also well use this module.
      Furthermore, we sketch a possible development strategy for extending the usability of this already robust simulation, in preparation for exciting future experiments.

      \section*{Acknowledgments} \label{SECT_acknowledgments}
      \par We would like to thank C. Andreopoulos, R. Hatcher, and M. Roda for valuable input and guidance during the \verb|GENIE| development process. We further thank L. Fields for helpful discussions on simulating the neutrino flux.
      We also thank R. Shrock for valuable comments on the manuscript.
      X.L. is supported by the STFC (UK) Grants No. ST/S003533/1 and ST/S003533/2.

      \newpage
      \begin{widetext}
        \begin{figure*}
          \centering
          \begin{subfigure}[b]{0.47\textwidth}
            \centering
            \includegraphics[width=\textwidth]{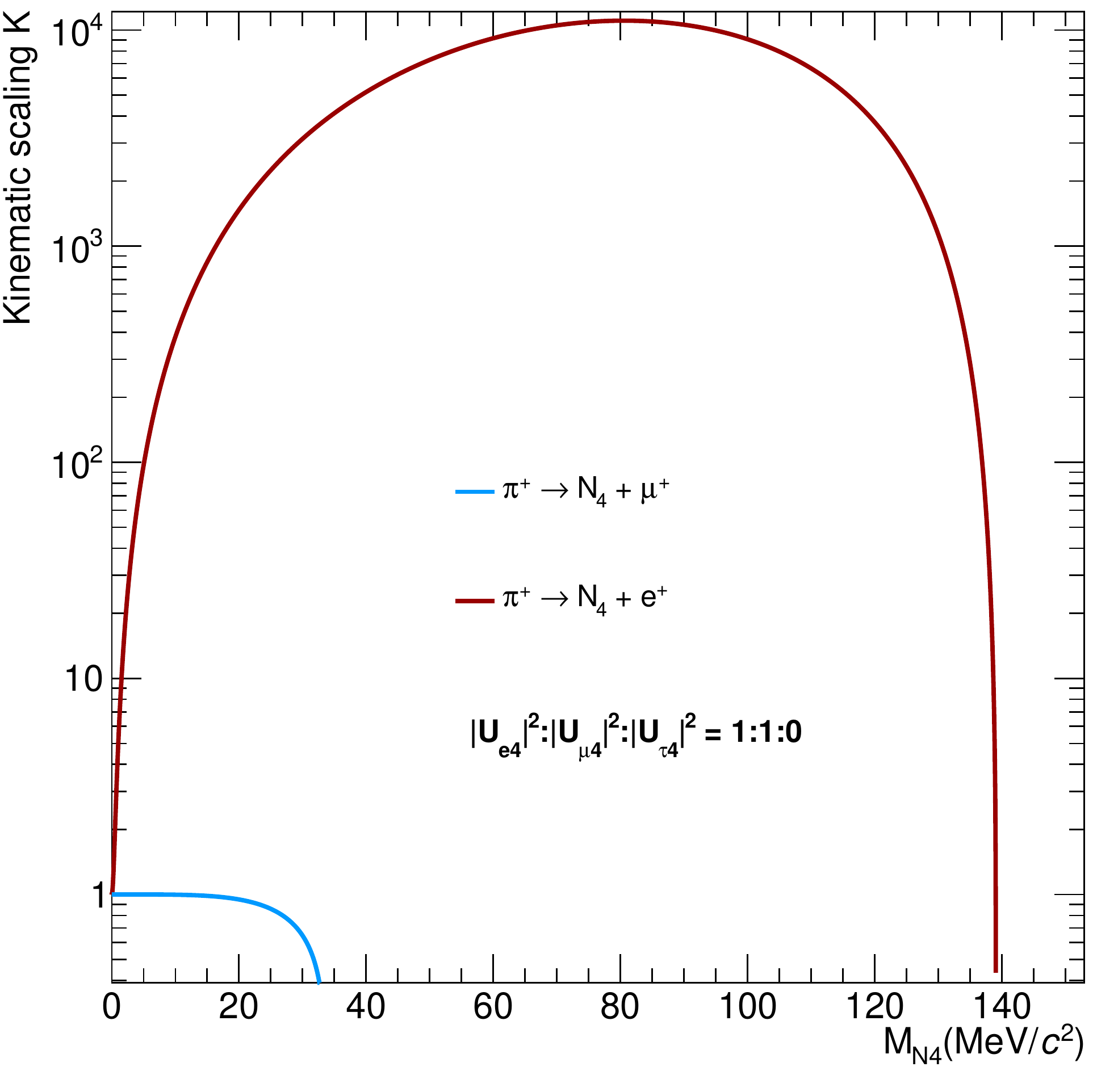}
          \end{subfigure}
          \hfill
          \begin{subfigure}[b]{0.47\textwidth}
            \centering
            \includegraphics[width=\textwidth]{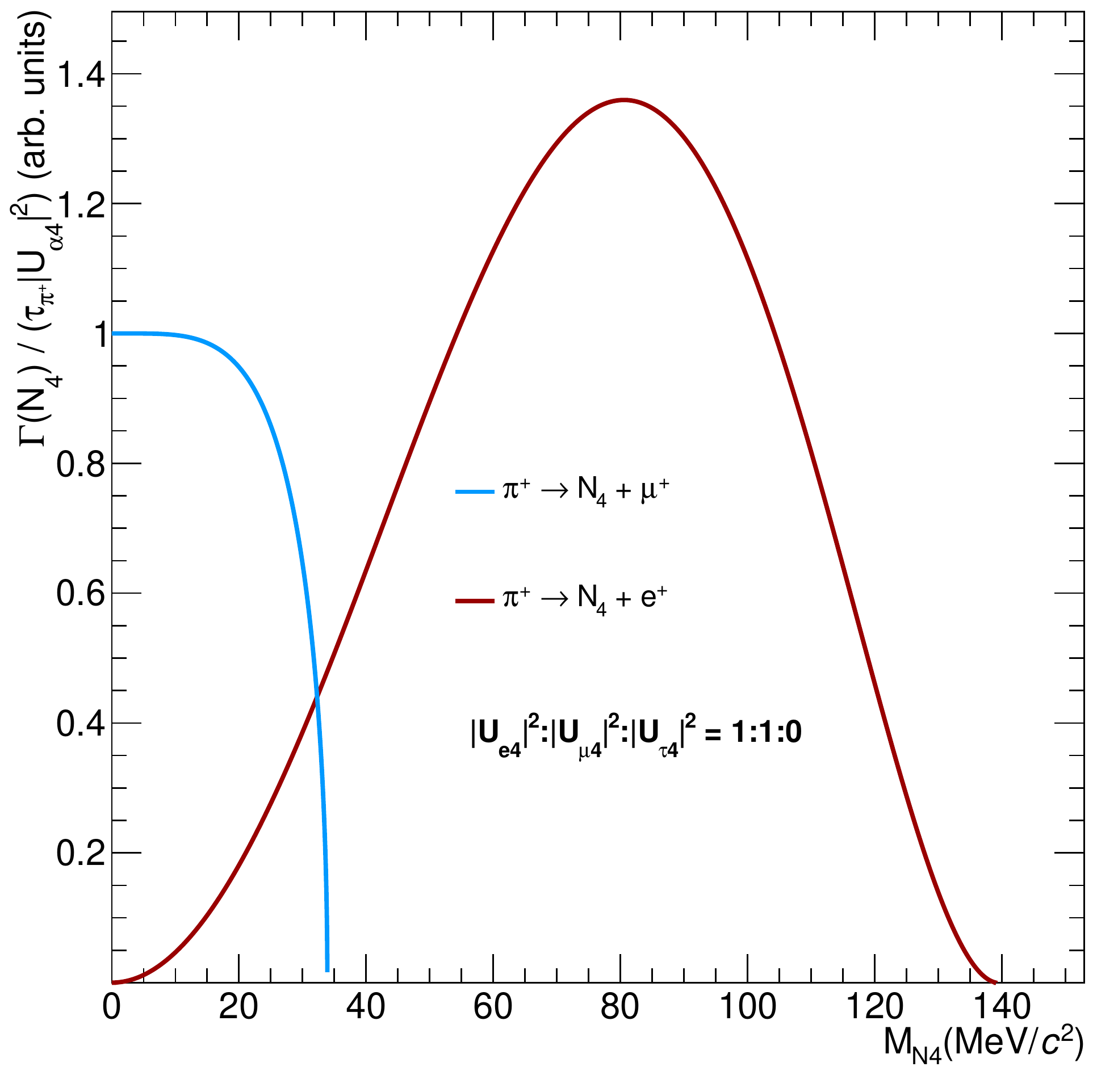}
          \end{subfigure}
          \caption{Kinematic scaling factor and $\mathfrak{B}_{\text{channel}}$ for the HNL production channels resulting from $\pi^{\pm}$ decay. See text for details.}
          \label{fig:piRates}
          \vspace*{-2em}
        \end{figure*}
      \end{widetext}
      \appendix
      \section{Code and algorithm details}
      \subsection{Quick guide to running \texttt{BeamHNL} in \texttt{GENIE}} \label{appdx: code} 
      A fork of \verb|GENIE| with the \verb|BeamHNL| module implementation may be found on GitHub \cite{HNLGENIE}.
      The configuration file has been designed with ease-of-use in batch jobs, such as might be submitted to a computing grid, in mind.
      For this reason, there exists a single file \verb|config/CommonHNL.xml| that houses all the options that the module's components need to know about in order to simulate the HNL production, propagation, and decay.
      A sample file may be found in \verb|src/contrib/beamhnl|, and explanations for the various options are available in \verb|config/NHLPrimaryVtxGenerator.xml|.
      In this section, we will briefly outline the main components.
      \begin{itemize}
      \item \verb|ParameterSpace| defines $M_{\textrm{N}4}$ in $\text{GeV}/c^{2}$, $\{\left|U_{\alpha 4}\right|^{2}\}$, and Dirac vs. Majorana nature of the HNL.
      \item \verb|InterestingChannels| enumerates the 10 decay channels (the 7 summarised in Table \ref{tab: decayChannels} and $N_{4} \rightarrow \pi^{\pm}\pi^{0}\ell^{\mp}, \pi^{0}\pi^{0}\nu$) kinematically accessible to an HNL below the kaon mass. Entries set to \verb|false| will be inhibited from entering the event record, whereas \verb|true| entries are treated as valid truth signal channels.
      \item \verb|CoordinateXForm| defines the unique translation and rotation vectors $\boldsymbol{T}, \boldsymbol{R}$ from NEAR to BEAM and from NEAR to USER coordinates. $\boldsymbol{T}$ is given in metres, with respect to the NEAR system, and $\boldsymbol{R}$ is a vector of 3 Euler angles, following the ``extrinsic $X-Z-X$" convention; that is, the rotation matrix $R$ is given as
        \begin{equation}
          R(\alpha,\beta,\gamma) = R_{\textrm{X}}(\alpha)R_{\textrm{Z}}(\beta)R_{\textrm{X}}(\gamma),
        \end{equation}
        where $X,Y,Z$ are the fixed NEAR axes.
      \item \verb|FluxCalc| provides switches for the user to enable/disable certain features; namely, the module's polarisation accept/reject weight, and whether the simulation should evaluate Eq. (\ref{eq: acceptance_correction}) assuming the parent to be perfectly focused (setting $\zeta_{+}$ to one-half the detector's angular opening, and $\zeta_{-} = 0$)
      \end{itemize}
      For the purposes of running this module, an input \verb|dk2nu|-like flux ROOT flat-tree and a ROOT geometry file describing the detector are required.
      Sample inputs that are module-compliant are supplied along with the source code in \verb|src/contrib/beamhnl| for the user to be able to run the module immediately.
      \par A folder \verb|flatDk2nus| contains two flat ROOT trees, corresponding to one \verb|dk2nu| flux file from NuMI in neutrino mode, and one in antineutrino mode \cite{NuMIBeamFlux}.
      It also contains scripts to produce ROOT flat-trees from \verb|dk2nu| flux files.
      These flat-trees can then be used as input of the \verb|BeamHNL| module.
      Detailed instructions on how to generate these are written in the \verb|README| file located inside the folder.
      \par Three ROOT macros \verb|makeBox.C|, \verb|makeCylinder.C|, and \verb|makeHexagon.C| along with three respective outputs are also inside the \verb|contrib/beamhnl| directory, which allow the user to make three different ROOT geometries of arbitrary dimensions and rotation with respect to the USER coordinate system.
      \begin{figure}
        \centering
        \includegraphics[width=\linewidth]{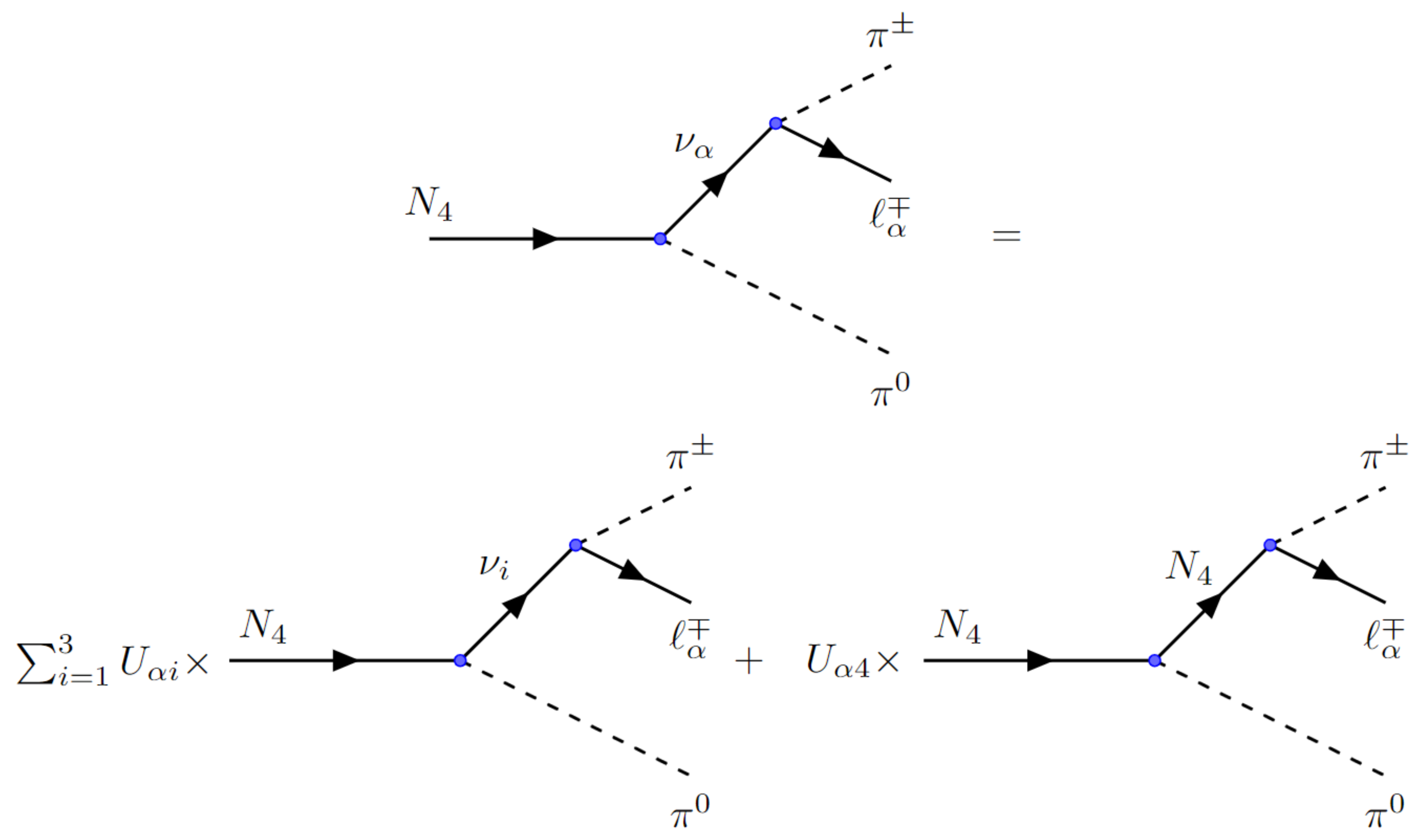}
        \caption{Feynman diagrams for the decay $N_{4} \rightarrow \pi^{\pm}+\pi^{0}+\ell^{\mp}$.}
        \label{fig:pipi0ell}
      \end{figure}
      \par To enable the \verb|BeamHNL| module, the user must first configure \verb|GENIE| appropriately by adding the following line to the \verb|configure| script (an example can be seen at the \verb|GENIE| website \cite{GENIEWebsite}):
      \begin{center}
        \verb|--enable-heavy-neutral-lepton \ |
      \end{center}
      After running \verb|make|, the \verb|BeamHNL| module is ready.
      The archetypal run command is
      \begin{center}
        \verb|gevgen_hnl -n <nSignalEvents>| \verb|-f <path/to/flux/dir> -g <path/to/geom/file.root>|
      \end{center}
      Detailed instructions can be referred to by passing the \verb|-h| flag.

      \subsection{HNL production and decay rates} \label{appdx: QM}
      The available production channels for an HNL with mass $M_{\textrm{N}4} \lesssim m_{\textrm{K}}$ are summarised in Table \ref{tab: prodChannels}. 
      \footnote[3]{
        Throughout this section, we will engage in a mild abuse of notation and write down such expressions as $K^{\pm} \rightarrow N_{4} + \mu^{\pm}$.
        The reader will notice that $N_{4}$ has been defined in Eq. (\ref{eq: fourth_nu_mixing}) as a \emph{mass eigenstate}, rather than a flavour eigenstate.
        In order to shorten notation, we will sacrifice formal correctness (i.e. $K^{\pm} \rightarrow \nu_{\mu} + \mu^{\pm},\,\,\nu_{\mu} = \sum_{i}U_{\mu i}\nu_{i} + U_{\mu 4}N_{4}$) and directly concentrate on the mixed-in element $U_{\alpha 4}N_{4}$, hoping that the implied association does not cause confusion.
      }
      For each channel, the threshold, kinematic scaling $\mathcal{K}$ and SM branching ratio are listed.
      The kinematic scaling $\mathcal{K}$ is defined through the decay width as in Eq. (\ref{eq: kineScaling})
      \begin{equation}
        \Gamma\left(P \rightarrow N_{4} + \ell_{\alpha} + ...\right) = \left|U_{\alpha 4}\right|^{2}\mathcal{K}\cdot\Gamma\left(P \rightarrow \nu + \ell_{\alpha} + ...\right)
      \end{equation}
      The decay width is then, for a given channel, written as
      \begin{equation}
        \Gamma\left(M_{\textrm{N}4},|U_{\alpha 4}|^{2}\right) = \mathcal{K}\left(M_{\textrm{N}4}\right)\cdot|U_{\alpha 4}|^{2},
      \end{equation}
      For each parent species ($\pi^{\pm}, K^{\pm}, K^{0}_{\textrm{L}}, \mu^{\pm}$) a series of scores $\{s_{i}\}, s_{i+1} = s_{i} + \Delta s$ is constructed, which is then used to determine the production channel for the HNL.
      The scores are calculated as
      \begin{equation}\label{eq: scores}
        \Delta s = \frac{\mathfrak{B}_{\text{channel}}}{\mathfrak{B}_{\text{tot}}},
      \end{equation}
      \begin{figure}
        \setlength{\belowcaptionskip}{-1pt}
        \centering
        \includegraphics[width=0.48\textwidth]{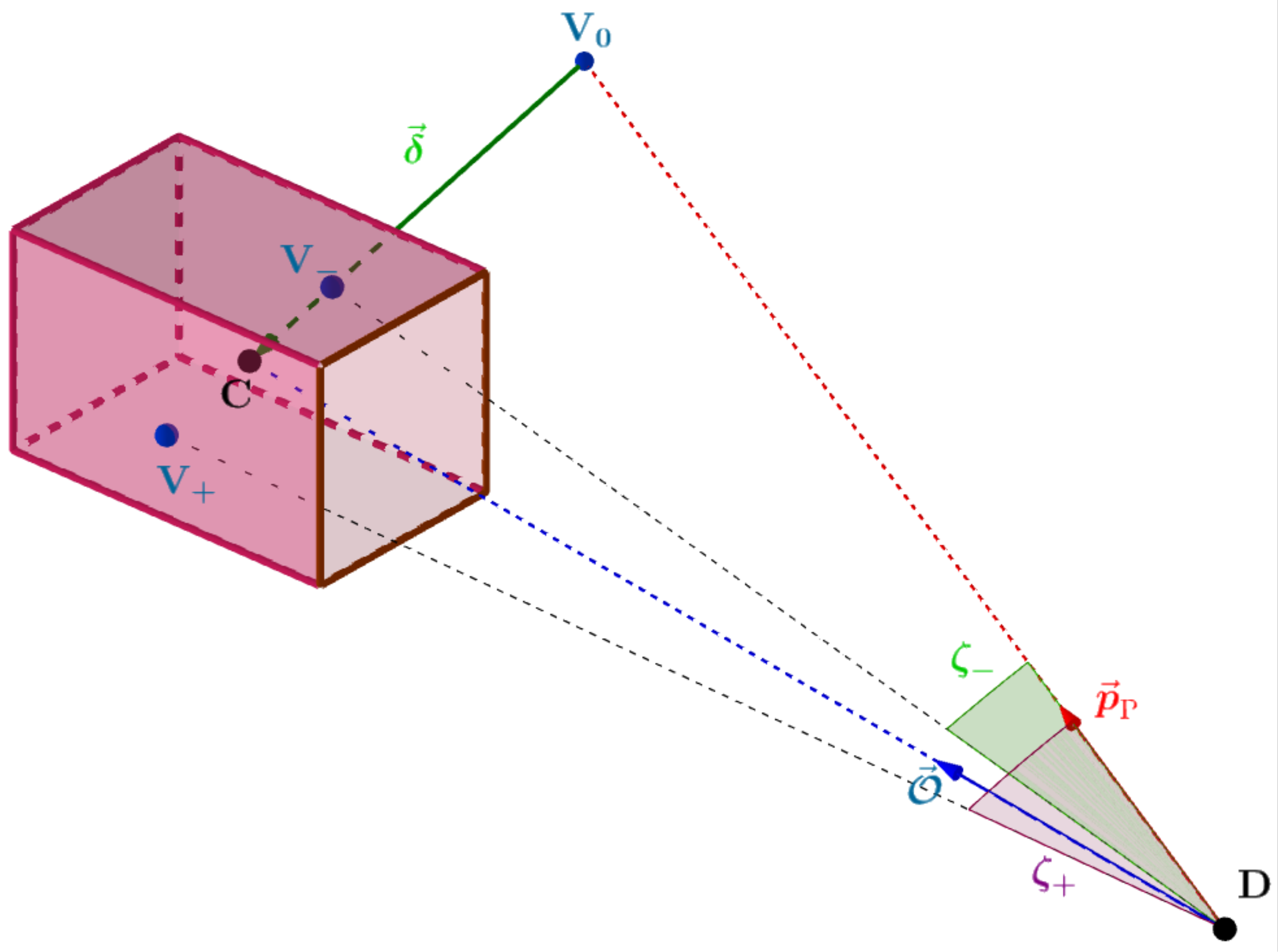}
        \caption{Calculation of deviation angles $\zeta_{\mp}$. The parent's momentum $\boldsymbol{p}_{\textrm{P}}$ is projected to the point $\textrm{V}_{0}$ such that $z_{\textrm{V}0} = z_{\textrm{C}}$, with $\textrm{C}$ the centre of the detector. The entry and exit points $\textrm{V}_{\mp}$ lie on the line $\epsilon: \boldsymbol{r}(u) = \boldsymbol{r}_{\textrm{V}0} + u\cdot\boldsymbol{\delta}$, where $\boldsymbol{\delta}$ is a sweep direction: $\boldsymbol{\delta} := \boldsymbol{r}_{\textrm{C}} - \boldsymbol{r}_{\textrm{V}0}$. The angles $\zeta_{\mp}$ are $\langle\boldsymbol{r}_{\mp}, \boldsymbol{p}_{\textrm{P}}\rangle$.}
        \label{fig:deviation}
      \end{figure}
      where $\mathfrak{B}_{\text{tot}}$ is the sum of $\mathfrak{B}_{\text{channel}}$ over all kinematically accessible HNL production channels (see \ref{tab: decayChannels}).
      \par We illustrate these steps for HNL production by pion decay, with the help of Fig. \ref{fig:piRates}.
      First, we calculate the kinematic scaling factor $\mathcal{K}_{\text{channel}}$ according to Table \ref{tab: prodChannels} (shown in the left panel).
      Afterwards, we multiply each  by the SM branching ratio ($\mathfrak{B}^{\text{(SM)}}_{\pi\rightarrow e} = 1.23\times 10^{-4}, \mathfrak{B}^{\text{(SM)}}_{\pi\rightarrow\mu} = 0.999877$) to obtain the $\mathfrak{B}_{\text{channel}}$ (shown in the right panel).
      \par The following definitions are used for the case of 2-body HNL production \cite{Shrock1981}:
      \begin{subequations}
        \begin{eqnarray}
          \delta(m,M) &\equiv& \delta_{M}^{m}:= \frac{m^{2}}{M^{2}},
          \\
          \lambda(x,y,z) &:=& x^{2} + y^{2} + z^{2} - 2(xy + yz + zx), \label{eq: kallen}
          \\
          f_{m}(x,y) &:=& x + y - (x-y)^{2},
          \\
          \rho(x,y) &:=& f_{m}(x,y)\cdot\lambda^{1/2}(x,y,1),
          \\
          \mathcal{P}_{\ell}(x,y) &:=& \frac{\rho(x,y)}{x(1-x)^{2}}.
        \end{eqnarray}
      \end{subequations}
      Calculations of the three-body HNL scaling factor for production have been done in the literature.
      We have used the helicity-summed scaling factors reported in \cite{Ballett2020} to construct through interpolation a scaling function $\mathcal{S}_{\textrm{P}\ell}$, with $P = K^{\pm}, K^{0}_{\textrm{L}}$ and $\ell = e,\mu$.
      For the special case of muon decay to HNL, we start from the known decay $\mu^{\pm} \rightarrow \nu_{\mu}\nu_{e}e^{\pm}$ and then ``promote" either the $\nu_{\mu}$ or $\nu_{e}$ to HNL, depending on the mixings $|U_{\textrm{e}4}|^{2}, |U_{\upmu 4}|^{2}$.
      \begin{figure}
        \centering
        \includegraphics[width=0.5\textwidth]{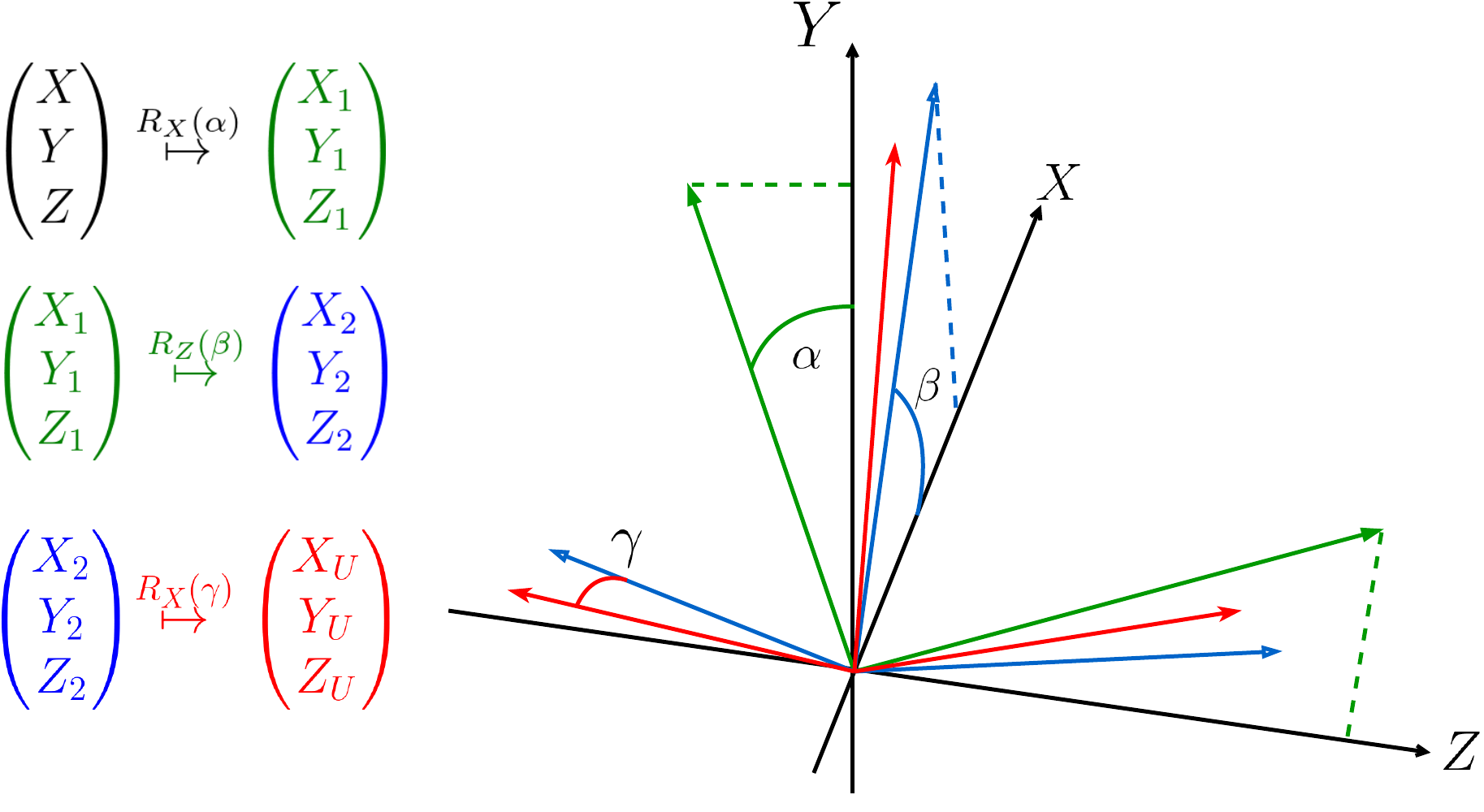}
        \caption{Extrinsic Euler rotations. The fixed system $(X,Y,Z)$ is rotated to the new system $(X_{\textrm{U}}, Y_{\textrm{U}}, Z_{\textrm{U}})$ by applying successive rotations first about $X$ (green), then about $Z$ (blue), and finally about $X$ again (red).}
        \label{fig:eulerAngles}
      \end{figure}
      \par We adopt the HNL decay widths in the context of an effective field theory describing interactions of HNL with mesons, as detailed in \cite{ColomaDUNEHNL}. 
      For the double-pion channels $N_{4} \rightarrow \pi^{\pm}\pi^{0}\ell^{\mp}, \pi^{0}\pi^{0}\nu$, whose thresholds for emission lie below the kaon mass, the decay widths in the literature have been noted to be dominated by the chain involving the emission of an on-shell $\rho$.
      Because $m_{\uprho} > m_{\textrm{K}}$, this argument cannot be applied in the case $M_{\textrm{N}4} < m_{\textrm{K}}$, and it is necessary to perform an explicit calculation to estimate the decay width resulting from double-pion channels.
      To that effect, we have used the Lagrangian for Dirac particles, made public by the authors of \cite{ColomaDUNEHNL} in the \verb|FeynRules| model database \cite{FeynRulesWebsite}, and extracted the double-differential decay rate over final-state particle energies using \verb|FeynRules| \cite{FeynRulesManual}, \verb|FeynArts| \cite{FeynArts} and \verb|FeynCalc| \cite{FeynCalc1, FeynCalc2, FeynCalc3}.
      The resulting expression for the two-dimensional $\textrm{d}^{2}\Gamma/\textrm{d}E_{\uppi}\textrm{d}E_{\ell}$ depends on the energy of the $\pi^{0}$ due to energy conservation in the HNL rest frame.
      Therefore, the integrated decay rate is obtained in runtime by numerically integrating the differential decay rate.
      \par As can be seen from the relevant tree-level Feynman diagrams in Fig. \ref{fig:pipi0ell}, the decay $N_{4} \rightarrow \pi^{+}\pi^{0}\ell^{-}$ is mediated by both light and heavy neutrinos.
      Compared to the left-hand diagrams, the right-hand diagram where $N_{4}$ enters both pion vertices is suppressed by a factor $\left|U_{\alpha 4}\right|^{2}/\left|U_{\alpha i}\right|^{2} \ll 1$, and is safely ignored.
      \par For the $\pi^{0}\pi^{0}\nu$ diagrams, the intermediate propagator is always a $N_{4}$, which renders the entire channel's decay width subleading to $\pi^{\pm}\pi^{0}\ell^{\mp}$.
      \par Comparing $N_{4} \rightarrow \pi^{\pm}\pi^{0}\ell_{\alpha}^{\pm}$, the $\alpha = e$ case has a larger decay width due to the larger phase space available to the final state. 
      The result is
      \begin{equation}
        \frac{\Gamma\left(N_{4}\rightarrow\pi^{+}\pi^{0}e^{-}\right)}{\sum_{\text{all lighter channels}}\Gamma} \lesssim 10^{-11},
      \end{equation}
      confirming the argument that the essential process for double-pion production is decays of HNL into on-shell $\rho$.
      The calculation proceeds similarly for Majorana HNL.
      \par To conserve computing time, we only simulate those channels that the user has explicitly defined in the configuration file as being ``interesting'' as a signal.
      For each run of the module, a C++ \verb|std::map<HNLDecayMode_t, double>| is constructed that contains the accessible and interesting channels and their decay widths. 
      These are then used to construct scores similarly to Eq. (\ref{eq: scores}), after which a standard Monte Carlo transformation method \cite{NumericalRecipes3rd} maps a uniform random number to a decay channel and fills the appropriate decay product list for use with \verb|GENIE|'s phase-space generator.
      We note that the HNL lifetime is computed from the full list of all kinematically available channels, regardless of whether these channels are ``interesting'' or not. 
      \par To calculate the decay rates, we make use of the following definitions found in \cite{ColomaDUNEHNL}:
      \begin{subequations}
        \begin{eqnarray}
          &B_{1}& := \frac{1}{4}\left(1 - 4\sin^{2}\theta_{\textrm{W}} + 8\sin^{4}\theta_{\textrm{W}}\right), 
          \\
          &B_{2}& := \frac{1}{2}\sin^{2}\theta_{\textrm{W}}\left(2\sin^{2}\theta_{\textrm{W}} - 1\right), 
          \\
          &\mathcal{C}_{\alpha}\left(M_{\textrm{N}4}\right)& := \sum_{\alpha}|U_{\alpha 4}|^{2}\cdot\big[F_{1}\left(m_{\alpha}/M_{\textrm{N}4}\right)B_{1} 
            \\
            \nonumber &&+ F_{2}\left(m_{\alpha}/M_{\textrm{N}4}\right)B_{2}\big], 
          \\
          &\mathcal{D}_{\alpha}\left(M_{\textrm{N}4}\right)& := |U_{\alpha 4}|^{2}\sin^{2}\theta_{\textrm{W}} 
          \\
          \nonumber &&\times \left[2F_{1}\left(M_{\textrm{N}4}\right) + F_{2}\left(M_{\textrm{N}4}\right)\right], 
          \\
          &\mathcal{G}\left(x,y\right)& := \lambda^{1/2}\left(x,y,1\right)
          \\
          \nonumber &&\times \left[1-y^{2} - x^{2}(2-x^{2}+y^{2})\right],
          \\ 
          &F_{1}(x) =& (1 - 14x^{2} - 2x^{4} - 12x^{6})\sqrt{1 - 4x^{2}} 
          \\
          \nonumber &&+ 12x^{4}(x^{4}-1)L(x),
          \\
          &F_{2}(x) =& 4x^{2}(2 + 10x^{2} - 12x^{4})\sqrt{1 - 4x^{2}} 
          \\
          \nonumber &&+ 24x^{4}(1 - 2x^{2} + 2x^{4})L(x),
          \\
          &L(x) =& \ln\left[\frac{1 - 3x^{2} - (1-x^{2})\sqrt{1 - 4x^{2}}}{x^{2}\left(1 + \sqrt{1 - 4x^{2}} \right)}\right],
        \end{eqnarray}
      \end{subequations}
      where $\theta_{\textrm{W}}$ is the Weinberg mixing angle, and $\lambda(x,y,z)$ is the K\"{a}ll\'{e}n function defined in Eq. (\ref{eq: kallen}).

      \subsection{Coordinate systems} \label{appdx: coords}
      A general detector can be both displaced and rotated arbitrarily with respect to the NEAR frame. 
      One can parametrise any such configuration by two vectors: one translation and one rotation.
      We have made the choice to use extrinsic Euler angles to write rotations (see Fig. \ref{fig:eulerAngles}): that is, the rotation matrix describing the transition between two coordinate systems $(X, Y, Z), (X_{\textrm{U}}, Y_{\textrm{U}}, Z_{\textrm{U}})$ centred around the same point is written as
      \begin{align}
        \begin{split}
          &R(\alpha, \beta, \gamma) = \\
          &= R_{\textrm{X}}(\gamma)R_{\textrm{Z}}(\beta)R_{\textrm{X}}(\alpha) \\
          &= \begin{pmatrix} 1 &0 &0 \\ 0 &c_{\gamma} &-s_{\gamma} \\ 0 &s_{\gamma} &c_{\gamma} \end{pmatrix} \begin{pmatrix} c_{\beta} &-s_{\beta} &0 \\ s_{\beta} &c_{\beta} &0 \\ 0 &0 &1 \end{pmatrix} \begin{pmatrix} 1 &0 &0 \\ 0 &c_{\alpha} &-s_{\alpha} \\ 0 &s_{\alpha} &c_{\alpha} \end{pmatrix} \\
          &= \begin{pmatrix} c_{\beta} &-c_{\alpha}s_{\beta} &s_{\alpha}s_{\beta} \\ s_{\beta}c_{\gamma} &c_{\alpha}c_{\beta}c_{\gamma} - s_{\alpha}s_{\gamma} &-s_{\alpha}c_{\beta}c_{\gamma} -c_{\alpha}s_{\gamma} \\ s_{\beta}s_{\gamma} &s_{\alpha}c_{\gamma} + c_{\alpha}c_{\beta}s_{\gamma} &c_{\alpha}c_{\gamma} - s_{\alpha}c_{\beta}s_{\gamma} \end{pmatrix},
        \end{split}
      \end{align}
      using $c_{\theta}, s_{\theta} \equiv \cos\theta, \sin\theta$.
      \subsection{Bookkeeping} \label{appdx: book}
      There are a certain number of quantities that must be kept track of during the simulation in order to correctly estimate the number of signal events, given a detector volume and number of protons on target \footnote[4]{By ``protons on target" we will account for the mean number of POT for a single HNL decay to \emph{signal} event. This information is calculated taking the input detector volume and beamline simulation; different inputs will yield different outputs.}. 
      These can be summarised in Fig. \ref{fig:POT_map}; we shall explain the steps taken forthwith.
      \par Suppose that the user has selected a channel $C$ as the channel of interest for a particular detector; for example, $N_{4} \rightarrow \mu^{-} \pi^{+}$. 
      We work our way backwards in order to obtain, \textit{for each signal event}, an estimate of the number of POT $N_{\text{POT}}$ that would result in one signal event occurring in the detector, and return this estimate as a weight attached to the \verb|EventRecord| describing the signal event, which makes POT counting a straightforward loop on the analysis side.
      \par First to be obtained is the expected number of \emph{total} HNL decays occurring in the detector, signal or not (for example, invisible decays $N_{4} \rightarrow \nu\nu\nu$ are almost never going to be considered signal events due to the inability to detect neutrinos in the final state directly), as
      \begin{equation}
        N_{H} = \frac{\sum_{i \in\text{all channels}}\Gamma_{i}}{\Gamma_{C}} \equiv \frac{\Gamma_{\text{tot}}}{\Gamma_{C}}.
      \end{equation}
      One then takes the detector geometry into account, by considering the HNL's lifetime $\tau = \hbar/\Gamma_{\text{tot}}$ and requiring that the HNL decays inside the detector volume.
      Suppose a beam of HNL of rest-frame lifetime $\tau$ and velocity $\beta c$ propagates along the $z$ axis, and the detector volume is in between the planes $z = z_{1}$ and $z = z_{2}$.
      The probability distribution for the HNL decay location is
      \begin{equation}
        p(z) = \frac{1}{\widetilde{N}}\exp\left(-\frac{z}{\beta c \gamma\tau}\right),
      \end{equation}
      where $\widetilde{N}$ is some dimensionful normalisation constant.
      The probability of decay inside the detector is then
      \begin{equation}
        P\left(z_{1} \leq z_{\text{decay}} \leq z_{2}\right) = \int_{z_{1}}^{z_{2}}\textrm{d}u\,p(u)
      \end{equation}
      which yields
      \begin{widetext}
        \begin{figure*}
          \setlength{\belowcaptionskip}{-5pt}
          \centering
          \includegraphics[width=\textwidth]{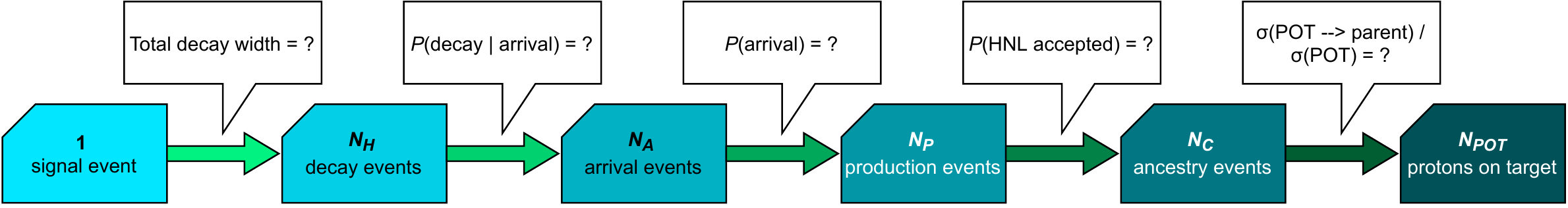}
          \caption{Sequence for estimating number of POT for each event, working backwards. See text for definitions.}
          \label{fig:POT_map}
        \end{figure*}
        \vspace*{-2em}
      \end{widetext}
      \begin{align} \label{eq: survival}
        \begin{split}
          P &= \frac{\beta\gamma c\tau}{\widetilde{N}}\exp\left(-\frac{z_{1}}{\beta\gamma c\tau}\right)\left[1 - \exp\left(-\frac{z_{2}-z_{1}}{\beta\gamma c\tau}\right)\right] \\
          &= P(\text{arrival}) \cdot P(\text{decay} | \text{arrival}).
        \end{split}
      \end{align}
      Equation (\ref{eq: survival}) states that the HNL must first survive long enough to reach the detector, and then decay promptly while inside it.
      \par We can generalise this to the full 3D picture.
      Let D, E, and X be the HNL production point, and the intersections of its trajectory with the detector at entry and exit, respectively.
      Then for every $N_{\textrm{P}}$ HNL emitted that could be accepted, $N_{\textrm{A}}$ will survive until the detector, and $N_{\textrm{H}}$ will decay without exiting the detector.
      These quantities are straightforwardly obtained :
      \begin{equation}\label{eq: lifetimeWeight}
        N_{\textrm{A}} = N_{\textrm{H}}\cdot\exp\left(\frac{\ell}{\beta c \gamma \tau}\right),
      \end{equation}
      \begin{equation}
        N_{\textrm{P}} = N_{\textrm{A}}\cdot\exp\left(\frac{L}{\beta c \gamma \tau}\right)
      \end{equation}
      where $\ell, L$ are the distances $|\boldsymbol{r}_{\textrm{X}} - \boldsymbol{r}_{\textrm{E}}|, |\boldsymbol{r}_{\textrm{D}} - \boldsymbol{r}_{\textrm{E}}|$.
      \par Next, we factor in those HNL that decayed before reaching the detector volume to get the total number $N_{\textrm{P}}$ of HNL produced in the accepted region, using once again Eq. (\ref{eq: lifetimeWeight}) but substituting $\ell = |\boldsymbol{r}_{\textrm{X}} - \boldsymbol{r}_{\textrm{E}}|$ for $L = |\boldsymbol{r}_{\textrm{E}} - \boldsymbol{r}_{\textrm{D}}|$.
      \par Further backward, we have the estimate of acceptance; only those HNL emitted in the correct angular region would intersect the detector.
      It is assumed that the decays of parents are isotropic in the parents' rest frame, which is correct for pseudoscalar mesons ($\pi^{\pm}, K^{\pm}, K^{0}$). 
      Furthermore, the detector is assumed to be sufficiently far away that the small angle approximation $\sin\theta \simeq \theta$ is valid.
      The size and position of the detector defines a window in the observer's frame, which is then transformed into a rest-frame window using the collimation-effect function $f:[0,\pi]\rightarrow[0,\pi]$ described in Fig. \ref{fig:collimation}. 
      The lab-frame emission angle can be interpreted as an angular deviation of the HNL's trajectory from the parent's momentum.
      The angular size of the detector is given by $\zeta_{+} - \zeta_{-}$, where the angles $\zeta_{\mp}$ are the minimum and maximum deviation angles for which the HNL's trajectory can intersect the detector.
      Figure \ref{fig:deviation} shows how these deviation angles are calculated.
      A ``sweep direction" $\boldsymbol{\delta}$ is constructed from the parent's momentum $\boldsymbol{p}_{\textrm{P}}$ and the detector centre C, and the points of entry $\textrm{V}_{-}$ and exit $\textrm{V}_{+}$ along this sweep are obtained, giving the deviation angles as the angles between the parent momentum and the vectors $\boldsymbol{r}_{-}, \boldsymbol{r}_{+}$.
      \par We thus estimate the acceptance correction $\mathcal{A}$ induced by the collimation effect of HNL becoming dominated by their parent's Lorentz boost, as 
      \begin{equation}\label{eq: acceptance_correction}
        \mathcal{A} = \frac{|\mathcal{I}_{\textrm{N}4}|}{|\mathcal{I}_{\nu}|},
      \end{equation}
      where $\mathcal{I}$ is the pre-image under $f$ of the angular opening $[\zeta_{-},\zeta_{+}]$. 
      For large HNL masses, it is $\mathcal{I}_{\textrm{N}4} = \mathcal{I}_{\textrm{F}} \sqcup \mathcal{I}_{\textrm{B}}$ where $\mathcal{I}_{\textrm{F},\textrm{B}}$ are the forward and backward rest-frame angular regions where HNL can be accepted.
      One can contrast this with the case of light neutrinos, where $f$ increases monotonously and only forward emitted neutrinos can reach the detector.
      In the end, the number of total HNL emitted (``ancestry events") is given by
      \begin{equation}
        N_{\textrm{C}} = \frac{N_{\textrm{P}}}{\sum_{\alpha=\textrm{e},\upmu,\uptau}\left|U_{\alpha 4}\right|^{2}}\cdot \frac{1}{\omega_{\text{det}}\mathcal{B}^{2}\mathcal{A}},
      \end{equation}
      where $\mathcal{B}, \mathcal{A}$ were defined in Eqs. \ref{eq: boostFactor}, \ref{eq: acceptance_correction} and $\omega_{\text{det}}$ is the angular size of the detector in the lab frame, with $D$ at the origin.
      The angular size of the detector in the parent's rest frame is then $\omega_{\text{det}}\mathcal{B}^{2}$, assuming its face is roughly perpendicular to the parent momentum; the acceptance correction $\mathcal{A}$ parametrises the intrinsic increase of the probability that a randomly chosen direction for HNL emission will be boosted such that the HNL gets accepted.
      The prefactor $(\sum_{\alpha} |U_{\alpha 4}|^{2})^{-1}$ is inserted to correct for the fact that, to conserve computing resources, we assume all parent decays result in an HNL.
      Finally, we estimate $N_{\text{POT}} = N_{\textrm{C}}\cdot\widetilde{N}(P)$, $\widetilde{N}(P) = n(\text{all particles})/n(P)$, where $n$ is the number of particles produced in a $p$+target interaction.
      For example, if in $n(\pi^{+}) = 0.4, n(K^{+}) = 0.1, n(K^{0}) \simeq n(K^{0}_{L}) = 0.05, n(p) = 0.15, n(\mu^{+}+\text{other}) \simeq 0.05$, then the relevant factors are
      \begin{equation}
        \widetilde{N}\left\{\pi^{+}, K^{+}, K^{0}_{\textrm{L}}, \mu^{+}, p\right\} = \left\{1.875, 7.5, 15, 5, 15\right\}.
      \end{equation}
      For practical purposes, the code expects the values $\widetilde{N}(P)$ as inputs from the user in the configuration file.
      \bibliographystyle{apsrev4-2}
      \bibliography{biblio.bib}

\begin{thebibliography}{159}%
\makeatletter
\providecommand \@ifxundefined [1]{%
 \@ifx{#1\undefined}
}%
\providecommand \@ifnum [1]{%
 \ifnum #1\expandafter \@firstoftwo
 \else \expandafter \@secondoftwo
 \fi
}%
\providecommand \@ifx [1]{%
 \ifx #1\expandafter \@firstoftwo
 \else \expandafter \@secondoftwo
 \fi
}%
\providecommand \natexlab [1]{#1}%
\providecommand \enquote  [1]{``#1''}%
\providecommand \bibnamefont  [1]{#1}%
\providecommand \bibfnamefont [1]{#1}%
\providecommand \citenamefont [1]{#1}%
\providecommand \href@noop [0]{\@secondoftwo}%
\providecommand \href [0]{\begingroup \@sanitize@url \@href}%
\providecommand \@href[1]{\@@startlink{#1}\@@href}%
\providecommand \@@href[1]{\endgroup#1\@@endlink}%
\providecommand \@sanitize@url [0]{\catcode `\\12\catcode `\$12\catcode
  `\&12\catcode `\#12\catcode `\^12\catcode `\_12\catcode `\%12\relax}%
\providecommand \@@startlink[1]{}%
\providecommand \@@endlink[0]{}%
\providecommand \url  [0]{\begingroup\@sanitize@url \@url }%
\providecommand \@url [1]{\endgroup\@href {#1}{\urlprefix }}%
\providecommand \urlprefix  [0]{URL }%
\providecommand \Eprint [0]{\href }%
\providecommand \doibase [0]{https://doi.org/}%
\providecommand \selectlanguage [0]{\@gobble}%
\providecommand \bibinfo  [0]{\@secondoftwo}%
\providecommand \bibfield  [0]{\@secondoftwo}%
\providecommand \translation [1]{[#1]}%
\providecommand \BibitemOpen [0]{}%
\providecommand \bibitemStop [0]{}%
\providecommand \bibitemNoStop [0]{.\EOS\space}%
\providecommand \EOS [0]{\spacefactor3000\relax}%
\providecommand \BibitemShut  [1]{\csname bibitem#1\endcsname}%
\let\auto@bib@innerbib\@empty
\bibitem [{\citenamefont {Pontecorvo}(1958)}]{PonteFirstOscPaper}%
  \BibitemOpen
  \bibfield  {author} {\bibinfo {author} {\bibfnamefont {B.}~\bibnamefont
  {Pontecorvo}},\ }\href {http://jetp.ras.ru/cgi-bin/e/index/e/6/2/p429?a=list}
  {\bibfield  {journal} {\bibinfo  {journal} {\JZHETP}\ }\textbf {\bibinfo
  {volume} {33}},\ \bibinfo {pages} {429} (\bibinfo {year} {1958})}\BibitemShut
  {NoStop}%
\bibitem [{\citenamefont {Pontecorvo}(1968)}]{PonteOscPaper}%
  \BibitemOpen
  \bibfield  {author} {\bibinfo {author} {\bibfnamefont {B.}~\bibnamefont
  {Pontecorvo}},\ }\href
  {http://jetp.ras.ru/cgi-bin/e/index/e/26/5/p984?a=list} {\bibfield  {journal}
  {\bibinfo  {journal} {\JZHETP}\ }\textbf {\bibinfo {volume} {26}},\ \bibinfo
  {pages} {984} (\bibinfo {year} {1968})}\BibitemShut {NoStop}%
\bibitem [{\citenamefont {Bilenky}(2016)}]{oscHistory}%
  \BibitemOpen
  \bibfield  {author} {\bibinfo {author} {\bibfnamefont {S.}~\bibnamefont
  {Bilenky}},\ }\href
  {https://doi.org/https://doi.org/10.1016/j.nuclphysb.2016.01.025} {\bibfield
  {journal} {\bibinfo  {journal} {\JNPB}\ }\textbf {\bibinfo {volume} {908}},\
  \bibinfo {pages} {2} (\bibinfo {year} {2016})}\BibitemShut {NoStop}%
\bibitem [{\citenamefont {Davis}\ \emph {et~al.}(1968)\citenamefont {Davis},
  \citenamefont {Harmer},\ and\ \citenamefont {Hoffman}}]{DavisSolarNeutrino}%
  \BibitemOpen
  \bibfield  {author} {\bibinfo {author} {\bibfnamefont {R.}~\bibnamefont
  {Davis}}, \bibinfo {author} {\bibfnamefont {D.~S.}\ \bibnamefont {Harmer}},\
  and\ \bibinfo {author} {\bibfnamefont {K.~C.}\ \bibnamefont {Hoffman}},\
  }\href {https://doi.org/10.1103/PhysRevLett.20.1205} {\bibfield  {journal}
  {\bibinfo  {journal} {\JPRL}\ }\textbf {\bibinfo {volume} {20}},\ \bibinfo
  {pages} {1205} (\bibinfo {year} {1968})}\BibitemShut {NoStop}%
\bibitem [{\citenamefont {Ahmad~\textit{et al}}(2002)}]{SNOOscDiscovery}%
  \BibitemOpen
  \bibfield  {author} {\bibinfo {author} {\bibfnamefont {Q.~R.}\ \bibnamefont
  {Ahmad~\textit{et al}}} (\bibinfo {collaboration} {{SNO Collaboration}}),\
  }\href {https://doi.org/10.1103/PhysRevLett.89.011301} {\bibfield  {journal}
  {\bibinfo  {journal} {\JPRL}\ }\textbf {\bibinfo {volume} {89}},\ \bibinfo
  {pages} {011301} (\bibinfo {year} {2002})}\BibitemShut {NoStop}%
\bibitem [{\citenamefont {Fukuda~\textit{et al}}(1998)}]{SKOscDiscovery}%
  \BibitemOpen
  \bibfield  {author} {\bibinfo {author} {\bibfnamefont {Y.}~\bibnamefont
  {Fukuda~\textit{et al}}} (\bibinfo {collaboration} {{Super-Kamiokande
  Collaboration}}),\ }\href {https://doi.org/10.1103/PhysRevLett.81.1562}
  {\bibfield  {journal} {\bibinfo  {journal} {\JPRL}\ }\textbf {\bibinfo
  {volume} {81}},\ \bibinfo {pages} {1562} (\bibinfo {year}
  {1998})}\BibitemShut {NoStop}%
\bibitem [{\citenamefont {Eguchi~\textit{et al}}(2003)}]{KLNDOscDiscovery}%
  \BibitemOpen
  \bibfield  {author} {\bibinfo {author} {\bibfnamefont {K.}~\bibnamefont
  {Eguchi~\textit{et al}}} (\bibinfo {collaboration} {{KamLAND
  Collaboration}}),\ }\href {https://doi.org/10.1103/PhysRevLett.90.021802}
  {\bibfield  {journal} {\bibinfo  {journal} {\JPRL}\ }\textbf {\bibinfo
  {volume} {90}},\ \bibinfo {pages} {021802} (\bibinfo {year}
  {2003})}\BibitemShut {NoStop}%
\bibitem [{\citenamefont {de~Salas~\textit{et al}}(2021)}]{FittingStatus2020}%
  \BibitemOpen
  \bibfield  {author} {\bibinfo {author} {\bibfnamefont {P.~F.}\ \bibnamefont
  {de~Salas~\textit{et al}}},\ }\href {https://doi.org/10.1007/jhep02(2021)071}
  {\bibfield  {journal} {\bibinfo  {journal} {\JJHEP}\ }\textbf {\bibinfo
  {volume} {02}},\ \bibinfo {pages} {71} (\bibinfo {year} {2021})}\BibitemShut
  {NoStop}%
\bibitem [{\citenamefont {Hampel~\textit{et al}}(1999)}]{GALLEX}%
  \BibitemOpen
  \bibfield  {author} {\bibinfo {author} {\bibfnamefont {W.}~\bibnamefont
  {Hampel~\textit{et al}}} (\bibinfo {collaboration} {{GALLEX
  Collaboration}}),\ }\href
  {https://doi.org/https://doi.org/10.1016/S0370-2693(98)01579-2} {\bibfield
  {journal} {\bibinfo  {journal} {\JPLB}\ }\textbf {\bibinfo {volume} {447}},\
  \bibinfo {pages} {127} (\bibinfo {year} {1999})}\BibitemShut {NoStop}%
\bibitem [{\citenamefont {Altmann~\textit{et al}}(2005)}]{GNO}%
  \BibitemOpen
  \bibfield  {author} {\bibinfo {author} {\bibfnamefont {M.}~\bibnamefont
  {Altmann~\textit{et al}}} (\bibinfo {collaboration} {{GNO Collaboration}}),\
  }\href {https://doi.org/10.1016/j.physletb.2005.04.068} {\bibfield  {journal}
  {\bibinfo  {journal} {\JPLB}\ }\textbf {\bibinfo {volume} {616}},\ \bibinfo
  {pages} {174} (\bibinfo {year} {2005})}\BibitemShut {NoStop}%
\bibitem [{\citenamefont {Fukuda~\textit{et al}}(1996)}]{Kamiokande}%
  \BibitemOpen
  \bibfield  {author} {\bibinfo {author} {\bibfnamefont {Y.}~\bibnamefont
  {Fukuda~\textit{et al}}},\ }\href
  {https://doi.org/10.1103/PhysRevLett.77.1683} {\bibfield  {journal} {\bibinfo
   {journal} {\JPRL}\ }\textbf {\bibinfo {volume} {77}},\ \bibinfo {pages}
  {1683} (\bibinfo {year} {1996})}\BibitemShut {NoStop}%
\bibitem [{\citenamefont {Agostini~\textit{et al}}(2020)}]{Borexino}%
  \BibitemOpen
  \bibfield  {author} {\bibinfo {author} {\bibfnamefont {M.}~\bibnamefont
  {Agostini~\textit{et al}}} (\bibinfo {collaboration} {{Borexino
  Collaboration}}),\ }\href {https://doi.org/10.1038/s41586-020-2934-0}
  {\bibfield  {journal} {\bibinfo  {journal} {\JNATURE}\ }\textbf {\bibinfo
  {volume} {587}},\ \bibinfo {pages} {577} (\bibinfo {year}
  {2020})}\BibitemShut {NoStop}%
\bibitem [{\citenamefont {Ahn~\textit{et al}}(2006)}]{K2K}%
  \BibitemOpen
  \bibfield  {author} {\bibinfo {author} {\bibfnamefont {M.~H.}\ \bibnamefont
  {Ahn~\textit{et al}}} (\bibinfo {collaboration} {{K2K Collaboration}}),\
  }\href {https://doi.org/10.1103/physrevd.74.072003} {\bibfield  {journal}
  {\bibinfo  {journal} {\JPRD}\ }\textbf {\bibinfo {volume} {74}},\ \bibinfo
  {pages} {072003} (\bibinfo {year} {2006})}\BibitemShut {NoStop}%
\bibitem [{\citenamefont {Adamson~\textit{et al}}(2014)}]{MINOS}%
  \BibitemOpen
  \bibfield  {author} {\bibinfo {author} {\bibfnamefont {P.}~\bibnamefont
  {Adamson~\textit{et al}}},\ }\href
  {https://doi.org/10.1103/physrevlett.112.191801} {\bibfield  {journal}
  {\bibinfo  {journal} {\JPRL}\ }\textbf {\bibinfo {volume} {112}},\ \bibinfo
  {pages} {191801} (\bibinfo {year} {2014})}\BibitemShut {NoStop}%
\bibitem [{\citenamefont {Agafonova~\textit{et al}}(2018)}]{OPERA}%
  \BibitemOpen
  \bibfield  {author} {\bibinfo {author} {\bibfnamefont {N.}~\bibnamefont
  {Agafonova~\textit{et al}}} (\bibinfo {collaboration} {{OPERA
  Collaboration}}),\ }\href {https://doi.org/10.1103/physrevlett.120.211801}
  {\bibfield  {journal} {\bibinfo  {journal} {\JPRL}\ }\textbf {\bibinfo
  {volume} {120}},\ \bibinfo {pages} {211801} (\bibinfo {year}
  {2018})}\BibitemShut {NoStop}%
\bibitem [{\citenamefont {Aguilar~\textit{et al}}(2001)}]{LSND}%
  \BibitemOpen
  \bibfield  {author} {\bibinfo {author} {\bibfnamefont {A.}~\bibnamefont
  {Aguilar~\textit{et al}}} (\bibinfo {collaboration} {{LSND Collaboration}}),\
  }\href {https://doi.org/10.1103/physrevd.64.112007} {\bibfield  {journal}
  {\bibinfo  {journal} {\JPRD}\ }\textbf {\bibinfo {volume} {64}},\ \bibinfo
  {pages} {112007} (\bibinfo {year} {2001})}\BibitemShut {NoStop}%
\bibitem [{\citenamefont {Armbruster~\textit{et al}}(2002)}]{KARMEN}%
  \BibitemOpen
  \bibfield  {author} {\bibinfo {author} {\bibfnamefont {B.}~\bibnamefont
  {Armbruster~\textit{et al}}} (\bibinfo {collaboration} {{KARMEN
  Collaboration}}),\ }\href {https://doi.org/10.1103/physrevd.65.112001}
  {\bibfield  {journal} {\bibinfo  {journal} {\JPRD}\ }\textbf {\bibinfo
  {volume} {65}},\ \bibinfo {pages} {112001} (\bibinfo {year}
  {2002})}\BibitemShut {NoStop}%
\bibitem [{\citenamefont {Aguilar-Arevalo~\textit{et al}}(2021)}]{miniBooNE}%
  \BibitemOpen
  \bibfield  {author} {\bibinfo {author} {\bibfnamefont {A.~A.}\ \bibnamefont
  {Aguilar-Arevalo~\textit{et al}}} (\bibinfo {collaboration} {{MiniBooNE
  Collaboration}}),\ }\href {https://doi.org/10.1103/physrevd.103.052002}
  {\bibfield  {journal} {\bibinfo  {journal} {\JPRD}\ }\textbf {\bibinfo
  {volume} {103}},\ \bibinfo {pages} {052002} (\bibinfo {year}
  {2021})}\BibitemShut {NoStop}%
\bibitem [{\citenamefont {Apollonio~\textit{et al}}(2003)}]{CHOOZ}%
  \BibitemOpen
  \bibfield  {author} {\bibinfo {author} {\bibfnamefont {M.}~\bibnamefont
  {Apollonio~\textit{et al}}},\ }\href
  {https://doi.org/10.1140/epjc/s2002-01127-9} {\bibfield  {journal} {\bibinfo
  {journal} {\JEPJC}\ }\textbf {\bibinfo {volume} {27}},\ \bibinfo {pages}
  {331} (\bibinfo {year} {2003})}\BibitemShut {NoStop}%
\bibitem [{\citenamefont {Boehm~\textit{et al}}(2001)}]{PaloVerde}%
  \BibitemOpen
  \bibfield  {author} {\bibinfo {author} {\bibfnamefont {F.}~\bibnamefont
  {Boehm~\textit{et al}}},\ }\href {https://doi.org/10.1103/physrevd.64.112001}
  {\bibfield  {journal} {\bibinfo  {journal} {\JPRD}\ }\textbf {\bibinfo
  {volume} {64}},\ \bibinfo {pages} {112001} (\bibinfo {year}
  {2001})}\BibitemShut {NoStop}%
\bibitem [{\citenamefont {Adamson~\textit{et al}}(2020)}]{MINOSPlus}%
  \BibitemOpen
  \bibfield  {author} {\bibinfo {author} {\bibfnamefont {P.}~\bibnamefont
  {Adamson~\textit{et al}}} (\bibinfo {collaboration} {{MINOS+
  Collaboration}}),\ }\href {https://doi.org/10.1103/physrevlett.125.131802}
  {\bibfield  {journal} {\bibinfo  {journal} {\JPRL}\ }\textbf {\bibinfo
  {volume} {125}},\ \bibinfo {pages} {131802} (\bibinfo {year}
  {2020})}\BibitemShut {NoStop}%
\bibitem [{\citenamefont {Abe~\textit{et al}}(2021)}]{T2KOscillation}%
  \BibitemOpen
  \bibfield  {author} {\bibinfo {author} {\bibfnamefont {K.}~\bibnamefont
  {Abe~\textit{et al}}} (\bibinfo {collaboration} {{T2K Collaboration}}),\
  }\href {https://doi.org/10.1103/physrevd.103.112008} {\bibfield  {journal}
  {\bibinfo  {journal} {\JPRD}\ }\textbf {\bibinfo {volume} {103}},\ \bibinfo
  {pages} {112008} (\bibinfo {year} {2021})}\BibitemShut {NoStop}%
\bibitem [{\citenamefont {Acero~\textit{et al}}(2019)}]{NOvAOscillation}%
  \BibitemOpen
  \bibfield  {author} {\bibinfo {author} {\bibfnamefont {M.~A.}\ \bibnamefont
  {Acero~\textit{et al}}} (\bibinfo {collaboration} {{NOvA Collaboration}}),\
  }\href {https://doi.org/10.1103/physrevlett.123.151803} {\bibfield  {journal}
  {\bibinfo  {journal} {\JPRL}\ }\textbf {\bibinfo {volume} {123}},\ \bibinfo
  {pages} {151803} (\bibinfo {year} {2019})}\BibitemShut {NoStop}%
\bibitem [{\citenamefont {Abratenko~\textit{et al}}(2021)}]{uBooNEExcess}%
  \BibitemOpen
  \bibfield  {author} {\bibinfo {author} {\bibfnamefont {P.}~\bibnamefont
  {Abratenko~\textit{et al}}} (\bibinfo {collaboration} {{MicroBooNE
  Collaboration}}),\ }\href {https://doi.org/10.1103/PhysRevLett.128.241801}
  {\bibfield  {journal} {\bibinfo  {journal} {\JPRL}\ }\textbf {\bibinfo
  {volume} {128}},\ \bibinfo {pages} {241801} (\bibinfo {year}
  {2021})}\BibitemShut {NoStop}%
\bibitem [{\citenamefont {Adey~\textit{et al}}(2019)}]{DayaBay}%
  \BibitemOpen
  \bibfield  {author} {\bibinfo {author} {\bibfnamefont {D.}~\bibnamefont
  {Adey~\textit{et al}}} (\bibinfo {collaboration} {{Daya Bay
  Collaboration}}),\ }\href {https://doi.org/10.1103/physrevd.100.052004}
  {\bibfield  {journal} {\bibinfo  {journal} {\JPRD}\ }\textbf {\bibinfo
  {volume} {100}},\ \bibinfo {pages} {052004} (\bibinfo {year}
  {2019})}\BibitemShut {NoStop}%
\bibitem [{\citenamefont {Bak~\textit{et al}}(2018)}]{RENO}%
  \BibitemOpen
  \bibfield  {author} {\bibinfo {author} {\bibfnamefont {G.}~\bibnamefont
  {Bak~\textit{et al}}} (\bibinfo {collaboration} {{RENO Collaboration}}),\
  }\href {https://doi.org/10.1103/physrevlett.121.201801} {\bibfield  {journal}
  {\bibinfo  {journal} {\JPRL}\ }\textbf {\bibinfo {volume} {121}},\ \bibinfo
  {pages} {201801} (\bibinfo {year} {2018})}\BibitemShut {NoStop}%
\bibitem [{\citenamefont {de~Kerret~\textit{et al}}(2020)}]{DoubleCHOOZ}%
  \BibitemOpen
  \bibfield  {author} {\bibinfo {author} {\bibfnamefont {H.}~\bibnamefont
  {de~Kerret~\textit{et al}}} (\bibinfo {collaboration} {{Double Chooz
  Collaboration}}),\ }\href {https://doi.org/10.1038/s41567-020-0831-y}
  {\bibfield  {journal} {\bibinfo  {journal} {\JNP}\ }\textbf {\bibinfo
  {volume} {16}},\ \bibinfo {pages} {558} (\bibinfo {year} {2020})}\BibitemShut
  {NoStop}%
\bibitem [{\citenamefont {Machado}\ \emph {et~al.}(2019)\citenamefont
  {Machado}, \citenamefont {Palamara},\ and\ \citenamefont
  {Schmitz}}]{SBNReview}%
  \BibitemOpen
  \bibfield  {author} {\bibinfo {author} {\bibfnamefont {P.~A.~N.}\
  \bibnamefont {Machado}}, \bibinfo {author} {\bibfnamefont {O.}~\bibnamefont
  {Palamara}},\ and\ \bibinfo {author} {\bibfnamefont {D.~W.}\ \bibnamefont
  {Schmitz}},\ }\href {https://doi.org/10.1146/annurev-nucl-101917-020949}
  {\bibfield  {journal} {\bibinfo  {journal} {\JANNREV}\ }\textbf {\bibinfo
  {volume} {69}},\ \bibinfo {pages} {363} (\bibinfo {year} {2019})}\BibitemShut
  {NoStop}%
\bibitem [{\citenamefont {Abi~\textit{et al}}(2020)}]{DUNEPhysicsOverview}%
  \BibitemOpen
  \bibfield  {author} {\bibinfo {author} {\bibfnamefont {B.}~\bibnamefont
  {Abi~\textit{et al}}},\ }\href {https://arxiv.org/abs/2002.03005} {\
  (\bibinfo {year} {2020})},\ \Eprint {https://arxiv.org/abs/2002.03005}
  {arXiv:2002.03005 [hep-ex]} \BibitemShut {NoStop}%
\bibitem [{\citenamefont {An~\textit{et al}}(2016)}]{JUNOOverview}%
  \BibitemOpen
  \bibfield  {author} {\bibinfo {author} {\bibfnamefont {F.}~\bibnamefont
  {An~\textit{et al}}},\ }\href {https://doi.org/10.1088/0954-3899/43/3/030401}
  {\bibfield  {journal} {\bibinfo  {journal} {\JJPG}\ }\textbf {\bibinfo
  {volume} {43}},\ \bibinfo {pages} {030401} (\bibinfo {year}
  {2016})}\BibitemShut {NoStop}%
\bibitem [{\citenamefont {Abe~\textit{et al}}(2014)}]{HyperKOverview}%
  \BibitemOpen
  \bibfield  {author} {\bibinfo {author} {\bibfnamefont {K.}~\bibnamefont
  {Abe~\textit{et al}}} (\bibinfo {collaboration} {{Hyper-Kamiokande Working
  Group}}),\ }\href {https://arxiv.org/abs/1412.4673} {\  (\bibinfo {year}
  {2014})},\ \Eprint {https://arxiv.org/abs/1412.4673} {arXiv:1412.4673
  [ins-det]} \BibitemShut {NoStop}%
\bibitem [{\citenamefont {Aker~\textit{et al}}(2022)}]{Katrin0.8eVNature}%
  \BibitemOpen
  \bibfield  {author} {\bibinfo {author} {\bibfnamefont {M.}~\bibnamefont
  {Aker~\textit{et al}}} (\bibinfo {collaboration} {{KATRIN Collaboration}}),\
  }\href {https://doi.org/10.1038/s41567-021-01463-1} {\bibfield  {journal}
  {\bibinfo  {journal} {\JNP}\ }\textbf {\bibinfo {volume} {18}},\ \bibinfo
  {pages} {160} (\bibinfo {year} {2022})}\BibitemShut {NoStop}%
\bibitem [{\citenamefont {Workman~\textit{et al}}(2022)}]{PDG2022}%
  \BibitemOpen
  \bibfield  {author} {\bibinfo {author} {\bibfnamefont {R.~L.}\ \bibnamefont
  {Workman~\textit{et al}}} (\bibinfo {collaboration} {Particle Data Group}),\
  }\href {https://doi.org/10.1093/ptep/ptac097} {\bibfield  {journal} {\bibinfo
   {journal} {\JPTEP}\ }\textbf {\bibinfo {volume} {2022}},\ \bibinfo {pages}
  {083C01} (\bibinfo {year} {2022})}\BibitemShut {NoStop}%
\bibitem [{\citenamefont {Formaggio}\ \emph {et~al.}(2021)\citenamefont
  {Formaggio}, \citenamefont {de~Gouvêa},\ and\ \citenamefont
  {Robertson}}]{FormaggioDirectMass}%
  \BibitemOpen
  \bibfield  {author} {\bibinfo {author} {\bibfnamefont {J.~A.}\ \bibnamefont
  {Formaggio}}, \bibinfo {author} {\bibfnamefont {A.~L.~C.}\ \bibnamefont
  {de~Gouvêa}},\ and\ \bibinfo {author} {\bibfnamefont {R.~G.~H.}\
  \bibnamefont {Robertson}},\ }\href
  {https://doi.org/10.1016/j.physrep.2021.02.002} {\bibfield  {journal}
  {\bibinfo  {journal} {\JPREP}\ }\textbf {\bibinfo {volume} {914}},\ \bibinfo
  {pages} {1} (\bibinfo {year} {2021})}\BibitemShut {NoStop}%
\bibitem [{\citenamefont {Miranda}\ and\ \citenamefont
  {Valle}(2016)}]{SeesawAndOscillations}%
  \BibitemOpen
  \bibfield  {author} {\bibinfo {author} {\bibfnamefont {O.}~\bibnamefont
  {Miranda}}\ and\ \bibinfo {author} {\bibfnamefont {J.}~\bibnamefont
  {Valle}},\ }\href {https://doi.org/10.1016/j.nuclphysb.2016.03.027}
  {\bibfield  {journal} {\bibinfo  {journal} {\JNUCB}\ }\textbf {\bibinfo
  {volume} {908}},\ \bibinfo {pages} {436} (\bibinfo {year}
  {2016})}\BibitemShut {NoStop}%
\bibitem [{\citenamefont {Abada~\textit{et al}}(2007)}]{Abada_ReviewOfSeesaws}%
  \BibitemOpen
  \bibfield  {author} {\bibinfo {author} {\bibfnamefont {A.}~\bibnamefont
  {Abada~\textit{et al}}},\ }\href
  {https://doi.org/10.1088/1126-6708/2007/12/061} {\bibfield  {journal}
  {\bibinfo  {journal} {\JJHEP}\ }\textbf {\bibinfo {volume} {12}},\ \bibinfo
  {pages} {061} (\bibinfo {year} {2007})}\BibitemShut {NoStop}%
\bibitem [{\citenamefont {Cai~\textit{et al}}(2018)}]{ColliderTestsOfSeesaws}%
  \BibitemOpen
  \bibfield  {author} {\bibinfo {author} {\bibfnamefont {Y.}~\bibnamefont
  {Cai~\textit{et al}}},\ }\href {https://doi.org/10.3389/fphy.2018.00040}
  {\bibfield  {journal} {\bibinfo  {journal} {\JFIP}\ }\textbf {\bibinfo
  {volume} {6}},\ \bibinfo {pages} {40} (\bibinfo {year} {2018})}\BibitemShut
  {NoStop}%
\bibitem [{\citenamefont {Asaka}\ and\ \citenamefont
  {Shaposhnikov}(2005)}]{AsakaPhysLetB}%
  \BibitemOpen
  \bibfield  {author} {\bibinfo {author} {\bibfnamefont {T.}~\bibnamefont
  {Asaka}}\ and\ \bibinfo {author} {\bibfnamefont {M.}~\bibnamefont
  {Shaposhnikov}},\ }\href {https://doi.org/10.1016/j.physletb.2005.06.020}
  {\bibfield  {journal} {\bibinfo  {journal} {\JPLB}\ }\textbf {\bibinfo
  {volume} {620}},\ \bibinfo {pages} {17} (\bibinfo {year} {2005})}\BibitemShut
  {NoStop}%
\bibitem [{\citenamefont {Brivio~\textit{et al}}(2019)}]{LeptogenesisTypeI}%
  \BibitemOpen
  \bibfield  {author} {\bibinfo {author} {\bibfnamefont {I.}~\bibnamefont
  {Brivio~\textit{et al}}},\ }\href {https://doi.org/10.1007/JHEP10(2019)059}
  {\bibfield  {journal} {\bibinfo  {journal} {\JJHEP}\ }\textbf {\bibinfo
  {volume} {10}},\ \bibinfo {pages} {59} (\bibinfo {year} {2019})}\BibitemShut
  {NoStop}%
\bibitem [{\citenamefont {Brdar~\textit{et al}}(2019)}]{BrdarTypeI}%
  \BibitemOpen
  \bibfield  {author} {\bibinfo {author} {\bibfnamefont {V.}~\bibnamefont
  {Brdar~\textit{et al}}},\ }\href
  {https://doi.org/10.1103/PhysRevD.100.075029} {\bibfield  {journal} {\bibinfo
   {journal} {\JPRD}\ }\textbf {\bibinfo {volume} {100}},\ \bibinfo {pages}
  {075029} (\bibinfo {year} {2019})}\BibitemShut {NoStop}%
\bibitem [{\citenamefont {Boyarsky~\textit{et al}}(2019)}]{Boyarsky2019}%
  \BibitemOpen
  \bibfield  {author} {\bibinfo {author} {\bibfnamefont {A.}~\bibnamefont
  {Boyarsky~\textit{et al}}},\ }\href
  {https://doi.org/10.1016/j.ppnp.2018.07.004} {\bibfield  {journal} {\bibinfo
  {journal} {\JPPNP}\ }\textbf {\bibinfo {volume} {104}},\ \bibinfo {pages} {1}
  (\bibinfo {year} {2019})}\BibitemShut {NoStop}%
\bibitem [{\citenamefont {Shaposhnikov}(2007)}]{Shaposhnikov_2007}%
  \BibitemOpen
  \bibfield  {author} {\bibinfo {author} {\bibfnamefont {M.}~\bibnamefont
  {Shaposhnikov}},\ }\href {https://doi.org/10.1016/j.nuclphysb.2006.11.003}
  {\bibfield  {journal} {\bibinfo  {journal} {\JNUCB}\ }\textbf {\bibinfo
  {volume} {763}},\ \bibinfo {pages} {49} (\bibinfo {year} {2007})}\BibitemShut
  {NoStop}%
\bibitem [{\citenamefont {Bernabéu~\textit{et al}}(1987)}]{BERNABEU1987303}%
  \BibitemOpen
  \bibfield  {author} {\bibinfo {author} {\bibfnamefont {J.}~\bibnamefont
  {Bernabéu~\textit{et al}}},\ }\href
  {https://doi.org/https://doi.org/10.1016/0370-2693(87)91100-2} {\bibfield
  {journal} {\bibinfo  {journal} {\JPLB}\ }\textbf {\bibinfo {volume} {187}},\
  \bibinfo {pages} {303} (\bibinfo {year} {1987})}\BibitemShut {NoStop}%
\bibitem [{\citenamefont {Deppisch}\ \emph {et~al.}(2015)\citenamefont
  {Deppisch}, \citenamefont {Bhupal~Dev},\ and\ \citenamefont
  {Pilaftsis}}]{NeutrinosAndColliders}%
  \BibitemOpen
  \bibfield  {author} {\bibinfo {author} {\bibfnamefont {F.~F.}\ \bibnamefont
  {Deppisch}}, \bibinfo {author} {\bibfnamefont {P.~S.}\ \bibnamefont
  {Bhupal~Dev}},\ and\ \bibinfo {author} {\bibfnamefont {A.}~\bibnamefont
  {Pilaftsis}},\ }\href {https://doi.org/10.1088/1367-2630/17/7/075019}
  {\bibfield  {journal} {\bibinfo  {journal} {\JNJP}\ }\textbf {\bibinfo
  {volume} {17}},\ \bibinfo {pages} {075019} (\bibinfo {year}
  {2015})}\BibitemShut {NoStop}%
\bibitem [{\citenamefont {Abazajian~\textit{et
  al}}(2012)}]{WhitePaperSterileNus}%
  \BibitemOpen
  \bibfield  {author} {\bibinfo {author} {\bibfnamefont {K.~N.}\ \bibnamefont
  {Abazajian~\textit{et al}}},\ }\href@noop {} {\  (\bibinfo {year} {2012})},\
  \Eprint {https://arxiv.org/abs/1204.5379} {arXiv:1204.5379 [hep-ph]}
  \BibitemShut {NoStop}%
\bibitem [{\citenamefont {Ibarra}\ \emph {et~al.}(2011)\citenamefont {Ibarra},
  \citenamefont {Molinaro},\ and\ \citenamefont {Petcov}}]{Ibarra2011}%
  \BibitemOpen
  \bibfield  {author} {\bibinfo {author} {\bibfnamefont {A.}~\bibnamefont
  {Ibarra}}, \bibinfo {author} {\bibfnamefont {E.}~\bibnamefont {Molinaro}},\
  and\ \bibinfo {author} {\bibfnamefont {S.~T.}\ \bibnamefont {Petcov}},\
  }\href {https://doi.org/10.1103/PhysRevD.84.013005} {\bibfield  {journal}
  {\bibinfo  {journal} {\JPRD}\ }\textbf {\bibinfo {volume} {84}},\ \bibinfo
  {pages} {013005} (\bibinfo {year} {2011})}\BibitemShut {NoStop}%
\bibitem [{\citenamefont {Fukugita}\ and\ \citenamefont
  {Yanagida}(1986)}]{BaryogenesisMixing}%
  \BibitemOpen
  \bibfield  {author} {\bibinfo {author} {\bibfnamefont {M.}~\bibnamefont
  {Fukugita}}\ and\ \bibinfo {author} {\bibfnamefont {T.}~\bibnamefont
  {Yanagida}},\ }\href
  {https://doi.org/https://doi.org/10.1016/0370-2693(86)91126-3} {\bibfield
  {journal} {\bibinfo  {journal} {\JPLB}\ }\textbf {\bibinfo {volume} {174}},\
  \bibinfo {pages} {45} (\bibinfo {year} {1986})}\BibitemShut {NoStop}%
\bibitem [{\citenamefont {Akhmedov}\ \emph {et~al.}(1998)\citenamefont
  {Akhmedov}, \citenamefont {Rubakov},\ and\ \citenamefont
  {Smirnov}}]{BaryogenesisNeutrinoOsc}%
  \BibitemOpen
  \bibfield  {author} {\bibinfo {author} {\bibfnamefont {E.~K.}\ \bibnamefont
  {Akhmedov}}, \bibinfo {author} {\bibfnamefont {V.~A.}\ \bibnamefont
  {Rubakov}},\ and\ \bibinfo {author} {\bibfnamefont {A.~Y.}\ \bibnamefont
  {Smirnov}},\ }\href {https://doi.org/10.1103/PhysRevLett.81.1359} {\bibfield
  {journal} {\bibinfo  {journal} {\JPRL}\ }\textbf {\bibinfo {volume} {81}},\
  \bibinfo {pages} {1359} (\bibinfo {year} {1998})}\BibitemShut {NoStop}%
\bibitem [{\citenamefont {Gorbunov}\ and\ \citenamefont
  {Shaposhnikov}(2007)}]{GorbunovNuMSM}%
  \BibitemOpen
  \bibfield  {author} {\bibinfo {author} {\bibfnamefont {D.}~\bibnamefont
  {Gorbunov}}\ and\ \bibinfo {author} {\bibfnamefont {M.}~\bibnamefont
  {Shaposhnikov}},\ }\href {https://doi.org/10.1088/1126-6708/2007/10/015}
  {\bibfield  {journal} {\bibinfo  {journal} {\JJHEP}\ }\textbf {\bibinfo
  {volume} {10}},\ \bibinfo {pages} {015} (\bibinfo {year} {2007})}\BibitemShut
  {NoStop}%
\bibitem [{\citenamefont {Bulbul~\textit{et al}}(2014)}]{3.5keVLine}%
  \BibitemOpen
  \bibfield  {author} {\bibinfo {author} {\bibfnamefont {E.}~\bibnamefont
  {Bulbul~\textit{et al}}},\ }\href
  {https://doi.org/10.1088/0004-637x/789/1/13} {\bibfield  {journal} {\bibinfo
  {journal} {\JAPJ}\ }\textbf {\bibinfo {volume} {789}},\ \bibinfo {pages} {13}
  (\bibinfo {year} {2014})}\BibitemShut {NoStop}%
\bibitem [{\citenamefont {Bhargava~\textit{et al}}(2020)}]{LineFail1}%
  \BibitemOpen
  \bibfield  {author} {\bibinfo {author} {\bibfnamefont {S.}~\bibnamefont
  {Bhargava~\textit{et al}}},\ }\href {https://doi.org/10.1093/mnras/staa1829}
  {\bibfield  {journal} {\bibinfo  {journal} {\JMNRAS}\ }\textbf {\bibinfo
  {volume} {497}},\ \bibinfo {pages} {656} (\bibinfo {year}
  {2020})}\BibitemShut {NoStop}%
\bibitem [{\citenamefont {Silich~\textit{et al}}(2021)}]{LineFail2}%
  \BibitemOpen
  \bibfield  {author} {\bibinfo {author} {\bibfnamefont {E.~M.}\ \bibnamefont
  {Silich~\textit{et al}}},\ }\href {https://doi.org/10.3847/1538-4357/ac043b}
  {\bibfield  {journal} {\bibinfo  {journal} {\JAPJ}\ }\textbf {\bibinfo
  {volume} {916}},\ \bibinfo {pages} {2} (\bibinfo {year} {2021})}\BibitemShut
  {NoStop}%
\bibitem [{\citenamefont {Abratenko~\textit{et al}}(2020)}]{uBooNEMuPiHNL}%
  \BibitemOpen
  \bibfield  {author} {\bibinfo {author} {\bibfnamefont {P.}~\bibnamefont
  {Abratenko~\textit{et al}}} (\bibinfo {collaboration} {{MicroBooNE
  Collaboration}}),\ }\href {https://doi.org/10.1103/PhysRevD.101.052001}
  {\bibfield  {journal} {\bibinfo  {journal} {\JPRD}\ }\textbf {\bibinfo
  {volume} {101}},\ \bibinfo {pages} {38} (\bibinfo {year} {2020})}\BibitemShut
  {NoStop}%
\bibitem [{\citenamefont {Porzio}(2019)}]{Porzio2019}%
  \BibitemOpen
  \bibfield  {author} {\bibinfo {author} {\bibfnamefont {S.~D.}\ \bibnamefont
  {Porzio}},\ }\emph {\bibinfo {title} {{Searches for Heavy Neutral Lepton
  Decays in the MicroBooNE Detector}}},\ \href
  {https://doi.org/10.2172/1576526} {Ph.D. thesis},\ \bibinfo  {school}
  {{Manchester University}} (\bibinfo {year} {2019})\BibitemShut {NoStop}%
\bibitem [{\citenamefont {Kelly}\ and\ \citenamefont
  {Machado}(2021)}]{HPSandHNL}%
  \BibitemOpen
  \bibfield  {author} {\bibinfo {author} {\bibfnamefont {K.~J.}\ \bibnamefont
  {Kelly}}\ and\ \bibinfo {author} {\bibfnamefont {P.~A.~N.}\ \bibnamefont
  {Machado}},\ }\href {https://doi.org/10.1103/PhysRevD.104.055015} {\bibfield
  {journal} {\bibinfo  {journal} {\JPRD}\ }\textbf {\bibinfo {volume} {104}},\
  \bibinfo {pages} {055015} (\bibinfo {year} {2021})}\BibitemShut {NoStop}%
\bibitem [{\citenamefont {Goodwin}(2022)}]{Goodwin2022}%
  \BibitemOpen
  \bibfield  {author} {\bibinfo {author} {\bibfnamefont {O.~R.~Y.}\
  \bibnamefont {Goodwin}},\ }\emph {\bibinfo {title} {{Search for Higgs Portal
  Scalars and Heavy Neutral Leptons Decaying in the MicroBooNE Detector}}},\
  \href {https://www.osti.gov/biblio/1872086} {Ph.D. thesis},\ \bibinfo
  {school} {{Manchester University}} (\bibinfo {year} {2022})\BibitemShut
  {NoStop}%
\bibitem [{\citenamefont {Abe~\textit{et al}}(2019)}]{T2KMainHNLSearch}%
  \BibitemOpen
  \bibfield  {author} {\bibinfo {author} {\bibfnamefont {K.}~\bibnamefont
  {Abe~\textit{et al}}} (\bibinfo {collaboration} {{T2K Collaboration}}),\
  }\href {https://doi.org/10.1103/PhysRevD.100.052006} {\bibfield  {journal}
  {\bibinfo  {journal} {\JPRD}\ }\textbf {\bibinfo {volume} {100}},\ \bibinfo
  {pages} {052006} (\bibinfo {year} {2019})}\BibitemShut {NoStop}%
\bibitem [{\citenamefont {Coloma~\textit{et
  al}}(2020{\natexlab{a}})}]{SKAtmHNL}%
  \BibitemOpen
  \bibfield  {author} {\bibinfo {author} {\bibfnamefont {P.}~\bibnamefont
  {Coloma~\textit{et al}}},\ }\href
  {https://doi.org/10.1140/epjc/s10052-020-7795-z} {\bibfield  {journal}
  {\bibinfo  {journal} {\JEPJC}\ }\textbf {\bibinfo {volume} {80}},\ \bibinfo
  {pages} {235} (\bibinfo {year} {2020}{\natexlab{a}})}\BibitemShut {NoStop}%
\bibitem [{\citenamefont {Acciarri~\textit{et al}}(2021)}]{Acciarri2021}%
  \BibitemOpen
  \bibfield  {author} {\bibinfo {author} {\bibfnamefont {R.}~\bibnamefont
  {Acciarri~\textit{et al}}} (\bibinfo {collaboration} {{ArgoNeuT
  Collaboration}}),\ }\href {https://doi.org/10.1103/PhysRevLett.127.121801}
  {\bibfield  {journal} {\bibinfo  {journal} {\JPRL}\ }\textbf {\bibinfo
  {volume} {127}},\ \bibinfo {pages} {121801} (\bibinfo {year}
  {2021})}\BibitemShut {NoStop}%
\bibitem [{\citenamefont {Aaij~\textit{et al}}(2021)}]{LHCbHNL}%
  \BibitemOpen
  \bibfield  {author} {\bibinfo {author} {\bibfnamefont {R.}~\bibnamefont
  {Aaij~\textit{et al}}} (\bibinfo {collaboration} {LHCb Collaboration}),\
  }\href {https://doi.org/10.1140/epjc/s10052-021-08973-5} {\bibfield
  {journal} {\bibinfo  {journal} {\JEPJC}\ }\textbf {\bibinfo {volume} {81}},\
  \bibinfo {pages} {248} (\bibinfo {year} {2021})}\BibitemShut {NoStop}%
\bibitem [{\citenamefont {Artamonov~\textit{et al}}(2015)}]{E949}%
  \BibitemOpen
  \bibfield  {author} {\bibinfo {author} {\bibfnamefont {A.~V.}\ \bibnamefont
  {Artamonov~\textit{et al}}} (\bibinfo {collaboration} {{E949
  Collaboration}}),\ }\href {https://doi.org/10.1103/physrevd.91.052001}
  {\bibfield  {journal} {\bibinfo  {journal} {\JPRD}\ }\textbf {\bibinfo
  {volume} {91}},\ \bibinfo {pages} {052001} (\bibinfo {year}
  {2015})}\BibitemShut {NoStop}%
\bibitem [{\citenamefont {Aguilar-Arevalo~\textit{et al}}(2018)}]{PIENU_Ue42}%
  \BibitemOpen
  \bibfield  {author} {\bibinfo {author} {\bibfnamefont {A.}~\bibnamefont
  {Aguilar-Arevalo~\textit{et al}}} (\bibinfo {collaboration} {{PIENU
  Collaboration}}),\ }\href {https://doi.org/10.1103/physrevd.97.072012}
  {\bibfield  {journal} {\bibinfo  {journal} {\JPRD}\ }\textbf {\bibinfo
  {volume} {97}},\ \bibinfo {pages} {072012} (\bibinfo {year}
  {2018})}\BibitemShut {NoStop}%
\bibitem [{\citenamefont {Aguilar-Arevalo~\textit{et al}}(2019)}]{PIENU_Um42}%
  \BibitemOpen
  \bibfield  {author} {\bibinfo {author} {\bibfnamefont {A.}~\bibnamefont
  {Aguilar-Arevalo~\textit{et al}}},\ }\href
  {https://doi.org/10.1016/j.physletb.2019.134980} {\bibfield  {journal}
  {\bibinfo  {journal} {\JPLB}\ }\textbf {\bibinfo {volume} {798}},\ \bibinfo
  {pages} {134980} (\bibinfo {year} {2019})}\BibitemShut {NoStop}%
\bibitem [{\citenamefont {Bernardi~\textit{et al}}(1986)}]{PS191_first}%
  \BibitemOpen
  \bibfield  {author} {\bibinfo {author} {\bibfnamefont {G.}~\bibnamefont
  {Bernardi~\textit{et al}}},\ }\href
  {https://doi.org/https://doi.org/10.1016/0370-2693(86)91602-3} {\bibfield
  {journal} {\bibinfo  {journal} {\JPLB}\ }\textbf {\bibinfo {volume} {166}},\
  \bibinfo {pages} {479} (\bibinfo {year} {1986})}\BibitemShut {NoStop}%
\bibitem [{\citenamefont {Bernardi~\textit{et al}}(1988)}]{PS191_second}%
  \BibitemOpen
  \bibfield  {author} {\bibinfo {author} {\bibfnamefont {G.}~\bibnamefont
  {Bernardi~\textit{et al}}},\ }\href
  {https://doi.org/https://doi.org/10.1016/0370-2693(88)90563-1} {\bibfield
  {journal} {\bibinfo  {journal} {\JPLB}\ }\textbf {\bibinfo {volume} {203}},\
  \bibinfo {pages} {332} (\bibinfo {year} {1988})}\BibitemShut {NoStop}%
\bibitem [{\citenamefont {Cortina Gil~\textrm{et al}}(2018)}]{NA62}%
  \BibitemOpen
  \bibfield  {author} {\bibinfo {author} {\bibfnamefont {E.}~\bibnamefont
  {Cortina Gil~\textrm{et al}}} (\bibinfo {collaboration} {{NA62
  Collaboration}}),\ }\href {https://doi.org/10.1016/j.physletb.2018.01.031}
  {\bibfield  {journal} {\bibinfo  {journal} {\JPLB}\ }\textbf {\bibinfo
  {volume} {778}},\ \bibinfo {pages} {137} (\bibinfo {year}
  {2018})}\BibitemShut {NoStop}%
\bibitem [{\citenamefont {Abdullahi~\textit{et
  al}}(2023)}]{SnowmassHNLOverview}%
  \BibitemOpen
  \bibfield  {author} {\bibinfo {author} {\bibfnamefont {A.~M.}\ \bibnamefont
  {Abdullahi~\textit{et al}}},\ }\href
  {https://doi.org/10.1088/1361-6471/ac98f9} {\bibfield  {journal} {\bibinfo
  {journal} {\JJPG}\ }\textbf {\bibinfo {volume} {50}},\ \bibinfo {pages}
  {020501} (\bibinfo {year} {2023})}\BibitemShut {NoStop}%
\bibitem [{\citenamefont {Abe~\textit{et al}}(2018)}]{HyperKDesignDoc}%
  \BibitemOpen
  \bibfield  {author} {\bibinfo {author} {\bibfnamefont {K.}~\bibnamefont
  {Abe~\textit{et al}}} (\bibinfo {collaboration} {{Hyper-Kamiokande
  proto-collaboration}}),\ }\href {https://arxiv.org/abs/1805.04163} {\
  (\bibinfo {year} {2018})},\ \Eprint {https://arxiv.org/abs/1805.04163}
  {arXiv:1805.04163 [ins-det]} \BibitemShut {NoStop}%
\bibitem [{\citenamefont {Abi~\textit{et al}}(2021)}]{DUNEBSMOverview}%
  \BibitemOpen
  \bibfield  {author} {\bibinfo {author} {\bibfnamefont {B.}~\bibnamefont
  {Abi~\textit{et al}}} (\bibinfo {collaboration} {{DUNE Collaboration}}),\
  }\href {https://doi.org/10.1140/epjc/s10052-021-09007-w} {\bibfield
  {journal} {\bibinfo  {journal} {\JEPJC}\ }\textbf {\bibinfo {volume} {81}},\
  \bibinfo {pages} {322} (\bibinfo {year} {2021})}\BibitemShut {NoStop}%
\bibitem [{\citenamefont {Breitbach~\textit{et al}}(2022)}]{DuneNDBSM}%
  \BibitemOpen
  \bibfield  {author} {\bibinfo {author} {\bibfnamefont {M.}~\bibnamefont
  {Breitbach~\textit{et al}}},\ }\href
  {https://doi.org/10.1007/JHEP01(2022)048} {\bibfield  {journal} {\bibinfo
  {journal} {\JJHEP}\ }\textbf {\bibinfo {volume} {01}},\ \bibinfo {pages} {48}
  (\bibinfo {year} {2022})}\BibitemShut {NoStop}%
\bibitem [{\citenamefont {Ballett}\ \emph {et~al.}(2017)\citenamefont
  {Ballett}, \citenamefont {Pascoli},\ and\ \citenamefont
  {Ross-Lonergan}}]{Ballett2017}%
  \BibitemOpen
  \bibfield  {author} {\bibinfo {author} {\bibfnamefont {P.}~\bibnamefont
  {Ballett}}, \bibinfo {author} {\bibfnamefont {S.}~\bibnamefont {Pascoli}},\
  and\ \bibinfo {author} {\bibfnamefont {M.}~\bibnamefont {Ross-Lonergan}},\
  }\href {https://doi.org/10.1007/JHEP04(2017)102} {\bibfield  {journal}
  {\bibinfo  {journal} {\JJHEP}\ }\textbf {\bibinfo {volume} {04}},\ \bibinfo
  {pages} {102} (\bibinfo {year} {2017})}\BibitemShut {NoStop}%
\bibitem [{\citenamefont {Ahdida~\textrm{et al}}(2019)}]{SHiPSensitivity}%
  \BibitemOpen
  \bibfield  {author} {\bibinfo {author} {\bibfnamefont {C.}~\bibnamefont
  {Ahdida~\textrm{et al}}} (\bibinfo {collaboration} {{SHiP Collaboration}}),\
  }\href {https://doi.org/10.1007/JHEP04(2019)077} {\bibfield  {journal}
  {\bibinfo  {journal} {\JJHEP}\ }\textbf {\bibinfo {volume} {04}},\ \bibinfo
  {pages} {77} (\bibinfo {year} {2019})}\BibitemShut {NoStop}%
\bibitem [{\citenamefont {Gorbunov~\textit{et al}}(2020)}]{SHiPKaons}%
  \BibitemOpen
  \bibfield  {author} {\bibinfo {author} {\bibfnamefont {D.}~\bibnamefont
  {Gorbunov~\textit{et al}}},\ }\href
  {https://doi.org/10.1016/j.physletb.2020.135817} {\bibfield  {journal}
  {\bibinfo  {journal} {\JPLB}\ }\textbf {\bibinfo {volume} {810}},\ \bibinfo
  {pages} {135817} (\bibinfo {year} {2020})}\BibitemShut {NoStop}%
\bibitem [{\citenamefont {Wang}\ and\ \citenamefont
  {Wang}(2020)}]{LeptonFDHNL}%
  \BibitemOpen
  \bibfield  {author} {\bibinfo {author} {\bibfnamefont {Z.~S.}\ \bibnamefont
  {Wang}}\ and\ \bibinfo {author} {\bibfnamefont {K.}~\bibnamefont {Wang}},\
  }\href {https://doi.org/10.1103/PhysRevD.101.075046} {\bibfield  {journal}
  {\bibinfo  {journal} {\JPRD}\ }\textbf {\bibinfo {volume} {101}},\ \bibinfo
  {pages} {075046} (\bibinfo {year} {2020})}\BibitemShut {NoStop}%
\bibitem [{\citenamefont {Feng~\textit{et al}}(2023)}]{SnowmassForwardLHCHNL}%
  \BibitemOpen
  \bibfield  {author} {\bibinfo {author} {\bibfnamefont {J.~L.}\ \bibnamefont
  {Feng~\textit{et al}}},\ }\href {https://doi.org/10.1088/1361-6471/ac865e}
  {\bibfield  {journal} {\bibinfo  {journal} {\JJPG}\ }\textbf {\bibinfo
  {volume} {50}},\ \bibinfo {pages} {030501} (\bibinfo {year}
  {2023})}\BibitemShut {NoStop}%
\bibitem [{\citenamefont {Cerci~\textit{et al}}(2022)}]{Cerci_2022}%
  \BibitemOpen
  \bibfield  {author} {\bibinfo {author} {\bibfnamefont {S.}~\bibnamefont
  {Cerci~\textit{et al}}},\ }\href {https://doi.org/10.1007/jhep06(2022)110}
  {\bibfield  {journal} {\bibinfo  {journal} {\JJHEP}\ }\textbf {\bibinfo
  {volume} {06}},\ \bibinfo {pages} {110} (\bibinfo {year} {2022})}\BibitemShut
  {NoStop}%
\bibitem [{\citenamefont {Kling}\ and\ \citenamefont
  {Trojanowski}(2018)}]{FASERHNL}%
  \BibitemOpen
  \bibfield  {author} {\bibinfo {author} {\bibfnamefont {F.}~\bibnamefont
  {Kling}}\ and\ \bibinfo {author} {\bibfnamefont {S.}~\bibnamefont
  {Trojanowski}},\ }\href {https://doi.org/10.1103/PhysRevD.97.095016}
  {\bibfield  {journal} {\bibinfo  {journal} {\JPRD}\ }\textbf {\bibinfo
  {volume} {97}},\ \bibinfo {pages} {095016} (\bibinfo {year}
  {2018})}\BibitemShut {NoStop}%
\bibitem [{\citenamefont {Bryman}\ and\ \citenamefont
  {Shrock}(2019)}]{BrymanModelIndependentBounds}%
  \BibitemOpen
  \bibfield  {author} {\bibinfo {author} {\bibfnamefont {D.}~\bibnamefont
  {Bryman}}\ and\ \bibinfo {author} {\bibfnamefont {R.}~\bibnamefont
  {Shrock}},\ }\href {https://doi.org/10.1103/physrevd.100.073011} {\bibfield
  {journal} {\bibinfo  {journal} {\JPRD}\ }\textbf {\bibinfo {volume} {100}},\
  \bibinfo {pages} {073011} (\bibinfo {year} {2019})}\BibitemShut {NoStop}%
\bibitem [{\citenamefont {Mosel}(2019)}]{UlrichOverview}%
  \BibitemOpen
  \bibfield  {author} {\bibinfo {author} {\bibfnamefont {U.}~\bibnamefont
  {Mosel}},\ }\href {https://doi.org/10.1088/1361-6471/ab3830} {\bibfield
  {journal} {\bibinfo  {journal} {\JJPG}\ }\textbf {\bibinfo {volume} {46}},\
  \bibinfo {pages} {113001} (\bibinfo {year} {2019})}\BibitemShut {NoStop}%
\bibitem [{\citenamefont {Campbell~\textit{et
  al}}(2022)}]{EventGeneratorsOverviewTHISISWHYWEDOHNLGENIE}%
  \BibitemOpen
  \bibfield  {author} {\bibinfo {author} {\bibfnamefont {J.~M.}\ \bibnamefont
  {Campbell~\textit{et al}}},\ }\href {https://arxiv.org/abs/2203.11110} {\
  (\bibinfo {year} {2022})},\ \Eprint {https://arxiv.org/abs/2203.11110}
  {arXiv:2203.11110 [hep-ph]} \BibitemShut {NoStop}%
\bibitem [{\citenamefont {Isaacson~\textit{et al}}(2022)}]{Achilles}%
  \BibitemOpen
  \bibfield  {author} {\bibinfo {author} {\bibfnamefont {J.}~\bibnamefont
  {Isaacson~\textit{et al}}},\ }\href
  {https://doi.org/10.1103/physrevd.105.096006} {\bibfield  {journal} {\bibinfo
   {journal} {\JPRD}\ }\textbf {\bibinfo {volume} {105}},\ \bibinfo {pages}
  {096006} (\bibinfo {year} {2022})}\BibitemShut {NoStop}%
\bibitem [{\citenamefont {Abdullahi~\textit{et al}}(2022)}]{DarkNews}%
  \BibitemOpen
  \bibfield  {author} {\bibinfo {author} {\bibfnamefont {A.~M.}\ \bibnamefont
  {Abdullahi~\textit{et al}}},\ }\href {https://arxiv.org/abs/2207.04137} {\
  (\bibinfo {year} {2022})},\ \Eprint {https://arxiv.org/abs/2207.04137}
  {arXiv:2207.04137 [hep-ph]} \BibitemShut {NoStop}%
\bibitem [{\citenamefont {Andreopoulos~\textit{et al}}(2010)}]{GENIEMainPaper}%
  \BibitemOpen
  \bibfield  {author} {\bibinfo {author} {\bibfnamefont {C.}~\bibnamefont
  {Andreopoulos~\textit{et al}}},\ }\href
  {https://doi.org/10.1016/j.nima.2009.12.009} {\bibfield  {journal} {\bibinfo
  {journal} {\JNIMA}\ }\textbf {\bibinfo {volume} {614}},\ \bibinfo {pages}
  {87} (\bibinfo {year} {2010})}\BibitemShut {NoStop}%
\bibitem [{\citenamefont {Andreopoulos~\textit{et al}}(2015)}]{GENIEManual}%
  \BibitemOpen
  \bibfield  {author} {\bibinfo {author} {\bibfnamefont {C.}~\bibnamefont
  {Andreopoulos~\textit{et al}}},\ }\href@noop {} {\  (\bibinfo {year}
  {2015})},\ \Eprint {https://arxiv.org/abs/1510.05494} {arXiv:1510.05494
  [hep-ph]} \BibitemShut {NoStop}%
\bibitem [{\citenamefont {Brun}\ and\ \citenamefont {Rademakers}(1997)}]{ROOT}%
  \BibitemOpen
  \bibfield  {author} {\bibinfo {author} {\bibfnamefont {R.}~\bibnamefont
  {Brun}}\ and\ \bibinfo {author} {\bibfnamefont {F.}~\bibnamefont
  {Rademakers}},\ }\href
  {https://doi.org/https://doi.org/10.1016/S0168-9002(97)00048-X} {\bibfield
  {journal} {\bibinfo  {journal} {\JNIMA}\ }\textbf {\bibinfo {volume} {389}},\
  \bibinfo {pages} {81} (\bibinfo {year} {1997})}\BibitemShut {NoStop}%
\bibitem [{ROO()}]{ROOTMaster}%
  \BibitemOpen
  \href {{https://doi.org/10.5281/zenodo.848818}} {\bibinfo {title} {{ROOT
  version 6.18/02}}}\BibitemShut {NoStop}%
\bibitem [{\citenamefont {Alvarez-Ruso~\textit{et al}}(2021)}]{GENIEv3Updates}%
  \BibitemOpen
  \bibfield  {author} {\bibinfo {author} {\bibfnamefont {L.}~\bibnamefont
  {Alvarez-Ruso~\textit{et al}}},\ }\href
  {https://doi.org/10.1140/epjs/s11734-021-00295-7} {\bibfield  {journal}
  {\bibinfo  {journal} {\JEPJST}\ }\textbf {\bibinfo {volume} {230}},\ \bibinfo
  {pages} {4449} (\bibinfo {year} {2021})}\BibitemShut {NoStop}%
\bibitem [{Dar()}]{DarkNeutrino}%
  \BibitemOpen
  \href
  {https://github.com/GENIE-MC/Generator/tree/master/src/Physics/DarkNeutrino}
  {\bibinfo {title} {{GENIE DarkNeutrino module (available from GENIE
  v3.2.0)}}}\BibitemShut {NoStop}%
\bibitem [{\citenamefont {Bertuzzo~\textit{et al}}(2018)}]{DarkNeutrinoPaper}%
  \BibitemOpen
  \bibfield  {author} {\bibinfo {author} {\bibfnamefont {E.}~\bibnamefont
  {Bertuzzo~\textit{et al}}},\ }\href
  {https://doi.org/10.1103/physrevlett.121.241801} {\bibfield  {journal}
  {\bibinfo  {journal} {\JPRL}\ }\textbf {\bibinfo {volume} {121}},\ \bibinfo
  {pages} {241801} (\bibinfo {year} {2018})}\BibitemShut {NoStop}%
\bibitem [{\citenamefont {Drewes}(2013)}]{Drewes2013}%
  \BibitemOpen
  \bibfield  {author} {\bibinfo {author} {\bibfnamefont {M.}~\bibnamefont
  {Drewes}},\ }\href {https://doi.org/10.1142/s0218301313300191} {\bibfield
  {journal} {\bibinfo  {journal} {\JIJMPE}\ }\textbf {\bibinfo {volume} {22}},\
  \bibinfo {pages} {1330019} (\bibinfo {year} {2013})}\BibitemShut {NoStop}%
\bibitem [{\citenamefont {Antusch~\textit{et
  al}}(2006)}]{AntuschNonUnitaryMixing}%
  \BibitemOpen
  \bibfield  {author} {\bibinfo {author} {\bibfnamefont {S.}~\bibnamefont
  {Antusch~\textit{et al}}},\ }\href
  {https://doi.org/10.1088/1126-6708/2006/10/084} {\bibfield  {journal}
  {\bibinfo  {journal} {\JJHEP}\ }\textbf {\bibinfo {volume} {10}},\ \bibinfo
  {pages} {084} (\bibinfo {year} {2006})}\BibitemShut {NoStop}%
\bibitem [{\citenamefont {Agostinho~\textit{et
  al}}(2018)}]{PMNSNonUnitarityTypeI}%
  \BibitemOpen
  \bibfield  {author} {\bibinfo {author} {\bibfnamefont {N.~R.}\ \bibnamefont
  {Agostinho~\textit{et al}}},\ }\href
  {https://doi.org/10.1140/epjc/s10052-018-6347-2} {\bibfield  {journal}
  {\bibinfo  {journal} {\JEPJC}\ }\textbf {\bibinfo {volume} {78}},\ \bibinfo
  {pages} {895} (\bibinfo {year} {2018})}\BibitemShut {NoStop}%
\bibitem [{\citenamefont {Soumya}(2022)}]{T2KNonUnitarity}%
  \BibitemOpen
  \bibfield  {author} {\bibinfo {author} {\bibfnamefont {C.}~\bibnamefont
  {Soumya}},\ }\href {https://doi.org/10.1103/PhysRevD.105.015012} {\bibfield
  {journal} {\bibinfo  {journal} {\JPRD}\ }\textbf {\bibinfo {volume} {105}},\
  \bibinfo {pages} {015012} (\bibinfo {year} {2022})}\BibitemShut {NoStop}%
\bibitem [{\citenamefont {Esteban~\textit{et al}}(2020)}]{NuFitPub}%
  \BibitemOpen
  \bibfield  {author} {\bibinfo {author} {\bibfnamefont {I.}~\bibnamefont
  {Esteban~\textit{et al}}},\ }\href {https://doi.org/10.1007/jhep09(2020)178}
  {\bibfield  {journal} {\bibinfo  {journal} {\JJHEP}\ }\textbf {\bibinfo
  {volume} {09}},\ \bibinfo {pages} {178} (\bibinfo {year} {2020})}\BibitemShut
  {NoStop}%
\bibitem [{\citenamefont {Chuliá}\ \emph {et~al.}(2021)\citenamefont
  {Chuliá}, \citenamefont {Srivastava},\ and\ \citenamefont
  {Vicente}}]{InverseDiracAndMajorana}%
  \BibitemOpen
  \bibfield  {author} {\bibinfo {author} {\bibfnamefont {S.~C.}\ \bibnamefont
  {Chuliá}}, \bibinfo {author} {\bibfnamefont {R.}~\bibnamefont
  {Srivastava}},\ and\ \bibinfo {author} {\bibfnamefont {A.}~\bibnamefont
  {Vicente}},\ }\href {https://doi.org/10.1007/JHEP03(2021)248} {\bibfield
  {journal} {\bibinfo  {journal} {\JJHEP}\ }\textbf {\bibinfo {volume} {03}},\
  \bibinfo {pages} {248} (\bibinfo {year} {2021})}\BibitemShut {NoStop}%
\bibitem [{\citenamefont {Borah}\ and\ \citenamefont
  {Karmakar}(2019)}]{LinearDiracSeesaw}%
  \BibitemOpen
  \bibfield  {author} {\bibinfo {author} {\bibfnamefont {D.}~\bibnamefont
  {Borah}}\ and\ \bibinfo {author} {\bibfnamefont {B.}~\bibnamefont
  {Karmakar}},\ }\href
  {https://doi.org/https://doi.org/10.1016/j.physletb.2018.12.006} {\bibfield
  {journal} {\bibinfo  {journal} {\JPLB}\ }\textbf {\bibinfo {volume} {789}},\
  \bibinfo {pages} {59} (\bibinfo {year} {2019})}\BibitemShut {NoStop}%
\bibitem [{\citenamefont {Tastet}\ and\ \citenamefont
  {Timiryasov}(2020)}]{TastetPol}%
  \BibitemOpen
  \bibfield  {author} {\bibinfo {author} {\bibfnamefont {J.-L.}\ \bibnamefont
  {Tastet}}\ and\ \bibinfo {author} {\bibfnamefont {I.}~\bibnamefont
  {Timiryasov}},\ }\href {https://doi.org/10.1007/JHEP04(2020)005} {\bibfield
  {journal} {\bibinfo  {journal} {\JJHEP}\ }\textbf {\bibinfo {volume} {04}},\
  \bibinfo {pages} {5} (\bibinfo {year} {2020})}\BibitemShut {NoStop}%
\bibitem [{\citenamefont {Anamiati}\ \emph {et~al.}(2016)\citenamefont
  {Anamiati}, \citenamefont {Hirsch},\ and\ \citenamefont {Nardi}}]{HNLOsc}%
  \BibitemOpen
  \bibfield  {author} {\bibinfo {author} {\bibfnamefont {G.}~\bibnamefont
  {Anamiati}}, \bibinfo {author} {\bibfnamefont {M.}~\bibnamefont {Hirsch}},\
  and\ \bibinfo {author} {\bibfnamefont {E.}~\bibnamefont {Nardi}},\ }\href
  {https://doi.org/10.1007/jhep10(2016)010} {\bibfield  {journal} {\bibinfo
  {journal} {\JJHEP}\ }\textbf {\bibinfo {volume} {10}},\ \bibinfo {pages} {10}
  (\bibinfo {year} {2016})}\BibitemShut {NoStop}%
\bibitem [{\citenamefont {Gustafson}\ \emph {et~al.}(2022)\citenamefont
  {Gustafson}, \citenamefont {Plestid},\ and\ \citenamefont
  {Shoemaker}}]{NDipoleUpscattering}%
  \BibitemOpen
  \bibfield  {author} {\bibinfo {author} {\bibfnamefont {R.~A.}\ \bibnamefont
  {Gustafson}}, \bibinfo {author} {\bibfnamefont {R.}~\bibnamefont {Plestid}},\
  and\ \bibinfo {author} {\bibfnamefont {I.~M.}\ \bibnamefont {Shoemaker}},\
  }\href {https://doi.org/10.1103/PhysRevD.106.095037} {\bibfield  {journal}
  {\bibinfo  {journal} {\JPRD}\ }\textbf {\bibinfo {volume} {106}},\ \bibinfo
  {pages} {095037} (\bibinfo {year} {2022})}\BibitemShut {NoStop}%
\bibitem [{\citenamefont {Magill~\textit{et al}}(2018)}]{DipoleHNL}%
  \BibitemOpen
  \bibfield  {author} {\bibinfo {author} {\bibfnamefont {G.}~\bibnamefont
  {Magill~\textit{et al}}},\ }\href
  {https://doi.org/10.1103/PhysRevD.98.115015} {\bibfield  {journal} {\bibinfo
  {journal} {\JPRD}\ }\textbf {\bibinfo {volume} {98}},\ \bibinfo {pages}
  {115015} (\bibinfo {year} {2018})}\BibitemShut {NoStop}%
\bibitem [{\citenamefont {Ovchynnikov}\ \emph {et~al.}(2022)\citenamefont
  {Ovchynnikov}, \citenamefont {Schwetz},\ and\ \citenamefont
  {Zhu}}]{DipoleNuTail}%
  \BibitemOpen
  \bibfield  {author} {\bibinfo {author} {\bibfnamefont {M.}~\bibnamefont
  {Ovchynnikov}}, \bibinfo {author} {\bibfnamefont {T.}~\bibnamefont
  {Schwetz}},\ and\ \bibinfo {author} {\bibfnamefont {J.-Y.}\ \bibnamefont
  {Zhu}},\ }\href {https://doi.org/10.48550/ARXIV.2210.13141} {\  (\bibinfo
  {year} {2022})},\ \Eprint {https://arxiv.org/abs/2210.13141}
  {arXiv:2210.13141 [hep-ph]} \BibitemShut {NoStop}%
\bibitem [{\citenamefont {Kopp~\textit{et al}}(2013)}]{SterileOscillations}%
  \BibitemOpen
  \bibfield  {author} {\bibinfo {author} {\bibfnamefont {J.}~\bibnamefont
  {Kopp~\textit{et al}}},\ }\href {https://doi.org/10.1007/jhep05(2013)050}
  {\bibfield  {journal} {\bibinfo  {journal} {\JJHEP}\ }\textbf {\bibinfo
  {volume} {05}},\ \bibinfo {pages} {50} (\bibinfo {year} {2013})}\BibitemShut
  {NoStop}%
\bibitem [{\citenamefont {Coloma~\textit{et al}}(2017)}]{IceCubeDoubleBangs}%
  \BibitemOpen
  \bibfield  {author} {\bibinfo {author} {\bibfnamefont {P.}~\bibnamefont
  {Coloma~\textit{et al}}},\ }\href
  {https://doi.org/10.1103/PhysRevLett.119.201804} {\bibfield  {journal}
  {\bibinfo  {journal} {\JPRL}\ }\textbf {\bibinfo {volume} {119}},\ \bibinfo
  {pages} {201804} (\bibinfo {year} {2017})}\BibitemShut {NoStop}%
\bibitem [{\citenamefont {Coloma~\textit{et
  al}}(2020{\natexlab{b}})}]{ColomaDUNEHNL}%
  \BibitemOpen
  \bibfield  {author} {\bibinfo {author} {\bibfnamefont {P.}~\bibnamefont
  {Coloma~\textit{et al}}},\ }\href
  {https://doi.org/10.1140/epjc/s10052-021-08861-y} {\bibfield  {journal}
  {\bibinfo  {journal} {\JEPJC}\ }\textbf {\bibinfo {volume} {81}},\ \bibinfo
  {pages} {78} (\bibinfo {year} {2020}{\natexlab{b}})}\BibitemShut {NoStop}%
\bibitem [{\citenamefont {Keung}\ and\ \citenamefont
  {Senjanovi\'{c}}(1983)}]{Colliders1}%
  \BibitemOpen
  \bibfield  {author} {\bibinfo {author} {\bibfnamefont {W.-Y.}\ \bibnamefont
  {Keung}}\ and\ \bibinfo {author} {\bibfnamefont {G.}~\bibnamefont
  {Senjanovi\'{c}}},\ }\href {https://doi.org/10.1103/PhysRevLett.50.1427}
  {\bibfield  {journal} {\bibinfo  {journal} {\JPRL}\ }\textbf {\bibinfo
  {volume} {50}},\ \bibinfo {pages} {1427} (\bibinfo {year}
  {1983})}\BibitemShut {NoStop}%
\bibitem [{\citenamefont {Hessler~\textit{et al}}(2015)}]{Colliders2}%
  \BibitemOpen
  \bibfield  {author} {\bibinfo {author} {\bibfnamefont {A.~G.}\ \bibnamefont
  {Hessler~\textit{et al}}},\ }\href
  {https://doi.org/10.1103/PhysRevD.91.115004} {\bibfield  {journal} {\bibinfo
  {journal} {\JPRD}\ }\textbf {\bibinfo {volume} {91}},\ \bibinfo {pages}
  {115004} (\bibinfo {year} {2015})}\BibitemShut {NoStop}%
\bibitem [{\citenamefont {Alva}\ \emph {et~al.}(2015)\citenamefont {Alva},
  \citenamefont {Han},\ and\ \citenamefont {Ruiz}}]{Colliders3}%
  \BibitemOpen
  \bibfield  {author} {\bibinfo {author} {\bibfnamefont {D.}~\bibnamefont
  {Alva}}, \bibinfo {author} {\bibfnamefont {T.}~\bibnamefont {Han}},\ and\
  \bibinfo {author} {\bibfnamefont {R.}~\bibnamefont {Ruiz}},\ }\href
  {https://doi.org/10.1007/jhep02(2015)072} {\bibfield  {journal} {\bibinfo
  {journal} {\JJHEP}\ }\textbf {\bibinfo {volume} {02}},\ \bibinfo {pages} {72}
  (\bibinfo {year} {2015})}\BibitemShut {NoStop}%
\bibitem [{\citenamefont {Fuks~\textit{et al}}(2021)}]{Colliders4}%
  \BibitemOpen
  \bibfield  {author} {\bibinfo {author} {\bibfnamefont {B.}~\bibnamefont
  {Fuks~\textit{et al}}},\ }\href {https://doi.org/10.1103/PhysRevD.103.055005}
  {\bibfield  {journal} {\bibinfo  {journal} {\JPRD}\ }\textbf {\bibinfo
  {volume} {103}},\ \bibinfo {pages} {055005} (\bibinfo {year}
  {2021})}\BibitemShut {NoStop}%
\bibitem [{\citenamefont {Bondarenko~\textit{et
  al}}(2018)}]{Bondarenko_HNLPheno}%
  \BibitemOpen
  \bibfield  {author} {\bibinfo {author} {\bibfnamefont {K.}~\bibnamefont
  {Bondarenko~\textit{et al}}},\ }\href
  {https://doi.org/10.1007/jhep11(2018)032} {\bibfield  {journal} {\bibinfo
  {journal} {\JJHEP}\ }\textbf {\bibinfo {volume} {11}},\ \bibinfo {pages} {32}
  (\bibinfo {year} {2018})}\BibitemShut {NoStop}%
\bibitem [{\citenamefont {Shrock}(1981)}]{Shrock1981}%
  \BibitemOpen
  \bibfield  {author} {\bibinfo {author} {\bibfnamefont {R.~E.}\ \bibnamefont
  {Shrock}},\ }\href {https://doi.org/10.1103/PhysRevD.24.1232} {\bibfield
  {journal} {\bibinfo  {journal} {\JPRD}\ }\textbf {\bibinfo {volume} {24}},\
  \bibinfo {pages} {1232} (\bibinfo {year} {1981})}\BibitemShut {NoStop}%
\bibitem [{\citenamefont {Alloul~\textit{et al}}(2014)}]{FeynRulesManual}%
  \BibitemOpen
  \bibfield  {author} {\bibinfo {author} {\bibfnamefont {A.}~\bibnamefont
  {Alloul~\textit{et al}}},\ }\href {https://doi.org/10.1016/j.cpc.2014.04.012}
  {\bibfield  {journal} {\bibinfo  {journal} {\JCPC}\ }\textbf {\bibinfo
  {volume} {185}},\ \bibinfo {pages} {2250} (\bibinfo {year}
  {2014})}\BibitemShut {NoStop}%
\bibitem [{\citenamefont {Hahn}(2001)}]{FeynArts}%
  \BibitemOpen
  \bibfield  {author} {\bibinfo {author} {\bibfnamefont {T.}~\bibnamefont
  {Hahn}},\ }\href {https://doi.org/10.1016/s0010-4655(01)00290-9} {\bibfield
  {journal} {\bibinfo  {journal} {\JCPC}\ }\textbf {\bibinfo {volume} {140}},\
  \bibinfo {pages} {418} (\bibinfo {year} {2001})}\BibitemShut {NoStop}%
\bibitem [{\citenamefont {Mertig}\ \emph {et~al.}(1991)\citenamefont {Mertig},
  \citenamefont {Böhm},\ and\ \citenamefont {Denner}}]{FeynCalc1}%
  \BibitemOpen
  \bibfield  {author} {\bibinfo {author} {\bibfnamefont {R.}~\bibnamefont
  {Mertig}}, \bibinfo {author} {\bibfnamefont {M.}~\bibnamefont {Böhm}},\ and\
  \bibinfo {author} {\bibfnamefont {A.}~\bibnamefont {Denner}},\ }\href
  {https://doi.org/https://doi.org/10.1016/0010-4655(91)90130-D} {\bibfield
  {journal} {\bibinfo  {journal} {\JCPC}\ }\textbf {\bibinfo {volume} {64}},\
  \bibinfo {pages} {345} (\bibinfo {year} {1991})}\BibitemShut {NoStop}%
\bibitem [{\citenamefont {Shtabovenko}\ \emph {et~al.}(2016)\citenamefont
  {Shtabovenko}, \citenamefont {Mertig},\ and\ \citenamefont
  {Orellana}}]{FeynCalc2}%
  \BibitemOpen
  \bibfield  {author} {\bibinfo {author} {\bibfnamefont {V.}~\bibnamefont
  {Shtabovenko}}, \bibinfo {author} {\bibfnamefont {R.}~\bibnamefont
  {Mertig}},\ and\ \bibinfo {author} {\bibfnamefont {F.}~\bibnamefont
  {Orellana}},\ }\href {https://doi.org/10.1016/j.cpc.2016.06.008} {\bibfield
  {journal} {\bibinfo  {journal} {\JCPC}\ }\textbf {\bibinfo {volume} {207}},\
  \bibinfo {pages} {432} (\bibinfo {year} {2016})}\BibitemShut {NoStop}%
\bibitem [{\citenamefont {Shtabovenko}\ \emph {et~al.}(2020)\citenamefont
  {Shtabovenko}, \citenamefont {Mertig},\ and\ \citenamefont
  {Orellana}}]{FeynCalc3}%
  \BibitemOpen
  \bibfield  {author} {\bibinfo {author} {\bibfnamefont {V.}~\bibnamefont
  {Shtabovenko}}, \bibinfo {author} {\bibfnamefont {R.}~\bibnamefont
  {Mertig}},\ and\ \bibinfo {author} {\bibfnamefont {F.}~\bibnamefont
  {Orellana}},\ }\href {https://doi.org/10.1016/j.cpc.2020.107478} {\bibfield
  {journal} {\bibinfo  {journal} {\JCPC}\ }\textbf {\bibinfo {volume} {256}},\
  \bibinfo {pages} {107478} (\bibinfo {year} {2020})}\BibitemShut {NoStop}%
\bibitem [{Fey()}]{FeynRulesWebsite}%
  \BibitemOpen
  \href {https://feynrules.irmp.ucl.ac.be/wiki/ModelDatabaseMainPage} {\bibinfo
  {title} {{FeynRules model database}}}\BibitemShut {NoStop}%
\bibitem [{\citenamefont {Buss~\textit{et al}}(2012)}]{GiBUU}%
  \BibitemOpen
  \bibfield  {author} {\bibinfo {author} {\bibfnamefont {O.}~\bibnamefont
  {Buss~\textit{et al}}},\ }\href
  {https://doi.org/10.1016/j.physrep.2011.12.001} {\bibfield  {journal}
  {\bibinfo  {journal} {\JPREP}\ }\textbf {\bibinfo {volume} {512}},\ \bibinfo
  {pages} {1} (\bibinfo {year} {2012})}\BibitemShut {NoStop}%
\bibitem [{\citenamefont {Golan}\ \emph {et~al.}(2012)\citenamefont {Golan},
  \citenamefont {Sobczyk},\ and\ \citenamefont {Żmuda}}]{NuWro}%
  \BibitemOpen
  \bibfield  {author} {\bibinfo {author} {\bibfnamefont {T.}~\bibnamefont
  {Golan}}, \bibinfo {author} {\bibfnamefont {J.~T.}\ \bibnamefont {Sobczyk}},\
  and\ \bibinfo {author} {\bibfnamefont {J.}~\bibnamefont {Żmuda}},\ }\href
  {https://doi.org/https://doi.org/10.1016/j.nuclphysbps.2012.09.136}
  {\bibfield  {journal} {\bibinfo  {journal} {\JNPB}\ }\textbf {\bibinfo
  {volume} {229-232}},\ \bibinfo {pages} {499} (\bibinfo {year} {2012})},\
  \bibinfo {note} {neutrino 2010}\BibitemShut {NoStop}%
\bibitem [{\citenamefont {Hayato}\ and\ \citenamefont
  {Pickering}(2021)}]{NEUT}%
  \BibitemOpen
  \bibfield  {author} {\bibinfo {author} {\bibfnamefont {Y.}~\bibnamefont
  {Hayato}}\ and\ \bibinfo {author} {\bibfnamefont {L.}~\bibnamefont
  {Pickering}},\ }\href {https://doi.org/10.1140/epjs/s11734-021-00287-7}
  {\bibfield  {journal} {\bibinfo  {journal} {\JEPJST}\ }\textbf {\bibinfo
  {volume} {230}},\ \bibinfo {pages} {4469} (\bibinfo {year}
  {2021})}\BibitemShut {NoStop}%
\bibitem [{\citenamefont {Young}\ and\ \citenamefont
  {Mohlenkamp}(2021)}]{DoubleSimpson}%
  \BibitemOpen
  \bibfield  {author} {\bibinfo {author} {\bibfnamefont {T.}~\bibnamefont
  {Young}}\ and\ \bibinfo {author} {\bibfnamefont {M.~J.}\ \bibnamefont
  {Mohlenkamp}},\ }\href
  {http://www.ohiouniversityfaculty.com/youngt/IntNumMeth/} {\bibinfo {title}
  {{Introduction to Numerical Methods and Matlab Programming for Engineers}}}
  (\bibinfo {year} {2021})\BibitemShut {NoStop}%
\bibitem [{\citenamefont {Ruiz}(2021)}]{Ruiz_2021}%
  \BibitemOpen
  \bibfield  {author} {\bibinfo {author} {\bibfnamefont {R.}~\bibnamefont
  {Ruiz}},\ }\href {https://doi.org/10.1103/physrevd.103.015022} {\bibfield
  {journal} {\bibinfo  {journal} {\JPRD}\ }\textbf {\bibinfo {volume} {103}},\
  \bibinfo {pages} {015022} (\bibinfo {year} {2021})}\BibitemShut {NoStop}%
\bibitem [{\citenamefont {Degrande~\textit{et al}}(2016)}]{Degrande_2016}%
  \BibitemOpen
  \bibfield  {author} {\bibinfo {author} {\bibfnamefont {C.}~\bibnamefont
  {Degrande~\textit{et al}}},\ }\href
  {https://doi.org/10.1103/physrevd.94.053002} {\bibfield  {journal} {\bibinfo
  {journal} {\JPRD}\ }\textbf {\bibinfo {volume} {94}},\ \bibinfo {pages}
  {053002} (\bibinfo {year} {2016})}\BibitemShut {NoStop}%
\bibitem [{\citenamefont {Gorkavenko}\ \emph {et~al.}(2021)\citenamefont
  {Gorkavenko}, \citenamefont {Borysenkova},\ and\ \citenamefont
  {Tsarenkova}}]{GorkavenkoPythiaHNL}%
  \BibitemOpen
  \bibfield  {author} {\bibinfo {author} {\bibfnamefont {V.~M.}\ \bibnamefont
  {Gorkavenko}}, \bibinfo {author} {\bibfnamefont {Y.~R.}\ \bibnamefont
  {Borysenkova}},\ and\ \bibinfo {author} {\bibfnamefont {M.~S.}\ \bibnamefont
  {Tsarenkova}},\ }\href {https://doi.org/10.1088/1361-6471/ac1394} {\bibfield
  {journal} {\bibinfo  {journal} {\JJPG}\ }\textbf {\bibinfo {volume} {48}},\
  \bibinfo {pages} {105001} (\bibinfo {year} {2021})}\BibitemShut {NoStop}%
\bibitem [{\citenamefont {Boschi}(2021)}]{NuShock}%
  \BibitemOpen
  \bibfield  {author} {\bibinfo {author} {\bibfnamefont {T.}~\bibnamefont
  {Boschi}},\ }\href {https://github.com/tboschi/NuShock} {\bibinfo {title}
  {{NuShock: Bunch of tools to study sensitivity of heavy neutrinos}}}
  (\bibinfo {year} {2021})\BibitemShut {NoStop}%
\bibitem [{\citenamefont {Ballett}\ \emph {et~al.}(2020)\citenamefont
  {Ballett}, \citenamefont {Boschi},\ and\ \citenamefont
  {Pascoli}}]{Ballett2020}%
  \BibitemOpen
  \bibfield  {author} {\bibinfo {author} {\bibfnamefont {P.}~\bibnamefont
  {Ballett}}, \bibinfo {author} {\bibfnamefont {T.}~\bibnamefont {Boschi}},\
  and\ \bibinfo {author} {\bibfnamefont {S.}~\bibnamefont {Pascoli}},\ }\href
  {https://doi.org/10.1007/JHEP03(2020)111} {\bibfield  {journal} {\bibinfo
  {journal} {\JJHEP}\ }\textbf {\bibinfo {volume} {11}},\ \bibinfo {pages}
  {111} (\bibinfo {year} {2020})}\BibitemShut {NoStop}%
\bibitem [{\citenamefont {Batell~\textit{et al}}(2021)}]{DarkQuestHNL}%
  \BibitemOpen
  \bibfield  {author} {\bibinfo {author} {\bibfnamefont {B.}~\bibnamefont
  {Batell~\textit{et al}}},\ }\href {https://doi.org/10.1007/jhep05(2021)049}
  {\bibfield  {journal} {\bibinfo  {journal} {\JJHEP}\ }\textbf {\bibinfo
  {volume} {05}},\ \bibinfo {pages} {49} (\bibinfo {year} {2021})}\BibitemShut
  {NoStop}%
\bibitem [{\citenamefont {Strait~\textit{et al}}(2016)}]{LBNF_CDR}%
  \BibitemOpen
  \bibfield  {author} {\bibinfo {author} {\bibfnamefont {J.}~\bibnamefont
  {Strait~\textit{et al}}},\ }\href {https://arxiv.org/abs/1601.05823} {\
  (\bibinfo {year} {2016})},\ \Eprint {https://arxiv.org/abs/1601.05823}
  {arXiv:1601.05823 [ins-det]} \BibitemShut {NoStop}%
\bibitem [{\citenamefont {Abud~\textit{et al}}(2021)}]{DUNE_ND_CDR}%
  \BibitemOpen
  \bibfield  {author} {\bibinfo {author} {\bibfnamefont {A.~A.}\ \bibnamefont
  {Abud~\textit{et al}}},\ }\href {https://doi.org/10.3390/instruments5040031}
  {\bibfield  {journal} {\bibinfo  {journal} {\JIN}\ }\textbf {\bibinfo
  {volume} {5}},\ \bibinfo {pages} {31} (\bibinfo {year} {2021})}\BibitemShut
  {NoStop}%
\bibitem [{\citenamefont {Aliaga~\textit{et al}}(2014)}]{TheMINERvADetector}%
  \BibitemOpen
  \bibfield  {author} {\bibinfo {author} {\bibfnamefont {L.}~\bibnamefont
  {Aliaga~\textit{et al}}},\ }\href
  {https://doi.org/https://doi.org/10.1016/j.nima.2013.12.053} {\bibfield
  {journal} {\bibinfo  {journal} {\JNIMA}\ }\textbf {\bibinfo {volume} {743}},\
  \bibinfo {pages} {130} (\bibinfo {year} {2014})}\BibitemShut {NoStop}%
\bibitem [{\citenamefont {Makariev}(2007)}]{NA49}%
  \BibitemOpen
  \bibfield  {author} {\bibinfo {author} {\bibfnamefont {M.}~\bibnamefont
  {Makariev}} (\bibinfo {collaboration} {NA49 Collaboration}),\ }\href
  {https://doi.org/10.1063/1.2733107} {\bibfield  {journal} {\bibinfo
  {journal} {\JAIPCP}\ }\textbf {\bibinfo {volume} {899}},\ \bibinfo {pages}
  {203} (\bibinfo {year} {2007})}\BibitemShut {NoStop}%
\bibitem [{\citenamefont {Aduszkiewicz~\textit{et al}}(2019)}]{NA61}%
  \BibitemOpen
  \bibfield  {author} {\bibinfo {author} {\bibfnamefont {A.}~\bibnamefont
  {Aduszkiewicz~\textit{et al}}} (\bibinfo {collaboration} {NA61/SHINE
  Collaboration}),\ }\href {https://doi.org/10.1103/PhysRevD.100.112001}
  {\bibfield  {journal} {\bibinfo  {journal} {\JPRD}\ }\textbf {\bibinfo
  {volume} {100}},\ \bibinfo {pages} {112001} (\bibinfo {year}
  {2019})}\BibitemShut {NoStop}%
\bibitem [{\citenamefont {Aliaga~\textit{et al}}(2016)}]{NuMIBeamFlux}%
  \BibitemOpen
  \bibfield  {author} {\bibinfo {author} {\bibfnamefont {L.}~\bibnamefont
  {Aliaga~\textit{et al}}} (\bibinfo {collaboration} {{MINERvA
  Collaboration}}),\ }\href {https://doi.org/10.1103/PhysRevD.94.092005}
  {\bibfield  {journal} {\bibinfo  {journal} {\JPRD}\ }\textbf {\bibinfo
  {volume} {94}},\ \bibinfo {pages} {092005} (\bibinfo {year}
  {2016})}\BibitemShut {NoStop}%
\bibitem [{\citenamefont {Hatcher}(2012)}]{Dk2NuProposal}%
  \BibitemOpen
  \bibfield  {author} {\bibinfo {author} {\bibfnamefont {R.}~\bibnamefont
  {Hatcher}},\ }\href
  {https://indico.fnal.gov/event/5499/sessions/9700/attachments/59207/70395/flux_ntuple.pdf}
  {\emph {\bibinfo {title} {{Proposal for a Unified "Flux" N-tuple Format}}}},\
  \bibinfo {type} {Tech. Rep.}\ (\bibinfo {year} {2012})\BibitemShut {NoStop}%
\bibitem [{\citenamefont {Arg\"{u}elles~\textit{et al}}(2020)}]{AtmoLLP}%
  \BibitemOpen
  \bibfield  {author} {\bibinfo {author} {\bibfnamefont {C.}~\bibnamefont
  {Arg\"{u}elles~\textit{et al}}},\ }\href
  {https://doi.org/10.1007/jhep02(2020)190} {\bibfield  {journal} {\bibinfo
  {journal} {\JJHEP}\ }\textbf {\bibinfo {volume} {02}},\ \bibinfo {pages}
  {190} (\bibinfo {year} {2020})}\BibitemShut {NoStop}%
\bibitem [{\citenamefont {Rubbia}(2022)}]{rubbia_2022}%
  \BibitemOpen
  \bibfield  {author} {\bibinfo {author} {\bibfnamefont {A.}~\bibnamefont
  {Rubbia}},\ }\href {https://doi.org/10.1017/9781009023429.006} {\emph
  {\bibinfo {title} {{Phenomenology of Particle Physics}}}}\ (\bibinfo
  {publisher} {{Cambridge University Press}},\ \bibinfo {year} {2022})\
  Chap.~\bibinfo {chapter} {5}, p.\ \bibinfo {pages} {144–183}\BibitemShut
  {NoStop}%
\bibitem [{DUN(2023)}]{DUNEFluxesWebsite}%
  \BibitemOpen
  \href {https://glaucus.crc.nd.edu/DUNEFluxes/} {\bibinfo {title} {{DUNE
  neutrino flux files generated with G4LBNF}}} (\bibinfo {year}
  {2023})\BibitemShut {NoStop}%
\bibitem [{\citenamefont {Balantekin}\ \emph {et~al.}(2019)\citenamefont
  {Balantekin}, \citenamefont {de~Gouv{\^{e}}a},\ and\ \citenamefont
  {Kayser}}]{BahaDiracVsMajPol}%
  \BibitemOpen
  \bibfield  {author} {\bibinfo {author} {\bibfnamefont {A.~B.}\ \bibnamefont
  {Balantekin}}, \bibinfo {author} {\bibfnamefont {A.}~\bibnamefont
  {de~Gouv{\^{e}}a}},\ and\ \bibinfo {author} {\bibfnamefont {B.}~\bibnamefont
  {Kayser}},\ }\href {https://doi.org/10.1016/j.physletb.2018.11.068}
  {\bibfield  {journal} {\bibinfo  {journal} {\JPLB}\ }\textbf {\bibinfo
  {volume} {789}},\ \bibinfo {pages} {488} (\bibinfo {year}
  {2019})}\BibitemShut {NoStop}%
\bibitem [{\citenamefont {Levy}(2018)}]{Levy}%
  \BibitemOpen
  \bibfield  {author} {\bibinfo {author} {\bibfnamefont {J.-M.}\ \bibnamefont
  {Levy}},\ }\href {http://arxiv.org/abs/1805.06419} {\  (\bibinfo {year}
  {2018})},\ \Eprint {https://arxiv.org/abs/1805.06419} {arXiv:1805.06419
  [hep-ph]} \BibitemShut {NoStop}%
\bibitem [{\citenamefont {de~Gouvêa~\textit{et
  al}}(2021)}]{HNL3BodyDecaysPol}%
  \BibitemOpen
  \bibfield  {author} {\bibinfo {author} {\bibfnamefont {A.}~\bibnamefont
  {de~Gouvêa~\textit{et al}}},\ }\href
  {https://doi.org/10.1103/PhysRevD.104.015038} {\bibfield  {journal} {\bibinfo
   {journal} {\JPRD}\ }\textbf {\bibinfo {volume} {104}},\ \bibinfo {pages}
  {015038} (\bibinfo {year} {2021})}\BibitemShut {NoStop}%
\bibitem [{\citenamefont {Chun~\textit{et al}}(2019)}]{Chun_2019}%
  \BibitemOpen
  \bibfield  {author} {\bibinfo {author} {\bibfnamefont {E.~J.}\ \bibnamefont
  {Chun~\textit{et al}}},\ }\href {https://doi.org/10.1103/physrevd.100.095022}
  {\bibfield  {journal} {\bibinfo  {journal} {\JPRD}\ }\textbf {\bibinfo
  {volume} {100}},\ \bibinfo {pages} {095022} (\bibinfo {year}
  {2019})}\BibitemShut {NoStop}%
\bibitem [{\citenamefont {Shrock}(1978)}]{ShrockBeamBuckets}%
  \BibitemOpen
  \bibfield  {author} {\bibinfo {author} {\bibfnamefont {R.~E.}\ \bibnamefont
  {Shrock}},\ }\href {https://doi.org/10.1103/PhysRevLett.40.1688} {\bibfield
  {journal} {\bibinfo  {journal} {\JPRL}\ }\textbf {\bibinfo {volume} {40}},\
  \bibinfo {pages} {1688} (\bibinfo {year} {1978})}\BibitemShut {NoStop}%
\bibitem [{\citenamefont {Gallas~\textit{et al}}(1995)}]{FNALFMMF}%
  \BibitemOpen
  \bibfield  {author} {\bibinfo {author} {\bibfnamefont {E.}~\bibnamefont
  {Gallas~\textit{et al}}} (\bibinfo {collaboration} {{FMMF Collaboration}}),\
  }\href {https://doi.org/10.1103/PhysRevD.52.6} {\bibfield  {journal}
  {\bibinfo  {journal} {\JPRD}\ }\textbf {\bibinfo {volume} {52}},\ \bibinfo
  {pages} {6} (\bibinfo {year} {1995})}\BibitemShut {NoStop}%
\bibitem [{\citenamefont {Das}\ \emph {et~al.}(2016)\citenamefont {Das},
  \citenamefont {Konar},\ and\ \citenamefont {Majhi}}]{ADasCollider}%
  \BibitemOpen
  \bibfield  {author} {\bibinfo {author} {\bibfnamefont {A.}~\bibnamefont
  {Das}}, \bibinfo {author} {\bibfnamefont {P.}~\bibnamefont {Konar}},\ and\
  \bibinfo {author} {\bibfnamefont {S.}~\bibnamefont {Majhi}},\ }\href
  {https://doi.org/10.1007/jhep06(2016)019} {\bibfield  {journal} {\bibinfo
  {journal} {\JJHEP}\ }\textbf {\bibinfo {volume} {06}},\ \bibinfo {pages} {19}
  (\bibinfo {year} {2016})}\BibitemShut {NoStop}%
\bibitem [{\citenamefont {Fischer}(2022)}]{IceCubeHNL}%
  \BibitemOpen
  \bibfield  {author} {\bibinfo {author} {\bibfnamefont {L.}~\bibnamefont
  {Fischer}} (\bibinfo {collaboration} {{IceCube Collaboration}}),\ }\href
  {https://doi.org/10.22323/1.414.0190} {\bibfield  {journal} {\bibinfo
  {journal} {\JPOS}\ }\textbf {\bibinfo {volume} {ICHEP2022}},\ \bibinfo
  {pages} {190} (\bibinfo {year} {2022})}\BibitemShut {NoStop}%
\bibitem [{\citenamefont {Bellini~\textit{et al}}(2013)}]{BorexinoHNL}%
  \BibitemOpen
  \bibfield  {author} {\bibinfo {author} {\bibfnamefont {G.}~\bibnamefont
  {Bellini~\textit{et al}}} (\bibinfo {collaboration} {{Borexino
  Collaboration}}),\ }\href {https://doi.org/10.1103/physrevd.88.072010}
  {\bibfield  {journal} {\bibinfo  {journal} {\JPRD}\ }\textbf {\bibinfo
  {volume} {88}},\ \bibinfo {pages} {072010} (\bibinfo {year}
  {2013})}\BibitemShut {NoStop}%
\bibitem [{\citenamefont {Hagner~\textit{et al}}(1995)}]{BugeyReactorHNL}%
  \BibitemOpen
  \bibfield  {author} {\bibinfo {author} {\bibfnamefont {C.}~\bibnamefont
  {Hagner~\textit{et al}}},\ }\href {https://doi.org/10.1103/PhysRevD.52.1343}
  {\bibfield  {journal} {\bibinfo  {journal} {\JPRD}\ }\textbf {\bibinfo
  {volume} {52}},\ \bibinfo {pages} {1343} (\bibinfo {year}
  {1995})}\BibitemShut {NoStop}%
\bibitem [{\citenamefont {Verstraeten}(2021)}]{SoLidHNL}%
  \BibitemOpen
  \bibfield  {author} {\bibinfo {author} {\bibfnamefont {M.}~\bibnamefont
  {Verstraeten}},\ }\emph {\bibinfo {title} {{Search for sterile neutrinos in
  the eV and MeV mass range with the SoLid detector}}},\ \href
  {https://inspirehep.net/literature/2031135} {Ph.D. thesis},\ \bibinfo
  {school} {{Antwerp University}} (\bibinfo {year} {2021})\BibitemShut
  {NoStop}%
\bibitem [{\citenamefont {de~Gouv{\^{e}}a}\ and\ \citenamefont
  {Kobach}(2016)}]{deGouveaGlobalConstraints}%
  \BibitemOpen
  \bibfield  {author} {\bibinfo {author} {\bibfnamefont {A.}~\bibnamefont
  {de~Gouv{\^{e}}a}}\ and\ \bibinfo {author} {\bibfnamefont {A.}~\bibnamefont
  {Kobach}},\ }\href {https://doi.org/10.1103/physrevd.93.033005} {\bibfield
  {journal} {\bibinfo  {journal} {\JPRD}\ }\textbf {\bibinfo {volume} {93}},\
  \bibinfo {pages} {033005} (\bibinfo {year} {2016})}\BibitemShut {NoStop}%
\bibitem [{\citenamefont {Cortina Gil~\textit{et al}}(2020)}]{NA62EDominance}%
  \BibitemOpen
  \bibfield  {author} {\bibinfo {author} {\bibfnamefont {E.}~\bibnamefont
  {Cortina Gil~\textit{et al}}} (\bibinfo {collaboration} {{NA62
  Collaboration}}),\ }\href
  {https://doi.org/https://doi.org/10.1016/j.physletb.2020.135599} {\bibfield
  {journal} {\bibinfo  {journal} {\JPLB}\ }\textbf {\bibinfo {volume} {807}},\
  \bibinfo {pages} {135599} (\bibinfo {year} {2020})}\BibitemShut {NoStop}%
\bibitem [{\citenamefont {Cortina Gil~\textit{et al}}(2021)}]{NA62MDominance}%
  \BibitemOpen
  \bibfield  {author} {\bibinfo {author} {\bibfnamefont {E.}~\bibnamefont
  {Cortina Gil~\textit{et al}}} (\bibinfo {collaboration} {{NA62
  Collaboration}}),\ }\href {https://doi.org/10.1016/j.physletb.2021.136259}
  {\bibfield  {journal} {\bibinfo  {journal} {\JPLB}\ }\textbf {\bibinfo
  {volume} {816}},\ \bibinfo {pages} {136259} (\bibinfo {year}
  {2021})}\BibitemShut {NoStop}%
\bibitem [{\citenamefont {Bondarenko~\textit{et al}}(2020)}]{BondarenkoBBN}%
  \BibitemOpen
  \bibfield  {author} {\bibinfo {author} {\bibfnamefont {K.}~\bibnamefont
  {Bondarenko~\textit{et al}}},\ }\href
  {https://doi.org/10.1007/jhep03(2020)118} {\bibfield  {journal} {\bibinfo
  {journal} {\JJHEP}\ }\textbf {\bibinfo {volume} {03}},\ \bibinfo {pages}
  {118} (\bibinfo {year} {2020})}\BibitemShut {NoStop}%
\bibitem [{\citenamefont {Boyarsky~\textit{et al}}(2021)}]{BoyarskyBBN}%
  \BibitemOpen
  \bibfield  {author} {\bibinfo {author} {\bibfnamefont {A.}~\bibnamefont
  {Boyarsky~\textit{et al}}},\ }\href
  {https://doi.org/10.1103/PhysRevD.104.023517} {\bibfield  {journal} {\bibinfo
   {journal} {\JPRD}\ }\textbf {\bibinfo {volume} {104}},\ \bibinfo {pages}
  {023517} (\bibinfo {year} {2021})}\BibitemShut {NoStop}%
\bibitem [{\citenamefont {Boyanovsky}(2014)}]{veeMultipleHNL}%
  \BibitemOpen
  \bibfield  {author} {\bibinfo {author} {\bibfnamefont {D.}~\bibnamefont
  {Boyanovsky}},\ }\href {https://doi.org/10.1103/physrevd.90.105024}
  {\bibfield  {journal} {\bibinfo  {journal} {\JPRD}\ }\textbf {\bibinfo
  {volume} {90}},\ \bibinfo {pages} {105024} (\bibinfo {year}
  {2014})}\BibitemShut {NoStop}%
\bibitem [{\citenamefont {Tastet}\ \emph {et~al.}(2021)\citenamefont {Tastet},
  \citenamefont {Ruchayskiy},\ and\ \citenamefont {Timiryasov}}]{MultipleHNL}%
  \BibitemOpen
  \bibfield  {author} {\bibinfo {author} {\bibfnamefont {J.-L.}\ \bibnamefont
  {Tastet}}, \bibinfo {author} {\bibfnamefont {O.}~\bibnamefont {Ruchayskiy}},\
  and\ \bibinfo {author} {\bibfnamefont {I.}~\bibnamefont {Timiryasov}},\
  }\href {https://doi.org/10.1007/jhep12(2021)182} {\bibfield  {journal}
  {\bibinfo  {journal} {\JJHEP}\ }\textbf {\bibinfo {volume} {12}},\ \bibinfo
  {pages} {182} (\bibinfo {year} {2021})}\BibitemShut {NoStop}%
\bibitem [{\citenamefont {Bondarenko~\textit{et
  al}}(2021)}]{Resurrection_belowKaon}%
  \BibitemOpen
  \bibfield  {author} {\bibinfo {author} {\bibfnamefont {K.}~\bibnamefont
  {Bondarenko~\textit{et al}}},\ }\href
  {https://doi.org/10.1007/jhep07(2021)193} {\bibfield  {journal} {\bibinfo
  {journal} {\JJHEP}\ }\textbf {\bibinfo {volume} {07}},\ \bibinfo {pages}
  {193} (\bibinfo {year} {2021})}\BibitemShut {NoStop}%
\bibitem [{\citenamefont {Drewes~\textit{et al}}(2018)}]{NA62MultipleHNL}%
  \BibitemOpen
  \bibfield  {author} {\bibinfo {author} {\bibfnamefont {M.}~\bibnamefont
  {Drewes~\textit{et al}}},\ }\href {https://doi.org/10.1007/jhep07(2018)105}
  {\bibfield  {journal} {\bibinfo  {journal} {\JJHEP}\ }\textbf {\bibinfo
  {volume} {07}},\ \bibinfo {pages} {105} (\bibinfo {year} {2018})}\BibitemShut
  {NoStop}%
\bibitem [{HNL(2023)}]{HNLGENIE}%
  \BibitemOpen
  \href {https://github.com/kjplows/Generator} {\bibinfo {title} {{GENIE
  BeamHNL module (expected GENIE release v3.4.0)}}} (\bibinfo {year}
  {2023})\BibitemShut {NoStop}%
\bibitem [{GEN(2023)}]{GENIEWebsite}%
  \BibitemOpen
  \href {http://www.genie-mc.org/} {\bibinfo {title} {{GENIE Event Generator \&
  Global Analysis of Neutrino}}} (\bibinfo {year} {2023})\BibitemShut {NoStop}%
\bibitem [{\citenamefont {Press~\textit{et al}}(2007)}]{NumericalRecipes3rd}%
  \BibitemOpen
  \bibfield  {author} {\bibinfo {author} {\bibfnamefont {W.~H.}\ \bibnamefont
  {Press~\textit{et al}}},\ }\href@noop {} {\emph {\bibinfo {title} {{Numerical
  recipes : the art of scientific computing}}}},\ \bibinfo {edition} {3rd}\
  ed.\ (\bibinfo  {publisher} {{Cambridge University Press}},\ \bibinfo {year}
  {2007})\BibitemShut {NoStop}%
\end{thebibliography}%

\end{document}